\documentclass[paper,12pt]{report}

\usepackage{a4wide}
\usepackage{amsthm,amssymb}
\usepackage{amsbsy,amsfonts,amsmath,amscd}
\usepackage{epic}
\usepackage{graphicx}
\usepackage{wrapfig}
\usepackage{array}
\usepackage{bm}
\usepackage{epsfig,multicol}
\usepackage{verbatim}
\usepackage{enumerate}
\newcommand{\ads}{AdS_5\times S^5}
\newcommand{\beq}{\begin{equation}}
\newcommand{\eeq}{\end{equation}}
\newcommand{\beqa}{\begin{eqnarray}}
\newcommand{\eeqa}{\end{eqnarray}}
\newcommand{\N}{\mathcal{N}}

\newcommand{\psua}{\mathfrak{psu}(2,2|4)}
\newcommand{\suzz}{\mathfrak{su}(2|2)}
\newcommand{\HS}{S}
\newcommand{\sfrac}[2]{{\textstyle\frac{#1}{#2}}}
\newcommand{\inddowns}[1]{^{\mathrm{\scriptscriptstyle #1}}}
\newcommand{\lr}[1]{\left(#1\right)}
\newcommand{\Lagr}{\mathcal{L}}
\newcommand{\fldW}{\mathcal{W}}

\newcommand{\fldA}{\mathcal{A}}
\newcommand{\Op}{\mathcal{O}}
\newcommand{\fldZ}{\mathcal{Z}}
\newcommand{\fldX}{\mathcal{X}}
\newcommand{\fldY}{\mathcal{Y}}
\newcommand{\fldU}{\mathcal{U}}
\newcommand{\fldV}{\mathcal{V}}
\newcommand{\fldF}{\mathcal{F}}
\newcommand{\cder}{\mathcal{D}}
\newcommand{\Tr}{\mathop{\rm Tr}}

\newcommand{\oa}{\mathbf{a}}
\newcommand{\ob}{\mathbf{b}}
\newcommand{\oc}{\mathbf{c}}
\newcommand{\od}{\mathbf{d}}
\newcommand{\gaba}{\gamma\inddowns{ABA}}
\newcommand{\M}{M}
\newcommand{\LoverlogM}{j}
\newcommand{\holegap}{c}
\newcommand{\magnonrho}{\rho_{\rm m}}

\newcommand{\korrho}{\rho_K}

\newcommand{\holerho}{\rho_{\rm h}}
\newcommand{\holebarrho}{\bar \rho_{\rm h}}

\newcommand{\oldtildesigma}{\hat{\sigma}}

\newtheorem{de}{Definition}

\begin{document}
\begin{flushright}\footnotesize
\texttt{\textbf{Imperial-TP-AR-2009-2}}\\
\vspace{0.5cm}
\end{flushright}
\setcounter{footnote}{0}

\begin{center}
{\Large\textbf{\mathversion{bold} Integrability and the AdS/CFT correspondence}}
\vspace{15mm}

{\sc Adam Rej}\\[1mm]

{\it Blackett Laboratory, Imperial College, London SW7 2AZ, U.K.}\\[3mm]
\texttt{adam.rej@imperial.ac.uk}\\[12mm]

\textbf{Abstract}\\[5mm]
\end{center}

\noindent{In this article we review the recently discovered asymptotic integrability in the planar\\ $\N=4$ SYM theory and discuss its breakdown  beyond the asymptotic region due to the wrapping interactions. We also discuss novel dynamical tests of the AdS/CFT correspondence one can perform in the special cases when the wrapping interactions may be neglected.}
\tableofcontents
\chapter{Introduction}
The AdS/CFT correspondence, proposed by Maldacena \cite{Maldacena:1997re} in 1997, stating that the string theory on the $AdS_5 \times S^5$ is dual to $\mathcal{N}=4$ supersymmetric Yang-Mills in four dimensions, has become one of the prime subjects of interest in gauge and string theory. Since it is a strong/weak coupling duality, it offers the unique possibility to investigate a four dimensional interacting gauge theory beyond perturbation theory. 

In the last few years, thanks to asymptotic integrability, a great progress has been made in confirming the above conjecture in the planar limit. This led, at the same time, to precious insights and to a deeper understanding of either of the related theories. Very briefly the history of the asymptotic integrability in the planar $\mathcal{N}=4$ SYM theory can be summarized as follows.

In the groundbreaking paper \cite{Minahan:2002ve}, the one-loop integrability of the dilatation operator in certain subsectors of the gauge side of the AdS/CFT correspondence was discovered. Later on \cite{Beisert:2003jj} the complete one-loop dilatation operator has been found and the corresponding one-loop Bethe equations were written down \cite{Beisert:2003yb}. After many non-trivial steps \cite{fussnote}, the form of the all-loop asymptotic Bethe equations (ABE) was conjectured \cite{Beisert:2005fw} up to the so called dressing factor \cite{Arutyunov:2004vx}, which only contributes starting from the four-loop order. Subsequently, relying on the crossing equation proposed in \cite{Janik:2006dc} and assuming certain transcendentality properties, it was possible to uniquely fix this factor \cite{Beisert:2006ez}. In this way the asymptotic spectrum of the planar $\mathcal{N}=4$ SYM theory has been completely determined. The asympoticity of these equations means that for a generic operator with $L$ constituent fields the corresponding anomalous dimension can be calculated correctly up to the $\mathcal{O}(g^{2L})$ order.

The appearance of the asymptotic integrability suggests that the spectral problem of the AdS/CFT correspondence is intimately related to the dynamics of some lower dimensional model (integrable one-dimensional spin chain? two-dimensional sigma model?). This relation is very surprising, in particular, from the perspective that both the string theory on $AdS_5 \times S^5$ as well as the $\N=4$ gauge theory are higher-dimensional quantum theories! It is the purpose of this review to explain how this link can be established for ``long'' operators. For operators of arbitrary length such a link cannot be established before revealing ``true'' nature of the lower dimensional model.

In this article we will lay all the necessary foundations (chapters 1-3) required to present the set of the asymptotic Bethe equations which, up to the wrapping order, determine the anomalous dimension of \textit{any} single trace operator of the planar $\N=4$ supersymmetric Yang-Mills theory. In chapter 4 we will discuss analytic properties of operators belonging to a particular subsector of the theory, the $\mathfrak{sl}(2)$ sector. The shortest of these operators describe the leading breaking of the Bjorken scaling and their leading anomalous dimension at unphysical negative values of the spin is determined by the celebrated BFKL equation. In chapter 5 we will discuss this relation and use the results of chapter 4 in order to check the veracity of the asymptotic Bethe equations beyond the asymptotic region. The asymptotic Bethe equations fail this test and therefore must be modified at and beyond the wrapping order. On the other hand, since the wrapping order is controlled by the length $L$, it is possible to study non-petrubative behaviour of anomalous dimensions in the limit $L \to \infty$. This question is the subject of chapter 6, where novel dynamical tests of the AdS/CFT correspondence  are presented that one can perform with help of the asymptotic Bethe equations. In chapter 7 we discuss an interesting observation relating the Bethe equations of the one of the subsectors of the full theory to the Bethe equations of the Hubbard model. This relationship, even though valid only to the first three orders of perturbation theory, suggests that a well-defined short range model may be capable to account for the wrapping interactions.\\
\\
\textit{Note added:} This work is based on the author's PhD thesis submitted to Humboldt University, Berlin.\\
\\
\textit{Note added:} After this review has been written, the Y-system for planar AdS/CFT correspondence \cite{Gromov:2009tv} and TBA equations  \cite{Bombardelli:2009ns}-\cite{Arutyunov:2009ur} have been proposed. Both sets of equations constitute a significant step forward in non-perturbative formulation of the spectral problem of the $\N=4$ gauge theory. We will, however, not discuss these recent proposals in this article.
\chapter{The $\N=4$ Super Yang-Mills Theory and the AdS/CFT Correspondence}
In this chapter we will briefly review the gauge theory side of the AdS/CFT duality and subsequently formulate the precise correspondence.  
\section{Gauge Theory Side}
\subsection{The $\N=4$ Super Yang-Mills Theory}
The $\N=4$ super Yang-Mills theory (SYM) is a maximally supersymmetric gauge theory in four dimensions without spin-two fields (gravitons). In what follows we will only review the aspects of this theory that are necessary to study the integrable structures appearing in the planar limit. For more details and references see  \cite{Beisert:2004ry}.

The constituents of the $\N=4$ gauge theory are: six scalar fields $\Phi_{m}$, four fermionic fields $\Psi_{\alpha a}$, $\dot\Psi_{\dot\alpha}^a$ and the gauge fields $A_{\mu}$. Anticipating the transformation rules under gauge transformations, we will regard the covariant derivative $\cder_{\mu}$,
\beq
\cder_\mu:=\partial_\mu-ig\fldA_\mu,
\qquad
\cder_\mu \fldW:=[\cder_\mu,\fldW]=
\partial_\mu \fldW-ig \fldA_\mu \fldW+ig\fldW\fldA_\mu,
\eeq
rather then the gauge field as the fundamental field. Here, $\fldW$ stands for a representative of the fundamental fields
\beq\label{fundfeld}
\fldW=(\cder_\mu,\Psi_{\alpha a},\dot\Psi_{\dot\alpha}^a,\Phi_{m}).
\eeq
The greek indices belong to the Lorenz algebra $\mathfrak{so}(3,1)=\mathfrak{su}(2) \times \mathfrak{su}(2)$, with  $\mu, \nu, \ldots$ running from $1$ to $4$ and $\alpha, \beta, \ldots$ as well as $\dot \alpha, \dot \beta, \ldots$ taking values $1,2$. The latin indices correspond to the R-symmetry algebra $\mathfrak{so}(6)=\mathfrak{su}(4)$. The scalar fields are assumed to be in the fundamental representation of $\mathfrak{so}(6)$ and thus $m,n, \ldots$ run from $1$ to $6$, whereas fermions transform in the spinor representation and the corresponding indices $a,b,\ldots$ take values $1,\ldots,4$. All fields are assumed to be in the adjoint representation of the gauge group $U(N)$, and under the local transformations $U(x)\in U(N)$ they transform as 
\beq
\fldW^i _{\,j} \mapsto U^i _{\, k} \, \fldW^k _{\ l} \lr{U^{-1}}^l _{\, j},\qquad
\lr{\fldA_\mu}^i _{\, j} \mapsto U^i _{\, k} \lr{\fldA_\mu}^k _{\ l} \lr{U^{-1}}^l _{\,j}-ig^{-1}\, \partial_\mu U^i_{\,k} \,\lr{U^{-1}}^k _{\ j}\,,
\eeq
with the repeating indices being summed over. In the following we will suppress the $U(N)$ indices. The field strength tensor $\fldF_{\mu\nu} $ is defined through
\beq
\fldF_{\mu\nu}=ig^{-1}[\cder_\mu,\cder_\nu]=
\partial_\mu \fldA_\nu-\partial_\nu \fldA_\mu-ig[\fldA_\mu,\fldA_\nu]
\eeq
and transforms canonically under local gauge transformations
\beq
\fldF_{\mu\nu}\mapsto U\fldF_{\mu\nu} U^{-1}\,.
\eeq

The Lagrangian of the $\N=4$ SYM theory is composed of terms of dimension four only
\beqa \label{lagrangefunktion}
\Lagr_{\mbox{\tiny YM}}(x,g)&=&\Tr \bigg(
\frac{1}{4} \fldF^{\mu\nu}\fldF_{\mu\nu}
+\frac{1}{2} \cder^\mu\Phi^n\cder_\mu\Phi_n
-\frac{1}{4}g^2  [\Phi^m,\Phi^n][\Phi_m,\Phi_n]
\nonumber \\ \nonumber
&&+ \dot\Psi^a_{\dot\alpha}\sigma_\mu^{\dot\alpha\beta}\cder^\mu\Psi_{\beta a}
-\frac{1}{2}ig \Psi_{\alpha a}\sigma_m^{ab}\varepsilon^{\alpha\beta}
[\Phi^m,\Psi_{\beta b}]
-\frac{1}{2}ig \dot\Psi^a_{\dot\alpha}
\sigma^m_{ab}\varepsilon^{\dot\alpha\dot\beta}
[\Phi_m,\dot\Psi^b_{\dot\beta}] \bigg).
\eeqa
In the above formula $\sigma^\mu$ and $\sigma^m$ denote the chiral gamma matrices in four und six dimensions. 

The equations of motion following from this Lagrangian are invariant under the $\N=4$ super Poincare symmetry algebra. The infinitesimal bosonic shifts can be parametrized by $s^\mu, \mu=0,1,2,3$ 
\beq \label{param1}
\delta_{s}=s^{\mu} P_{\mu}\,,
\eeq
and under this transformation the set \eqref{fundfeld} transforms as
\beqa \label{supertrans1}
\delta_{s} \cder_\mu &=&ig s^\nu\fldF_{\mu\nu},
\nonumber \\
\delta_{s} \Phi_m &=&s^\mu \cder_\mu\Phi_m,
\nonumber \\
\delta_{s} \Psi_{\alpha a}&=&s^\mu \cder_\mu \Psi_{\alpha a}
,
\nonumber \\
\delta_{s} \dot\Psi^a_{\dot\alpha}&=&s^\mu \cder_\mu \dot\Psi^a_{\dot\alpha}.
\eeqa
Here, we have assumed an adjoint action of the generators on the fields
\beq \label{adjwirk}
\delta_{s} \fldW := \left[\delta_{s}, \fldW \right] \,.
\eeq

The infinitesimal fermionic translations are parametrized by Grassman variables
\beq \label{param2}
\delta_{w}=w^{\alpha}_a Q^a_{\alpha}\,,
\eeq 
and the corresponding field transformations read
\beqa \label{supertrans2}
\delta_{w} \cder_\mu &=&
 ig w^\alpha_{a}\varepsilon_{\alpha\beta}\sigma_\mu^{\beta\dot\gamma}\dot\Psi^a_{\dot\gamma},
\nonumber \\
\delta_{w} \Phi_m &=&
w^\alpha_{a}\sigma_m^{ab}\Psi_{\alpha b},
\nonumber \\
\delta_{w} \Psi_{\alpha a}&=&
\frac{1}{2} ig \sigma^m_{ab}\sigma_n^{bc}\varepsilon_{\alpha\beta}w^\beta_{c} [\Phi_m,\Phi^n]-\frac{1}{2} \sigma^\mu_{\alpha\dot\beta}\varepsilon^{\dot\beta\dot\gamma}\sigma^\nu_{\dot\gamma\delta}
w^\delta_{a}\fldF_{\mu\nu}
\nonumber \\
\delta_{w} \dot\Psi^a_{\dot\alpha}&=&\,\sigma_n^{ab}\sigma^\mu_{\dot\alpha\beta}
w^\beta_{b}\cder_\mu\Phi^n\,.
\eeqa
The transformation laws under the action of $\bar{Q}$ can be obtained by replacing
\beq
\Psi \mapsto \dot \Psi\,, \qquad \qquad w \mapsto \dot w
\eeq
in \eqref{supertrans2}.\\

Using the explicit form of the parametrizations \eqref{param1}, \eqref{param2} and with help of the equations of motion  one can derive from \eqref{supertrans1}, \eqref{supertrans2} and the respective conjugated transformations the following algebraic relations between the generators
\beq
\begin{array}{crclcrcl}
\{Q^a_{\alpha},Q^b_{\beta}\}&=&
-2ig \varepsilon_{\alpha\beta}\sigma_m^{ab}\Phi^m\,,
&\qquad&
[P_\mu,Q^a_\alpha]&=&
-ig \varepsilon_{\alpha\beta}\sigma^{\beta\dot\gamma}_\mu \dot\Psi^a_{\dot\gamma}\,,
\\[4pt]
\{\bar{Q}_{\dot\alpha a},\bar{Q}_{\dot\beta b}\}&=&
-2ig \varepsilon_{\dot\alpha\dot\beta}\sigma^m_{ab}\Phi_m\,,
&\qquad&
[P_\mu,\bar{Q}_{\dot\alpha a}]&=&
-ig \varepsilon_{\dot\alpha\dot\beta}\sigma^{\dot\beta\gamma}_\mu \Psi_{\gamma a}\,,
\\
\{Q^a_{\alpha},\bar{Q}_{b\dot\beta}\}&=&
2 \delta^a_b\sigma^\mu_{\alpha\dot\beta} P_\mu\,,
&&
[P_\mu,P_\nu]&=& -ig\fldF_{\mu\nu}\,.
\end{array}
\eeq
Hereafter, we will use the chiral notation only
\beq
\fldW=(\cder_{\dot\alpha\beta},
\Phi_{ab},
\Psi_{\alpha b},
\dot\Psi^b_{\dot\alpha},
\fldF_{\alpha\beta},
\dot\fldF_{\dot\alpha\dot\beta})\,.
\eeq
The relation to the canonical notation is given by
\beqa \label{neudefinition}
\cder_\mu &\sim& \sigma_{\mu}^{\dot \alpha\beta} \cder_{\dot\alpha\beta},
\nonumber \\
\fldF_{\mu\nu} &\sim&
\sigma_\mu^{\alpha\dot\gamma}\varepsilon_{\dot\gamma\dot\delta}
\sigma_\nu^{\dot\delta\beta}
\fldF_{\alpha\beta}+
\sigma_\mu^{\dot\alpha\gamma}\varepsilon_{\gamma\delta}
\sigma_\nu^{\delta\dot\beta}
\dot\fldF_{\dot\alpha\dot\beta},
\nonumber \\
\Phi_m &\sim& \sigma_{m}^{ba}\Phi_{ab}.
\eeqa
In a similar way we redefine the translation operator $P_{\mu}$ (and subsequently also the conformal boost operator $K^{\mu}$)
\beq
P_{\mu} \sim \sigma_{\mu}^{\dot \alpha \beta} P_{\dot\alpha\beta} \,.
\eeq
The advantage of this change of notation is the occurrence of the ``fermionic'' indices of $\mathfrak{su}(2)\times \mathfrak{su}(2)$ and $\mathfrak{su}(4)$ only.
\subsection{The Conformal and Superconformal Symmtery}

The $\N=4$ gauge theory posses a further classical symmetry, the conformal symmetry \cite{Sohnius:1981sn}. What, however, distinguishes this field theory from other massless field theories is  the preservation of the conformal symmetry after the quantization. A simple consequence of this fact is the vanishing of the beta function
\beq
\beta=\mu\,\frac{\partial g}{\partial \mu}=0\,.
\eeq
Thus, the charge does not get renormalized and the momentum-energy tensor remains traceless. The only divergent quantities are the wave functions of the fields meaning that the scaling dimensions of the operators receive quantum corrections. Indeed, the conformal symmetry severely constraints the form of the two-point correlations functions. While for any two local scalar fields $\hat{\mathcal{O}}_A(x_1)$, $\hat{\mathcal{O}}_B(x_2)$ the Poincar\'e symmetry implies that
\beq
\langle \hat{\mathcal{O}}_A(x_1) \hat{\mathcal{O}}_B(x_2)\rangle= f_{AB}(x_1-x_2)\, ,
\eeq
with $f(x)$ being an arbitrary scalar function, the conformal symmetry constraint its form to
\beq \label{zweipunkt}
\langle \hat{\mathcal{O}}_A(x_1) \hat{\mathcal{O}}_B(x_2)\rangle= \frac{C_{AB}(g)}{{\vert x_1-x_2 \vert}^{2\Delta(g)}}\,.
\eeq
The quantity $\Delta(g)$ is the scaling dimension of an operator and generically receives quantum corrections
\beq
\Delta(g)=\Delta_0+\gamma(g)\, .
\eeq
Here, $\Delta_0$ denotes the canonical dimension and $\gamma(g)$ the anomalous part. Standard arguments of the renormalization theory relate $\gamma(g)$ of an operator to the corresponding wave function renormalization  $\frac{d}{d \mu} \log{Z}$. The conformal symmetry also constraints up to a constant  the form of the three-point function
\beq \label{dreipunkt}
\langle \hat{\mathcal{O}}_A(x_1)\,\hat{\mathcal{O}}_B(x_2)\, \hat{\mathcal{O}}_C(x_3)\rangle=
\frac{C_{ABC}(g)}
{|x_{12}|^{\Delta_A+\Delta_B-\Delta_C}\,
 |x_{23}|^{\Delta_B+\Delta_C-\Delta_A}\,
 |x_{31}|^{\Delta_C+\Delta_A-\Delta_B}}\,.
\eeq
The correlation functions of four and more fields are not fully determined by the conformal symmetry.

Although the dimension of any operator may be found directly from the correlation functions \eqref{zweipunkt} and \eqref{dreipunkt} it is much more advantageous, as it will become clear later, to consider the dilatation operator. The latter is one of the generators of the conformal algebra and its eigenvalues are precisely the scaling dimensions 
\beq
D\,\hat{\mathcal{O}}_A(x)= \Delta(g)\, \hat{\mathcal{O}}_A(x) \,.
\eeq
The two spacetime symmetry algebras, the conformal algebra and the supersymmetry algebra (together with the $\mathfrak{so}(6)$ flavor algebra), combine to the superconformal algebra $\psua$. The generators and the structure relations of this algebra are presented and discussed in chapter \ref{sec:integrabilitaet}. It is the unparalleled amount of the symmetries at the quantum level that is responsible for many interesting properties of the $\N=4$ gauge theory. 
\subsection{The 't Hooft Limit} \label{sec:tHooft}
In the renowned article of 't Hooft \cite{'t Hooft:1973jz} a novel limit for gauge theories with the $U(N)$ gauge groups was proposed, namely $N \mapsto \infty$. In this section we will briefly discuss this limit for the $\N=4$ SYM theory.

\begin{figure}
\begin{center}
\includegraphics[scale=0.7]{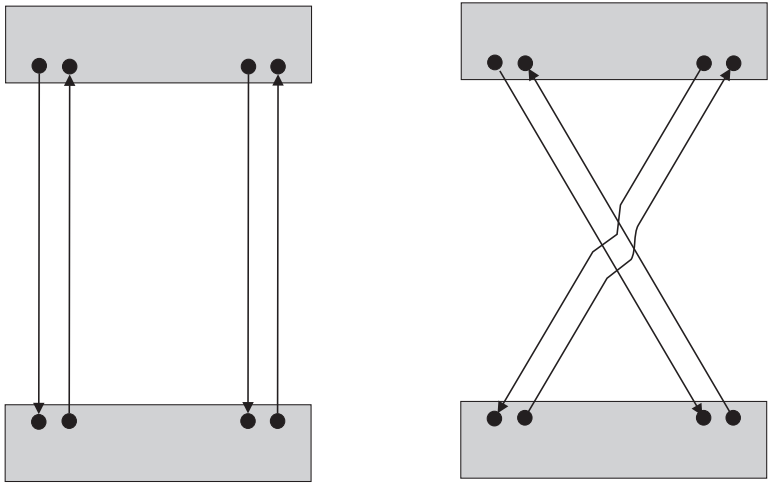}
\caption{An example of a planar and non-planar Feynman diagram. The grey box represents a single-trace operator composed of  two elementary fields.}\label{fig:pun}
\end{center}
\end{figure}

A common normalization of the actions of the Yang-Mills theories is 
\beq \label{N4SYMTwirkung}
S=\frac{2}{g^2_{\mbox{\tiny YM}}}\int d^4 x\,\Lagr_{\mbox{\tiny YM}}(x,1).
\eeq
For the purpose of the large $N$ expansion, it is convenient to rescale all fields with $g=\frac{1}{4\pi} g_{\mbox{\tiny YM}} \sqrt{N}$,
\beq
\fldW \mapsto g\,\fldW \,,
\eeq
leading to
\beq
S=N \int \frac{d^4 x}{8\pi^2}\,\Lagr_{\mbox{\tiny YM}}(x,g).
\eeq
All fields in the theory are assumed to be in a hermitian adjoint representation of the gauge group $U(N)$ and thus can be represented by $N \times N$ hermitian matrices $\fldW^i_{\phantom{\,}j}$. To each upper index one assigns an incoming and to each lower index an outgoing arrow\footnote{The reader should recall that the adjoint representation of $U(N)$ can be constructed from the tensor product of the fundamental and the antifundamental representation.} 
\begin{center}
\setlength{\unitlength}{1pt}
\small\thicklines
\begin{picture}(50,20)(0,-20)
\put(0,0){\makebox(0,0)[t]{$\fldW^i_{\,j}$}}
\put(60,-3){\vector(-1,0){40}}
\put(20,-9){\vector(1,0){40}}
\end{picture}
\end{center}
In this notation every propagator is depicted by two parallel lines. For gauge invariant operators all indices must be contracted resulting in the ``fat'' Feynman diagrams (see figure  \ref{fig:pun}). In this case the contribution of each diagram can be written as \cite{'t Hooft:1973jz}
\beq
\# \,N^{2-2g_e} \, (g^2)^{\ell}\,,
\eeq
where $\#$ is a number, $g_e$ denotes the genus of the surface spanned by the diagram and $\ell$ counts the loop oder. Therefore any physical quantity $\eta$ must admit the following expansion
\beq \label{gNentwicklung}
\eta=\eta_0+\sum^{\infty}_{j=1} g^j \sum^\infty _{g_e=0} \frac{1}{N^{2g_e-c}} \eta_{(j,g_e)}\,,
\eeq
with $\eta_0$ being the classical contribution. This applies, in particular, to the scaling dimension $\Delta(g,N)$ of the operators. 

An interesting limit, as may be seen from \eqref{gNentwicklung}, emerges when $N\to \infty$, $g_{\mbox{\tiny YM}}\to 0$ and $g^2=\frac{g^2_{\mbox{\tiny YM}}\,N}{16 \pi^2}=\mbox{const}$. This limit is called the planar limit, due to the fact that only the planar diagrams, that is with $g_e=0$ (see figure \ref{fig:pun}), contribute.
\subsection{The Physical Operators} \label{sec:physop}
The basis of local physical operators of the theory is spanned by multiple products of the single-trace operators
\beq \label{mehrfachespuren}
\hat{\mathcal{O}}=\Tr (\Omega_1 \Omega_2 \ldots \Omega_n)\Tr (\Omega_{n+1} \ldots \Omega_{n+m})\ldots \, \, .
\eeq
Taking into account that the covariant derivative must always act on a field one finds
\beq \label{Feldermenge}
\Omega_j \in (\cder^k \fldF,\cder^k \Psi,\cder^k \Phi,\cder^k \dot \Phi,\cder^k \dot \fldF) \, .
\eeq
It can be shown, see e.g. \cite{Beisert:2003jj}, that this set of fields, under the assumption that the Lorenz indices may not be contracted, is irreducible with respect to Bianchi identities, equations of motion, etc. 

In the planar limit the interactions between operators in different traces in \eqref{mehrfachespuren} are suppressed and it is sufficient to consider the single-trace operators
\beq \label{einzelspur}
\Tr (\Omega_1 \Omega_2 \ldots \Omega_p)\,.
\eeq
In this article we will confine ourselves to this case.
\section{Formulation of the AdS/CFT Correspondence} \label{sec:adscftformulierung}
The AdS/CFT correspondence states that the $\N=4$ SYM theory is dynamically equivalent to the IIB superstring theory on $AdS_5\times S^5$ if the parameters of the both theories are identified as follows
\beq \label{paramident}
g_s=\frac{4\pi g^2}{N}, \qquad \frac{R^2}{\alpha'}=4\pi g \,.
\eeq
In this formula $R$ denotes the radii of $AdS_5$ and $S^5$ and $g_s$ stands for the string coupling constant. Stated more precisely, to each gauge invariant operator of the $\N=4$ gauge theory there corresponds a dual string state such that all associated physical quantities coincide. 

Let $\vert \mathcal{O} \rangle$ be an arbitrary state of the string theory and $E(g_s, \frac{R^2}{\alpha'})$ the eigenvalue of this state with respect to the string Hamiltonian $\hat{H}$ defined as an operator conjugated to the $AdS_5$ time variable
\beq
H \vert \mathcal{O} \rangle = E(g_s, \frac{R^2}{\alpha'}) \vert \mathcal{O} \rangle\,.
\eeq
By comparing the string and gauge theory symmetry algebras and their representations, one concludes that the string Hamiltonian should correspond to the dilatation operator on the gauge theory side. Therefore, there should a exist a state $\mathcal{O}(x)$ such that
\beq
D\,\mathcal{O}(x)= \Delta(g,N)\, \mathcal{O}(x)
\eeq
and
\beq \label{EgD}
E(g_s, \frac{R^2}{\alpha'})=\Delta(g,N).
\eeq
In the planar limit this statement reduces to
\beq \label{EgDp}
E(0, \frac{R^2}{\alpha'})=\Delta(g,\infty).
\eeq
According to the above formula scaling dimensions of the planar gauge theory should be identified with energies of the free string theory! Unfortunately, the quantization of string theory on $\ads$ product space has not been understood so far and therefore a direct verification of \eqref{EgD} and \eqref{EgDp} is not feasible. Despite this fact, it is possible to test this equality in certain special cases, some of which we will discuss later.

The strongest version of the AdS/CFT correspondence states that the duality between the both theories holds for arbitrary values of $N$ and $g \sim \lambda=g^2_{\rm YM} N$ in \eqref{paramident}. In particular thus for small gauge groups, as for example $SU(2)$ or $SU(3)$, and for strongly coupled string theory. A more modest formulation claims the equivalence in the planar limit $N \to \infty, \,g_{\rm YM} \to 0$ only. One also cannot exclude that the equivalence of the both theories holds only asymptotically so that the $\Op\lr{\frac{1}{N}}$ expansion of the gauge and string theory agrees and the discrepancy of the results may be seen only after resummation.
\chapter{Foundations of Integrability}
The integrability of a physical system has established itself as an important concept in the mechanics and quantum mechanics since it usually leads to deep insights into the dynamics of the system. Although it does not necessarily imply  solvability, integrability puts severe constraints on the physical processes excluding in particular any chaotic behavior. 
\section{Integrability}
In this section we will introduce the concept of classical integrability and discuss its possible extensions to the quantum case.

Let $H(\bf{q},\bf{p})$ be a Hamiltonian of a system with $N$-dimensional phase space. The time evolution of the system is then determined through the Hamilton equations
\beqa
\dot q_j= \frac{\partial H}{\partial p_j} \qquad j=1,\ldots, N \,,\nonumber \\
\dot p_j=-\frac{\partial H}{\partial q_j} \qquad j=1,\ldots, N\,.
\eeqa
For the Hamilton mechanics the definition of the integrability may be formulated in the following way.
\begin{de}\label{klint}
A system is integrable iff there exist $N$ algebraically independent constants of motion, that is $N$ functions $(L_1,\ldots,L_n)$, which satisfy the following two conditions
\begin{enumerate}[i)]
\item \begin{center} $\forall i,j: \qquad \{L_i,L_j\}=0$,\end{center}
\item \begin{center}$\forall i: \qquad \{H,L_i\}=0\,.$ \end{center}
\end{enumerate}
\end{de}
Here, $\{\,,\}$ denotes the Poisson bracket. A common way to quantize a system is to replace the phase space by a Hilbert space, the variables $\bf{q}$, $\bf{p}$ and the Hamiltonian $H$ by the operators
\beq
\mathbf{q} \mapsto \hat{\mathbf{q}}\ , \qquad \mathbf{p} \mapsto \hat{\mathbf{p}}\ , \qquad  H\mapsto \hat{H}\,,
\eeq
and the Poisson bracket through the commutator
\beq
\{\,,\} \mapsto [\,,]\,.
\eeq 
Consequently, one could try to formulate the quantum definition of integrability as follows.
\begin{de} \label{intdef}
A quantized system is integrable iff there exist $N$ independent operators, $\hat{L}_1,\ldots, \hat{L}_N$, that commute with each other and with the Hamiltonian $\hat{H}$.
\end{de}
Unfortunately, this definition, as we will explain below, is not precise enough; see also the discussion in \cite{Sutherland:BM}.

Let  $\hat{L}$ be one of the constants of motion and let us assume that the spectrum of the system is not degenerated. It follows from $[\hat{H},\hat{L}]=0$ that $\hat{H}$ and $\hat{L}$ have common eigenvectors
\beq
\hat{H} \vert \Psi_j \rangle= E_j \vert \Psi_j \rangle, \qquad \hat{L} \vert \Psi_j \rangle= \eta_j \vert \Psi_j \rangle, \qquad j=1,\ldots,N \,,
\eeq
where $N$ is the dimension of the corresponding Hilbert space. Therefore the operator $\hat{L}$ can be decomposed into the sum of the projectors
\beq \label{LdurchPj}
\hat{L}=\sum^N _{j=1} \eta_j\,\hat{P}_j,\qquad \mbox{with}\qquad \hat{P}_j=\vert \Psi_j \rangle \otimes \langle \Psi_j \vert \,.
\eeq
On the other hand, each projector $P_j$ can be related to the Hamiltonian as follows
\beq
\hat{P}_k= \prod^N _{j=1, j \neq k} \frac{\hat{H}-E_j}{E_k-E_j} \,.
\eeq
Putting this formula into the decomposition \eqref{LdurchPj} one finds
\beq
\hat{L}=\sum^N _{j=1} \hat{H}^{j-1} \sum^N _{l=1} F_{jl}\lr{E_1, \ldots, E_n} \,\eta_l
\eeq
Any two commuting operators are thus algebraically dependent and the set of operators $\{H,L_1,\ldots,L_n\}$ can be at most linearly independent. This, however, is not much of a constraint on the quantum system since for example $\{\hat{H},\hat{H}^2,\hat{H}+\hat{H}^3\}$ satisfy this condition. There are also models known for which the existence of the additional commuting operators $\hat{L}_{j}$ was shown without immediate consequences for the understanding of the physics of the system.
 
A different possibility to define the quantum integrability offer systems which support scattering of particles. These need not to be physical, sometimes physical processes can be simply interpreted as scattering of certain particles. In what follows, we will discuss such one-dimensional systems.

Let $1$ and $2$ be two well-separated particles with asymptotic momenta $p_1$ and $p_2$. The ``incoming'' wave function is then given by
\beq
\Psi_{\mbox{\tiny in}}(x_1,x_2) \sim \exp{i (p_1 x_1+p_2 x_2)} \qquad x_1 \ll x_2\,.
\eeq
The both momenta satisfy two conservation conditions
\begin{enumerate}
\item $\mbox{the conservation of energy, e.g.}\qquad \quad E=\frac{1}{2}\lr{p^2_1+p^2_2}=const, \qquad \mbox{and}$

\item $\mbox{the conservation of momentum}\qquad \quad P=p_1+p_2=const\,.$
\end{enumerate}
Therefore, after the scattering process the outgoing momenta need to be a permutation of the incoming
\beq
p'_1=p_2, \qquad p'_2=p_1 \,.
\eeq
The complete asymptotic wave function can be written as
\beq \label{zkawf}
\Psi_{\mbox{\tiny asymp}}(x_1,x_2) \sim \exp{i (p_1 x_1+p_2 x_2)}+S_{1,2} (p_1,p_2)\exp{i (p_2 x_1+p_1 x_2)} \qquad x_1 \ll x_2 \,.
\eeq
The quantity $S(p_1,p_2)$ defines the corresponding scattering matrix (S-matrix) and can be determined using the Hamiltonian. For theories with the underlying translation invariance, the S-matrix depends only on the difference of the both momenta $S(p_1-p_2)$. In the case of three and more particles the conservation conditions 1. and 2. are not sufficient to determine the state of the system after scattering. Physically, this corresponds to the diffractive three- and many-body processes. The $n$-particle wave function in the asymptotic region, 
\beq
x_1 \ll x_2 \ll \ldots \ll x_n\,,
\eeq
can be represented as a sum of the diffractive and the non-diffractive part 
\beq \label{nbplusb}
\Psi_{\mbox{\tiny asymp}}(x_1,x_2,\ldots,x_n) \sim \sum_{\sigma\in \Pi_n} \Psi(\sigma) \exp{i(p_{\sigma(1)}x_1+\ldots+p_{\sigma(n)}x_n)}+\Psi_{\mbox{\tiny diffractive}}(x_1,x_2,\ldots,x_n)\,,
\eeq
where $\sigma$ denotes an element of the permutation group $\Pi_n$. The non-diffractive part (the Bethe ansatz) describes a sequence of consecutive two-body processes. Since the wave function for every two-body process can be represented as in \eqref{zkawf}, the coefficients $\Psi(\sigma)$ and $\Psi(\sigma')$ that correspond to two permutations $\sigma$ and $\sigma'$ related by a transposition,
\beqa
\sigma=\{\sigma(1),\sigma(2),\ldots,\sigma(i),\sigma(i+1),\ldots,\sigma(n)\} \\
\sigma'=\{\sigma(1),\sigma(2),\ldots,\sigma(i+1),\sigma(i),\ldots,\sigma(n)\}\,,
\eeqa
must satisfy the following relation
\beq \label{psisigmaspsisigma}
\frac{\Psi(\sigma')}{\Psi(\sigma)}= S_{i,i+1} (p_{\sigma(i)},p_{\sigma(i+1)}) \,.
\eeq
Consequently, the non-diffractive part of the wave function is fully determined by the two-body S-matrix. This fact allows for the following definition of the quantum integrability (see \cite{Sutherland:BM}).
\begin{de} \label{intdurchstreuu}
A quantum system that supports scattering is integrable iff the scattering of the particles is non-diffractive
\beq
\forall n: \qquad \Psi_{\mbox{\tiny diffractive}}(x_1,x_2,\ldots,x_n)=0 \,.
\eeq
\end{de}
In particular, this implies that no particles are being created or annihilated in a scattering process. 

The vanishing of the diffractive part of the wave function is an indication of existence of further conserved quantities (charges). A general three-body process is inelastic. If, however, in addition to the conservation laws 1. and 2. there exists a supplementary condition, e.g.
\beq
Q_3= p^3_1+p^3_2+\ldots+p^3_n=\textrm{const}\,,
\eeq
then the outgoing momenta must be again a permutation of the incoming ones. A generic $n$-body scattering process is non-diffractive if there exist $n$ conserved, algebraically independent and symmetric in all momenta quantities $Q_r=Q_r(p_1,\ldots,p_n), \ r=1,...,n$. The first two of these charges are usually assumed to correspond to the total momentum and the total energy of the system. The higher charges $Q_3,\ldots,Q_n$ are eigenvalues of the generators $\hat{Q}_3,\ldots,\hat{Q}_n$ of a hidden symmetry which is not manifest at the level of the Hamiltonian or the equations of motion. No general method is known how to construct the generators $Q_r, \ r \geq 3$ for a generic integrable system. For integrable spin chains, however, which we will consider in this article one can accomplish this by constructing the transfer matrix or with help of the so called ``boost'' operator, see \cite{Grabowski:1994rb}. 

Yet another possibility of defining the quantum integrability provides the Lax pair. Let $\hat{L}$ and $\hat{A}$ be two $N \times N$ matrices, with the  elements being functions of the operators of the system in question such that
\beq \label{LPdef}
\frac{d \hat{L}}{dt}=i (\hat{A}\,\hat{L}-\hat{L}\,\hat{A})
\eeq
is satisfied. The pair $(\hat{L},\hat{A})$ is called the Lax pair. From the relation \eqref{LPdef} it follows that the time evolution of $\hat{L}(t)$ is implemented by a unitary transformation generated by $\hat{A}$
\beq
\hat{L}(t)=U(\hat{A},t)\,\hat{L} (0)\,U^{\dagger}(\hat{A},t) \,.
\eeq
Upon introducing the determinant, $\hat{D}(\lambda)=\det \lr{\hat{L}(t)-\lambda\,\mathbb{I}}$, one finds that it is time-independent
\beqa
\hat{D}(\lambda)=\det \lr{\hat{L}(t)-\lambda\,\mathbb{I}}&=&\det \lr{U(\hat{A},t) \big[\hat{L} (0)\,-\lambda\,\mathbb{I}\big] U^{\dagger}(\hat{A},t)} \nonumber \\
&=&\det \lr{\hat{L}(0)-\lambda\,\mathbb{I}}=\hat{D}(0)\,.
\eeqa
Calogero has shown \cite{Calogero:1975ii} that the operator $\hat{D}(\lambda)$ can be unambiguously defined and that the following additional relations are satisfied
\beq \label{Drel}
\left[\hat{H},\hat{D}(\lambda)\right]=0, \qquad \left[\hat{D}(\lambda), \hat{D}(\lambda')\right]=0 \,.
\eeq
Moreover, $D(\lambda)$ is according to the definition a polynomial of the $N$-th order in $\lambda$
\beq \label{Dentw}
\hat{D}(\lambda)=\sum^N _{j=0} \lambda^j \,\hat{Q}_j \,.
\eeq
Substituting the above formula into \eqref{Drel}, one finds the commutation relations
\beq
[\hat{H},\hat{Q}_j]=0 \qquad [\hat{Q}_j,\hat{Q}_k]=0 \qquad \quad  j,k=1,...,N \,.
\eeq
Consequently, the existence of a Lax pair allows to show that the system is integrable in the sense of the definition \ref{intdef}, and it provides a method to construct all the higher charges. Unfortunately, no general procedure is known how to determine the corresponding $\hat{L}$ nd $\hat{A}$ matrices. It is worth mentioning that a similar method was used to show \cite{Bena:2003wd} that the classical equations of motion of the IIB superstring theory on $\ads$  are integrable. This, in turn, allowed to construct the algebraic curve of the AdS/CFT correspondence \cite{Kazakov:2004qf}-\cite{Beisert:2005bm}.

All of the attempts to define the quantum integrability discussed above are not general enough to be considered as the final definition. Together, however, they portray plausibly what integrability means at the quantum level. Since in this article we will mainly discuss spin chains, it is convenient to assume the definition \ref{intdurchstreuu} as the criterion of integrability.
\section{The S-matrix and the Yang-Baxter Equation} \label{sec:dieSmatrix}
In this section we will generalize the concept of the S-matrix to the scattering processes with different type of particles. We will also formulate consistency conditions that S-matrix of an integrable system must satisfy.

\begin{figure}
\begin{center}
\includegraphics[scale=0.8]{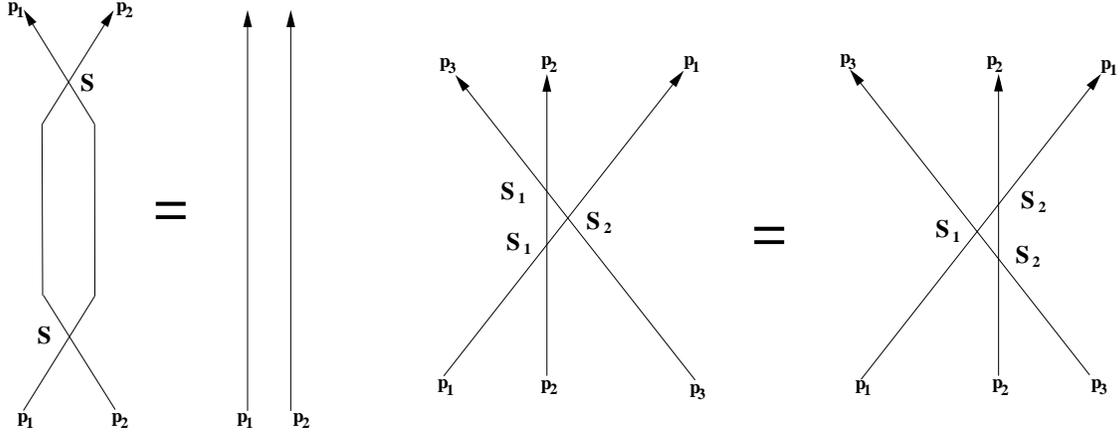}
\end{center}
\caption{A schematic representation of the relations \eqref{Squadrat} and \eqref{SYBG}.} \label{UpYBG}
\end{figure}

Let us consider an integrable system with $n$ particles that according to their ``flavor'' can be divided into $l$ groups. By flavor we mean a value of some additional charge, e.g. spin, electrical charge, etc. Let $1$ and $2$ be two such particles that carry the flavors $f_1, f_2$ and the momenta $k_1,k_2$ respectively. The following two scattering channels are possible\footnote{Here we do not consider flavor-changing processes.}
\beq
\{f'_1,f'_2,k'_1,k'_2\}=\{f_2,f_1,k_2,k_1\}, \qquad \{f'_1,f'_2,k'_1,k'_2\}=\{f_1,f_2,k_2,k_1\} \,.
\eeq
The first scattering channel corresponds to the transition, whereas the second to the reflection of the particles. The definition of the corresponding two-body S-matrix must incorporate the additional quantum numbers $f_1\ \mbox{and} \ f_2$
\beq
S_{12}(p_1,p_2) \mapsto S^{f'_1\,f'_2}_{f_1\,f_2} (p_1,p_2) \,.
\eeq
A scattering process of $n$ particles may be decomposed into a sequence of two-body scatterings. Since the latter are non-diffractive, the momenta of the outgoing particles must be a permutation of the incoming ones. Any outgoing configuration of the particles may be, however, achieved in many physically distinct ways similarly as the ordered set $\{2,3,1\}$ can be obtained by different sequences of transpositions
\beq \label{perms}
\{2,3,1\}=Z_1\,Z_2\,Z_1\,Z_2 \, \{1,2,3\}=Z_2\,Z_1 \, \{1,2,3\}=\ldots \,.
\eeq
Here, $Z_i$ denotes a transposition of the two neighboring sites $i$ and $i+1$. Using the property
\beq \label{P1}
Z^2_i=\mathbb{I}\,,
\eeq
the both sequences in \eqref{perms} are equivalent if
\beq \label{P2}
Z_1\,Z_2\,Z_1=Z_2\,Z_1\,Z_2 \,.
\eeq
The S-matrix must be consistently defined with respect to the identities \eqref{P1} and \eqref{P2}. In the case of \eqref{P1} this implies that
\beq \label{Squadrat}
\sum_{a,b} S^{f'_1\,f'_2}_{b\,a} (p_2,p_1)\,S^{a\,b}_{f_1\,f_2} (p_1,p_2)=\delta^{f'_1}_{f_1}\delta^{f'_2}_{f_2}\,,
\eeq
where the sum runs over the two possible intermediate states $\{a,b\}=\{f_1,f_2\}$ and $\{a,b\}=\{f_2,f_1\}$. Physically, equation \eqref{Squadrat} implies that two subsequent scattering processes of the same particles are equivalent to no scattering at all, see figure \ref{UpYBG}. The identity \eqref{P2}, on the other hand, is reflected in the following cubic relation between S-matrices
\beq \label{SYBG}
S^{f'_1\,f'_2}_{b\,c} (p_2,p_3)\,S^{f'_3\,c}_{a\,f_3} (p_1,p_3)\,S^{a\,b}_{f_1\,f_2} (p_1,p_2)=S^{f'_3\,f'_2}_{\tilde{c}\,\tilde{a}} (p_1,p_2)\,S^{\tilde{c}\,f'_1}_{f_1\, \tilde{b}} (p_1,p_3)\,S^{\tilde{a} \,\tilde{b}}_{f_2\,f_3} (p_2,p_3)\,,
\eeq
with the indices $a,b,c,\tilde{a},\tilde{b}\,\mbox{and} \, \tilde{c}$ being summed over\footnote{In the presence of fermionic particles both 
\eqref{Squadrat} and \eqref{SYBG} must be supplemented with extra minus signs.}. The relation \eqref{SYBG} is the celebrated Yang-Baxter equations. It implies that in a non-diffractive scattering process the sequence of the two-body scattering processes does not matter, see figure \ref{UpYBG}. 

The formulas \eqref{Squadrat} and \eqref{SYBG} are sufficient\footnote{It should be stressed, however, that these two relations do not imply non-diffractive scattering. Indeed, examples are known where the two-body S-matrices satisfy both conditions, but the system is not integrable.} to define consistently the scattering of $n$ particles. This follows from the fact that the permutation group of $n$ elements, $\Pi_n$, may be defined with help of the 2-cycles $Z_i,\ i=1,...,n-1$ introduced above. Two arbitrary $Z_i$ and $Z_j$ obey
\beq \label{ZiZj}
(Z_i\,Z_j)^{p(i,j)}=\mathbb{I}\,,
\eeq
where
\beq
p(i,j)= \left\{ \begin{array}{ll} 1 & \ i=j \\ 3 & \  \vert i-j \vert=1 \\2 & \ \vert i-j \vert >1 \end{array}\right. \,.
\eeq
The first two cases correspond to the identities \eqref{Squadrat} and \eqref{SYBG}. When $\vert i-j \vert >1$, it follows from the formula \eqref{ZiZj} that 2-cycles commute, which is trivially satisfied by the S-matrices
\beq
S^{f'_i \, f'_{i+1}}_{f_i \, f_{i+1}} (p_i,p_{i+1})\,S^{f'_j \, f'_{j+1}}_{f_j \, f_{j+1}}(p_j,p_{j+1})=S^{f'_j \, f'_{j+1}}_{f_j \, f_{j+1}}(p_j,p_{j+1})\,S^{f'_i \, f'_{i+1}}_{f_i \, f_{i+1}}(p_i,p_{i+1}) \,.
\eeq
\section{Integrable Spin Chains} \label{sec:isk}
In this section we will apply the previously discussed concepts to the case of spin chains. The latter constitute an important subgroup among the integrable models. Moreover, the asymptotic integrability in the $\N=4$ gauge theory discussed in chapter \ref{sec:integrabilitaet} is based on identifying the dilation operator with a Hamiltonian of an integrable long-range spin chain. 

The Hilbert space of a spin chain of length $L$ is a tensor product of $L$ local quantum spaces $V$
\beq
\mathcal{H}=V \otimes V \otimes \ldots \otimes V \,.
\eeq
Each single quantum space $V$ is a module of the algebra $\mathcal{A}$ \footnote{In the following we will only consider spin chains for which the the algebra $\mathcal{A}$ coincides with the symmetry algebra of the Hamiltonian}. A spin (particle) is an element of this module
\beq
\rho_A \in V \,,
\eeq
where $A$ denotes the index with respect to the algebra $\mathcal{A}$. On the Hilbert space $\mathcal{H}$ one defines the Hamiltonian $\hat{H}$,
\beq \label{Hspinkette}
\hat{H}=\sum^{L-1}_{j=1} \hat{H}_{i,i+1}\,.
\eeq
The operator $\hat{H}_{i,i+1}$ acts only on the $i$-th and the $(i+1)$-th quantum spaces
\beq
\ldots \otimes \underbrace{V \otimes V}_{\ i\quad \ \ i+1} \otimes \ldots \,.
\eeq
Additionally, the Hamiltonian must commute with the generators of the symmetry algebra
\beq
\left[\mathcal{H},\mathcal{A}\right]=0\,.
\eeq
In general one can impose two types of the boundary conditions on the spin chains
\begin{enumerate}[i)]
\item open boundary conditions if the last spin $L$ interacts only with preceding $(L-1)$-th spin, and
\item closed boundary conditions if the last spin $L$ is assumed to interact with both the $(L-1)$-th spin and the first spin in the chain.
\end{enumerate}
The second case corresponds to a spin chain on a circle and consequently every local quantity $\Theta_j$ must be related to $\Theta_{j+L}$ and the domain of summation in \eqref{Hspinkette} is extended to include $L$. In this article we will confine ourselves to periodic boundary conditions for which $\Theta_j=\Theta_{j+L}$.

The quantum integrability of a spin chain is understood in the sense of definition \ref{intdef} or \ref{intdurchstreuu}. Actually, in this case it is fairly simple to prove the equivalence of the both definitions. The operators
\beq
\hat{Q}_r: \mathcal{H} \mapsto \mathcal{H} \qquad \qquad r=3,\ldots,L \,,
\eeq
which commute with the Hamiltonian $\hat{H}$ are non-local and they act on $r$ neighboring sites, that is $\hat{Q}_r$ acts on
\beq
\ldots \otimes \underbrace{V \otimes V \otimes \ldots \otimes V}_{i\quad \  i+1 \qquad \ i+r-1} \otimes \ldots \,.
\eeq
A famous example of an integrable spin chain is the $\mathfrak{su}(2)$ Heisenberg chain, for which the spin takes only two values and the corresponding quantum space is assumed to be
\beq
V=\mathbb{C}^2\,.
\eeq
The basis in $V$ is spanned by two states: spin $+\frac{1}{2}$ and spin $-\frac{1}{2}$ particles, which we will denote by $\vert \uparrow \,\rangle$ and $\vert \downarrow \,\rangle$ in what follows. The Hamiltonian can be written in the following form
\beq \label{HSKH}
\hat{H}=\sum^{L}_{i=1} \frac{1}{4} \left(\overrightarrow{\sigma}_i \cdot \overrightarrow{\sigma} _{i+1} -1\right)\,,
\eeq
with the corresponding ground state being composed of $L$ spin-up states
\beq
\hat{H} \, \vert \uparrow \, \rangle \otimes \ldots \vert \uparrow \, \rangle =0\,.
\eeq
Excited states are obtained by replacing some $+\frac{1}{2}$ spins by the $-\frac{1}{2}$ particles\footnote{We assume here that the number of the spin-up particles $n_{\uparrow}$ is bigger or equal then the number of the spin-down particles $n_{\downarrow}=L-n_{\uparrow}$. If $n_{\downarrow}>n_{\uparrow}$ one should take $\vert \downarrow \, \rangle \otimes \ldots \vert \downarrow \, \rangle$ as the reference vacuum.}. For $M$ excitations (magnons) the basis is spanned by
\beq
\vert x_1,\ldots, x_M \rangle = \underbrace{\vert \uparrow \, \rangle \otimes \ldots \otimes \vert \uparrow \, \rangle}_{x_1-1} \otimes \vert \downarrow \, \rangle \otimes \underbrace{\vert \uparrow \, \rangle \otimes \ldots \otimes \vert \uparrow \, \rangle}_{x_2-x_1-1} \otimes \vert \downarrow \, \rangle \otimes \ldots
\eeq
\beq
1 \leq x_1 < x_2 < \ldots < x_M \leq L, \quad \qquad x_1,\ldots ,x_M=1,2,... \ .
\eeq
Consequently, each eigenvector $\vert \Psi \rangle$ can be decomposed in this basis
\beq \label{HSKZ}
\vert \Psi \rangle = \sum_{1\leq x_1< \ldots <x_M \leq L} \Psi(x_1,\ldots,x_M) \vert x_1,\ldots, x_M \rangle \,.
\eeq
It is here where the integrability plays its profound role and constraints the coefficients $\Psi(x_1,\ldots,x_M)$ to be determined purely by the Bethe ansatz\footnote{One can also reverse this logic and show that the state \eqref{HSKZ} together with the ansatz \eqref{HSKBA} is under the condition \eqref{HSKSM} an eigenstate of the Hamiltonian \eqref{HSKH}. Moreover, it can also be proven \cite{Faddeev:1996iy} that all $2^L$ states can be parametrized like this.}. Explicitly, 
\beq \label{HSKBA}
\Psi(x_1,\ldots,x_M)=\sum_{\sigma \in \Pi_M} \psi(\sigma)\exp(i \sum^M_{j=1} p_{\sigma(j)}\,x_j)\,,
\eeq
with the scattering of two magnons, in accordance with \eqref{psisigmaspsisigma}, being determined by the S-matrix
\beq \label{HSKSM}
\frac{\psi(\sigma')}{\psi(\sigma)}=S_{j,j+1}(p_{\sigma(j)},p_{\sigma(j+1)})=-\frac{1+e^{i\,p_{\sigma(j)}+i\,p_{\sigma(j+1)}}-2\,e^{ip_{\sigma(j+1)}}}{1+e^{i\,p_{\sigma(j)}+i\,p_{\sigma(j+1)}}-2\,e^{ip_{\sigma(j)}}}\,.
\eeq
It is convenient to introduce the so called rapidities
\beq
u_k= \frac{1}{2} \cot \frac{p_k}{2} \qquad \qquad k=1,2,\ldots,M\,,
\eeq
which allow to re-write the S-matrix elements in the algebraic form
\beq
S_{j,j+1}(u_{\sigma(j)},u_{\sigma(j+1)})=\frac{u_{\sigma(j)}-u_{\sigma(j+1)}-i}{u_{\sigma(j)}-u_{\sigma(j+1)}+i} \,.
\eeq
The periodic boundary conditions imply that the wave functions must obey
\beq
\Psi(x_1,\ldots,x_M)=\Psi(x_2,\ldots,x_M,x_1+L)\,,
\eeq
leading to the celebrated Bethe equations
\beq \label{XXX12bg}
\lr{\frac{u_k+\frac{i}{2}}{u_k-\frac{i}{2}}}^L=\prod^M_{j=1,j\neq k} \frac{u_k-u_j+i}{u_k-u_j-i} \qquad \qquad k=1,2,\ldots,M\,.
\eeq
In the case of the Heisenberg spin chain, the excitations are all of the same type once the reference vacuum has been chosen. For the spin chains with generic underlying symmetry algebras,  the module $V$ is generally $l$-dimensional. Let us consider such a generic spin chain with $K_1$ particles of type 1, $K_2$ particles of type 2, etc.  so that
\beq
K_1 \geq K_2 \geq \ldots \geq K_l, \qquad \qquad K_1+K_2+\ldots+K_l=L\,.
\eeq
Physical choice corresponds to picking up the most numerous particles (of type 1) as the reference vacuum\footnote{Hereby we assume that this choice is consistent with respect to the Hamiltonian.}
\beq
\vert 0 \rangle_{\rm I}= \underbrace{\vert 1 \rangle \otimes \ldots \otimes \vert 1 \rangle}_{L}\,,
\eeq
and to consider the remaining particles as the excitations on this vacuum. This choice, however, breaks the underlying symmetry of the module. The excitations transform under the residual symmetry. Under the assumption that the spin chain is integrable an arbitrary eigenstate of the Hamiltonian can be represented by a partial state $\vert A_1 \ldots A_k\rangle_{\rm I}$ (with $A_i=2,3,\ldots,l$) and a set of momenta
$\{p_1,\ldots,p_k\}$
\beq \label{vbawf}
\vert \Phi \rangle= \mathbf{\hat{S}}^{\rm I} \vert A_1 \ldots A_k\rangle_{\rm I}\,,
\eeq
with
\beq
\vert A_1 \ldots A_k\rangle_{\rm I}= \sum_{1\leq i_1 \leq i_2 \leq \ldots \leq i_k \leq L} e^{i p_1 i_1} e^{i p_2 i_2} \ldots e^{i p_k i_k} \vert A_1, \ldots ,A_k ; i_1, \ldots i_k \rangle_{\rm I}\,,
\eeq
and
\beq
\vert A_1, \ldots, A_k; i_1,\ldots, i_k \rangle_{\rm I} = \underbrace{\vert 1 \, \rangle \otimes \ldots \otimes \vert 1 \, \rangle}_{i_1-1} \otimes \vert A_1 \, \rangle \otimes \underbrace{\vert 1 \, \rangle \otimes \ldots \otimes \vert 1 \, \rangle}_{i_2-i_1-1} \otimes \vert A_2 \, \rangle \otimes \ldots\,.
\eeq
The operator $\mathbf{\hat{S}}^{I}$ is the many-body scattering matrix of the first level, which due to the integrability can be written as a sum over all possible permutations
\beq \label{SIzerlegung}
\mathbf{\hat{S}}^{\rm I}=\sum_{\sigma \in \Pi_k} \hat{S}^{\rm I}_{\sigma}\,,
\eeq
with the permutation  $\sigma$ corresponding to a given sequence of the two-body processes
\beq \label{SsigmaIzerlegung}
\hat{S}_{\sigma}^{\rm I}=\prod_{(i,j)\in \sigma} \hat{S}^{\rm I}_{ij}\,, \qquad \qquad \vert A'_2 A'_1 \rangle_{\rm I} = \sum_{A_1,A_2} ({S}^{\rm I}_{12})^{A'_2 A'_1}_{A_1 A_2} \vert A_1 A_2 \rangle_{\rm I} \,.
\eeq
It should be stressed that the S-matrix (scattering operator) $\hat{S}^{\rm I}_{12}$ is a $(l-1)^2 \times (l-1)^2$-dimensional matrix governing the scattering of two excitations of different flavors. The above construction, however, can be repeated after the most numerous excitations (of type 2) are chosen to be reference vacuum of the spin chain with $M=L-K_1$ sites
\beq
\vert 0 \rangle _{\rm II}=\vert \underbrace{2 \ldots 2}_{M} \rangle_{\rm I}\,.
\eeq
The remaining excitations $B_i=3,4,\ldots l$ on this vacuum scatter with the S-matrix of the second level $\mathbf{\hat{S}}^{\rm II}$, which can be decomposed similarly to \eqref{SIzerlegung} and \eqref{SsigmaIzerlegung}. Repeating this procedure $(l-1)$ times results in clearing entirely the spin chain from excitations. It should be noted that the action of the S-matrix on the first level vacuum is trivial, whereas the action on the vacuum of the second level results in an non-trivial scalar phase\footnote{Here, we exclude the processes $2+2\mapsto A_1+A_2$, with $A_1\neq A_2\neq 2$.}. In the first approximation (the first level spin chain) one assumes all of the excitations $A_i, i=2,\ldots, l$ to scatter with the same phase as the particles of type $2$. The error committed in this way is partially compensated by considering scattering processes on the vacuum of the second level, where the remaining $(l-3)$ excitations scatter with a phase different from the one for the preceding level. The requirement of consistency between the first and the second level scattering allows to find explicitly the form of the second level S-matrix. Recursive application of this method allows to determine the S-matrices of the $k$-th level, with $k=1,2,\ldots, l-1$. The periodicity conditions for each of these spin chains lead to the nested Bethe equations. This method is known in the literature as the nested Bethe ansatz \cite{Yang:1967bm}.  

The scattering in generic integrable spin chains with arbitrary values of spin or with non-compact symmetry algebras can be uniformly described with the concept of the R-matrix. In the following we will only consider systems with the underlying translation invariance. Let $\rho_A$ and $\rho_B \in V$ be two particles with the spectral parameter $u_1$ and $u_2$ respectively. The R-matrix is defined as an operator that commutes the elements of the corresponding modules
\beq \label{Rmatrix}
\tilde{\rho}_{\tilde{B}}(u_2)\otimes \tilde{\rho}_{\tilde{A}}(u_1)=R^{A\,B}_{\tilde{A}\,\tilde{B}}(u_1-u_2)\, \rho_A(u_1) \otimes \rho_B(u_2)\,.
\eeq
Since commuting the modules twice should be equivalent to the action of the identity operator, we impose
\beq
{R}^{\tilde{B}\,\tilde{A}}_{C\,D}(u_2-u_1)\,{R}^{A\,B}_{\tilde{A}\,\tilde{B}}(u_1-u_2)=\delta^A_C \, \delta^B_D\,.
\eeq
Similarly to \eqref{SYBG}, the consistency of the scattering processes requires the R-matrix to obey the Yang-Baxter equation
\beq
{R}^{\tilde{A}\,\tilde{B}}_{F\,G} (u_2-u_3)\,{R}^{\tilde{C}\,G}_{E\,C} (u_1-u_3)\,{R}^{E\,F}_{A\,B} (u_1-u_2)={R}^{\tilde{C}\,\tilde{B}}_{G\,E} (u_1-u_2)\,{R}^{G\,\tilde{A}}_{A\, F} (u_1-u_3)\,{R}^{E \,F}_{B\,C} (u_2-u_3)\,.
\eeq
Finally, the R-matrices acting on separated modules commute with each other. To a spin chain with $L$ lattice sites one can associate the monodromy matrix
\beq
\left\{{T}_e(u)\right\}^{A\,A_1\,A_2\,\ldots\,A_L}_{B\,B_1\,B_2\,\ldots\,B_L}={R}^{A\,A_L}_{C_{L-1}\,B_L}(u)\,{R}^{C_{L-1}\,A_{L-1}}_{C_{L-2}\,B_{L-1}}(u)\,\ldots\,{R}^{C_1\,A_1}_{B\,B_1}(u)\,,
\eeq
which can be intuitively interpreted as a quantity describing an external ``ghost'' particle circulating around the spin chain. The external quantum space $V_e$ is also assumed to be a module of the symmetry algebra $\mathcal{A}$, though this module must not necessarily coincide with the physical one $V$. The trace of the monodromy matrix with respect to the external quantum space
\beq \label{grossetm1}
\hat{T}(u)=\left\{{T}_e(u)\right\}^{A\,A_1\,A_2\,\ldots\,A_L}_{A\,B_1\,B_2\,\ldots\,B_L}
\eeq
defines the transfer matrix. If the external quantum space  coincides with the physical one, the transfer matrix becomes a generating matrix of the higher conserved charges
\beq \label{grossetm2}
\hat{T}^{(f)}(u)=\hat{U}\,\exp{i\sum^\infty_{r=2} u^{r-1} \hat{Q}_r}\,,
\eeq
where $\hat{U}$ stands for the translation operator
\beq \label{Verschiebungsoperator}
\hat{U}=\hat{T}^{(f)}(0)=P_{1,2}\, P_{2,3}\, \ldots \, P_{L-1,L}\,.
\eeq
Here, $P_{a,b}$ denotes the permutation operator between the $a$-th and $b$-th vector space. In particular the Hamiltonian can be represented as follows
\beq
\hat{H}=\hat{Q}_2=\frac{1}{i}\lr{\lr{\hat{T}^{(f)}(u)}^{-1}\,\frac{d}{du} \hat{T}^{(f)}(u)}\bigg\vert_{u=0} \,.
\eeq
As an example, let us consider the $XXX_{\frac{s}{2}}$ spin chain with the $\mathfrak{sl}(2)$ symmetry algebra. The modules $V$ can be classified according to the value of the spin $s/2$. Let $\vert \Psi \rangle$ be an eigenvector of the Hamiltonian corresponding to $M$ excitations with the rapidities $u_i\ i=1,\ldots, M$. The state $\vert \Psi \rangle$ is also an eigenstate of the fundamental transfer matrix \eqref{grossetm2}
\beq
\hat{T}^{(f)}(u) \vert \Psi \rangle= T^{(f)}(u,u_1,\ldots,u_M) \vert \Psi \rangle \,,
\eeq
with
\beq
T^{(f)}(u,u_1,\ldots,u_M)=U\,\exp{i\sum^\infty_{r=2} u^{r-1} Q_r}\,,
\eeq
and
\beq
U=\prod^M_{j=1} \frac{u_j+i\,\frac{\vert s \vert}{2}}{u_j-i\,\frac{\vert s \vert}{2}}, \qquad \qquad Q_r=\frac{i}{r-1} \sum^M_{j=1} \lr{\frac{1}{(u_j+i\,\frac{\vert s \vert}{2})^{r-1}}-\frac{1}{(u_j-i\,\frac{\vert s \vert}{2})^{r-1}}}\,.
\eeq
For the $XXX_{\frac{s}{2}}$ spin chains $\hat{T}^{(f)}(u)$ is a polynomial of the order $L$ in $u$. On the other hand, the eigenvalue $T^{(f)}(u,u_1,\ldots,u_M)$ exhibits poles that depend on the Bethe roots $u_j, j=1,\ldots, M$. Requiring these poles to cancel for arbitrary value of $L$, one finds the Bethe equations
\beq \label{xxxsd2}
\lr{\frac{u_k+i\,\frac{s}{2}}{u_k-i\,\frac{s}{2}}}^L=\prod^M_{j=1,j\neq k} \frac{u_k-u_j+i}{u_k-u_j-i} \qquad \qquad k=1,2,\ldots,M\,.
\eeq
The case $s=1$ corresponds to the Heisenberg spin chain discussed above.

For the $XXX_{\frac{s}{2}}$ spin chains it is convenient to introduce beside the fundamental transfer matrix also the auxiliary transfer matrix $\hat{T}^{(a)}(u)$ by choosing the external quantum space to be $V_e=\mathbb{C}^2$. It was shown in \cite{Baxter} that there exist an operator  $\hat{Q}(u)$ such that
\beq \label{QQTQ}
[\hat{Q}(u),\hat{Q}(v)]=[\hat{Q}(u),\hat{T}^{(a)}(v)]=0 
\eeq
and
\beq \label{OBG}
(u+i\frac{s}{2})^L\,\hat{Q}(u+i)+(u-i\frac{s}{2})^L\,\hat{Q}(u-i)=\hat{T}^{(a)}(u)\,\hat{Q}(u)\,.
\eeq\
According to the relation \eqref{QQTQ} both $\hat{T}^{(a)}(v)$ and $\hat{Q}(u)$ can be diagonalized simultaneously. Let $\vert \Phi \rangle$ be such eigenstate parametrized by the Bethe roots $u_i,\ i=1,\ldots,M$. Equation \eqref{OBG} then becomes a functional equation relating the eigenvalue of the auxiliary transfer matrix to the eigenvalue of the Baxter operator $\hat{Q}(u)$,
\beq \label{BG}
(u+i\frac{s}{2})^L\,Q(u+i)+(u-i\frac{s}{2})^L\,Q(u-i)=T^{(a)}(u)\,Q(u)\,.
\eeq
Here, we have suppressed the explicit dependence of the eigenvalues on $\{u_i\}$. The equation \eqref{BG} is called in the literature the Baxter equation. Under the following ansatz
\beq
Q(u)=\prod^M_{j=1} (u-u_j)
\eeq
one re-derives the Bethe equations \eqref{xxxsd2}. 
\chapter{Integrability in the $\N=4$ Super Yang-Mills Theory}\label{sec:integrabilitaet}
The asymptotic integrability in the planar $\N=4$ SYM theory allows to compare observables on the both side of the correspondence that are not affected by the wrapping interactions and therefore paves the way for the novel dynamical tests of the correspondence. In this chapter we will present the building blocks of the asymptotic integrability and discuss the asymptotic spectral equations.
 
\section{The $\psua$ Super Spin Chain}
\subsection{Lie Superalgebras and Superconformal Alegbras}
In this section we briefly review essential definitions and results of the Lie superalgebras, which we will use in what follows.

A Lie superalgebra $\mathcal{A}$ is an algebra with a $\mathbb{Z}_2$ grading equipped with a multiplication that obeys the following conditions
\begin{enumerate}
\item skew-symmetry
\beq \label{sab1}
\forall a,b \in \mathcal{A}: \qquad (a,b)=-(-1)^{g(a)g(b)} (b,a) \, ,
\eeq
\item generalized Jacobi identity
\beq \label{sab2}
\forall a,b,c \in \mathcal{A}: \qquad (-1)^{g(a)g(c)}(a,(b,c))+(-1)^{g(a)g(b)}(b,(c,a))+(-1)^{g(c)g(b)}(c,(a,b)) \, ,
\eeq
\end{enumerate}
where $g(.)$ denotes the grade of an element\footnote{In what follows the multiplication of two fermionic elements will be denoted by $\{\,,\,\}$.}. The subset
\beq
\mathcal{A}_0 = \{a \in \mathcal{A}: g(a)=0 \}
\eeq
is in view of the conditions \eqref{sab1} and \eqref{sab2} an ordinary Lie algebra. The fermionc component of $\mathcal{A}$ 
\beq
\mathcal{A}_1 = \{a \in \mathcal{A}: g(a)=1 \}
\eeq
can be interpreted as a linear representation of $\mathcal{A}_0$ since
\beq
(\mathcal{A}_0,\mathcal{A}_1) \in \mathcal{A}_1 \, .
\eeq
\begin{de}
A simple Lie superalgebra $\mathcal{A}$ for which the representation of its even subalgebra $\mathcal{A}_0$ on its odd part $\mathcal{A}_1$ is completely reducible is called a classical Lie superalgebra.
\end{de}
It can be proven that for classical Lie superalgebras the representation of $\mathcal{A}_0$ on $\mathcal{A}_1$ is either\begin{center}
\begin{enumerate}[i)]
\item irreducible, or
\item is a direct sum of two irreducible representations of $\mathcal{A}_0$.
\end{enumerate}
\end{center}
This feature allows to define Lie superalgebras of, correspondingly,  the first and the second kind.
\begin{de}
A classical Lie superalgebra $\mathcal{A}$ equipped with a non-degenerate bilinear invariant form is called a basic Lie superalgebra.
\end{de}
Similarly to the case of the classical semi-simple Lie algebras one can introduce for the basic Lie superalgebras the Cartan subalgebra $\mathcal{H}$ and the root system $\{E_{\alpha(i)}\},i=1,\ldots, n$ such that
\beqa
\qquad (H_i,H_j)&=&0\,, \nonumber \\
\qquad \left(H_i,E_{\pm \alpha(j)}\right)&=&\pm M_{ij}\,E_{\pm \alpha(j)}\,, \nonumber \\
\qquad \left(E_{\alpha(i)},E_{-\alpha(j)}\right)&=&\delta_{ij}\,H_i \, ,
\eeqa
where $M_{ij}$ is the Cartan matrix in the $\{H_i\}$ basis of the Cartan subalgebra. However, in contradistinction to the case of Lie algebras, the choice of the Dynkin diagram is not unique. This is due to the fact that there exist a residual freedom in the choice of the odd (fermionic) roots. A complete classfication of the fundamental classical Lie superalgebras was given in \cite{Kac:1977em}.

A superconformal algebra is defined to be a Lie superalgebra $\mathcal{A}$ for which the even part $\mathcal{A}_0$ contains the conformal subalgebra $\mathfrak{so}(d,2)$ that is spinorially represented on the odd component $\mathcal{A}_1$.
\subsection{The $\psua$ Superconformal Algebra and Its Representations} \label{sec:psualgebra}
The $\mathfrak{su}(2,2 \vert 4)$ Lie superalgebra is a superalgebra of the first type, which in the Kac classification corresponds to the $A(3,3)$ Lie superalgebra, see \cite{Minwalla:1997ka} for a pedagogical discussion of this issue.  The bosonic (even) component is spaned by $\mathfrak{su}(2,2)\oplus \mathfrak{su}(4)\oplus u(1)$. Taking the isomorphism $\mathfrak{su}(2,2) \simeq \mathfrak{so}(4,2)$ into account, one concludes that $\mathfrak{su}(2,2 \vert 4)$ is a superconformal algebra in four dimensions (the spinor representation of the $\mathfrak{so}(4,2)$ on the odd component of the superalgebra will be explicitly constructed below). The set of generators of $\mathfrak{su}(2,2,\vert 4)$ is spaned by
\beq
\{L^\alpha _{\beta}, \bar{L}^{\dot \alpha}_{\dot \beta}, P_{\dot \alpha \beta}, K_{\alpha \dot \beta}, D; R^a _b; C\, \vert \ Q^a_{\alpha}, \bar{Q}_{a \dot \alpha}, S^\alpha _a, S^{\dot \alpha a} \} \,.
\eeq
Here, the generators carry ``fermionic'' indices of $\mathfrak{su}(2)$ and $\mathfrak{su}(4)$ respectively and are assumed to transform canonically. The non-vanishing commutation relations of the $\mathfrak{su}(2,2 \vert 4)$ are given by
\beq \label{erstekr}
\left\{\big[D,P_{\dot \alpha \beta}\big]\,;-\big[D,K_{\alpha \dot \beta}\big]\right\}=\big\{P_{\dot \alpha \beta} \,; K_{\alpha \dot \beta} \big\} \,,
\eeq
\beq
\left\{\big[D,Q^a_\alpha\big]\,;\big[D,S^\alpha_b\big]\,;-\big[D,\bar{Q}_{\dot \alpha a}\big]\,;-\big[D,\bar{S}^{a \dot \beta}\big]\right\}=\frac{1}{2} \big\{Q^a_\alpha\ \,; S^\alpha_b \,; \bar{Q}_{\dot \alpha a} \,; \bar{S}^{a \dot \beta}\big\}\,,
\eeq
\beq
\big[K^{\alpha \dot\beta},P_{\dot\gamma\delta}\big]=
  \delta_{\dot\gamma}^{\dot\beta} L^{\alpha} _{\delta}
  +\delta_\gamma^\alpha \bar{L}^{\dot\beta} _{\dot\delta}
  +\delta_\gamma^\alpha\delta_{\dot\delta}^{\dot\beta} D\,,
\eeq
\beq
\begin{array}{cccccccc}
\{\bar{Q}_{\dot\alpha a},Q^b_\beta\}&=&
  \delta_a^b P_{\dot\alpha\beta}\,,
&\qquad&
\{\bar{S}^{a\dot\alpha},S^\beta_b\}&=&
  \delta^a_b K^{\beta\dot\alpha}\,,
\\[3pt]
[S^\alpha _a,P_{\dot\beta\gamma}]&=&
   \delta^\alpha_\gamma \bar{Q}_{\dot\beta a}\,,
&\qquad&
[K^{\alpha\dot\beta},\bar{Q}_{\dot\gamma c}]&=&
  \delta^{\dot\beta}_{\dot\gamma} S^\alpha _c\,,
\\[3pt]
[\bar{S}^{a\dot\alpha},P_{\dot\beta\gamma}]&=&
  \delta^{\dot\alpha}_{\dot\beta} Q^a _{\gamma}\,,
&\qquad&
[K^{\alpha\dot\beta},Q^c_\gamma]&=&
  \delta^\alpha_\gamma \bar{S}^{c\dot\beta}\,,
\end{array}
\eeq
\beqa \label{SQKR}
\{S^\alpha _a,Q^b_\beta\}&=&
  \delta^b_a L^\alpha _\beta
  +\delta_\beta^\alpha R^b _a
  +\frac{1}{2} \delta_a^b \delta_\beta^\alpha (D-C)\,,
  \\[3pt] \label{SBQBKR}
\{\bar{S}^{a\dot \alpha},\bar{Q}_{\dot\beta b}\}&=&
  \delta^a_b \bar{L}^{\dot\alpha} _{\dot\beta}
  -\delta_{\dot\beta}^{\dot\alpha} R^a _b
  +\frac{1}{2} \delta^a_b \delta_{\dot\beta}^{\dot\alpha} (D+C) \label{letztekr}\,.
\eeqa

It follows from the above relations that the generator $C$ plays the role of the central charge and consequently that  $\mathfrak{su}(2,2 \vert 4)$ is reducible. The irreducible part of this algebra, $\psua$, can be obtained by considering representations with vanishing central charge. It should be stressed, however, that the $\psua$ algebra contrary to the $\mathfrak{su}(2,2 \vert 4)$ does not possess the defining  $8\times8$ matrix representation.

The above commutation relations are also valid after quantization with all the generators except for $L^\alpha _\beta$, $\bar{L}^{\dot \alpha}_{\dot \beta}$ and $R^a _b$ receiving quantum corrections
\beq
\psua \ni J: \quad \qquad J \mapsto \hat{J}(g) \,.
\eeq
At weak coupling one expects the following expansion
\beq \label{Genentwicklung}
\hat{J}(g)=\sum^\infty _{j=1} \hat{J}_{2j}\,g^{2j} \,.
\eeq
At the one-loop level there is a particularly useful choice of the corresponding Dynkin diagram of  the $\mathfrak{su}(2,2 \vert 4)$ superalgebra, see figure \ref{fig:beautydiagram}. The positive and negative roots that correspond to this diagram are given by
\beqa
J^+&\in&\{K^{\alpha\dot \beta},S^\alpha _a,
\bar{S}^{a\dot\alpha},L^\alpha _\beta\,(\alpha < \beta),
\bar{L}^{\dot\alpha}_{\beta}\,(\dot\alpha<\dot\beta),R^a _b\,(a<b)\},
\\
J^0 &\in& \{L^\alpha _\beta\, (\alpha=\beta),
\bar{L}^{\dot\alpha} _{\dot\beta}\, (\dot\alpha=\dot\beta),
R^a _b\,(a=b),D,C\}, \label{cartanal}
\\
J^-&\in& \{P_{\alpha \dot \beta},Q^a _\alpha,\bar{Q}_{\dot\alpha a},
L^\alpha _\beta\,(\alpha>\beta),
\bar{L}^{\dot\alpha} _{\beta}\,(\dot\alpha>\dot\beta),R^a _b\,(a>b)\} \,. \label{senkoperatoren}
\eeqa
 \\

In the $\N=4$ SYM theory only non-compact and infinite representations of the superconformal algebra are of physical relevance, which is due to the presence of derivatives of fields (each field can be differentiated infinitely many times). Any state in the theory can be identified through the following set of Dynkin labels
\beq
\{\Delta,s_1,s_2,q_1,p,q_2,B,L\}\,,
\eeq
where $\Delta$ is the eigenvalue of the dilatation operator, the weights $[s_1,s_2]$ classify the spinor representations of the Lorentz algebra $\mathfrak{so}(3,1)=\mathfrak{su}(2)\times \mathfrak{su}(2)$ and  $[q_1,p,q_2]$ correspond to the flavor algebra $\mathfrak{so}(6)$. The relation of these labels to the eigenvalues of the Cartan algebra elements \eqref{cartanal} is given by the following formulas
\beq
s_1 = L^2_2-L^1_1,
\qquad
s_2 = \bar{L}^2_2-\bar{L}^1_1,
\eeq
\beq
q_1 = R^2_2-R^1_1,
\quad
p = R^3_3-R^2_2,
\quad
q_2 = R^4_4-R^3_3\,.
\eeq
The remaining quantities $B$ and $L$ are not related to the weights of the $\mathfrak{su}(2,2\vert 4)$ superalgebra and can be seen as eigenvalues of the external automorphisms. The length $L$ counts the number of fields in the trace \eqref{einzelspur} and is equal to $1$ independently of the value of $k$ for each field in \eqref{Feldermenge}. The hypercharge $B$, on the other hand, measures the hyperspin of the multiplet \eqref{Feldermenge} and is a multiple of $\sfrac{1}{2}$.
All physical operators of the theory may be classified into highest weight multiplets. Each such multiplet is defined by the highest weight state (the primary field) $\hat{\Op}=\vert \Op \rangle$
\beq
\forall \hat{J}^+:\qquad \hat{J}^+ \vert \Op \rangle=0 \,.
\eeq
Other states of a given multiplet can be obtained by collective action of the lowering operators $\hat{J}_-$ on $\vert \Op \rangle$
\beq
\hat{J}^{-}_1 \hat{J}^{-}_2 \ldots \hat{J}^{-}_k \vert \Op \rangle \,.
\eeq
An important example of a highest weight multiplet constitute the $\frac{1}{2}$-BPS operators
\beq
\vert \fldZ \rangle^L = \Tr \left(\fldZ^L \right), \qquad \fldZ=\Phi_{34}\,.
\eeq
These highest weight states are additionally annihilated by half of the supersymmetry generators and consequently their scaling dimension is protected and does not receive quantum corrections, see discussion in section \ref{sec:interpol}.

For the purpose of representing fields it is very convenient to use the oscillator realization of the $\psua$. The oscillators are defined as follows
\begin{enumerate}[a)]
\item bosonic $(\oa^\alpha,\oa^\dagger_\alpha)$ corresponding to the one copy of the $\mathfrak{su}(2)$ subalgebra\,,
\item bosonic $(\ob^{\dot \alpha},\ob^\dagger_{\dot \alpha})$ corresponding to the second copy of the $\mathfrak{su}(2)$ subalgebra\,,
\item fermionic $(\oc^a,\oc^\dagger_a)$ of the $\mathfrak{su}(4)$ subalgebra\,,
\end{enumerate}
and are assumed to obey the following commutation relations
\beq \label{abckr}
[\oa^{\alpha}, \oa^\dagger_{\beta}]=\delta^\alpha_\beta \qquad [\ob^{\dot \alpha}, \ob^\dagger_{\dot \beta}]=\delta^{\dot \alpha}_{\dot \beta} \qquad \{\oc^{a}, \oc^\dagger_{b}\}=\delta^a_b \,.
\eeq
The elements of the $\psua$ Lie superalgebra for $g=0$ can be then represented through 
\beq
\begin{array}{cccccccc}
L^{\alpha}_{\beta}&=&\oa^\dagger_{\beta}\oa^{\alpha}
-\frac{1}{2} \delta^\alpha_\beta\oa^\dagger_{\gamma}\oa^{\gamma}\,, &\qquad&
\bar{L}^{\dot\alpha}_{\dot\beta}&=&\ob^\dagger_{\dot\beta}\ob^{\dot\alpha}
-\frac{1}{2} \delta^{\dot\alpha}_{\dot\beta}\ob^\dagger_{\dot\gamma}\ob^{\dot\gamma}\,,
\\
\\
D&=&
1+\frac{1}{2} \oa^\dagger_{\gamma}\oa^{\gamma}
+\frac{1}{2} \ob^\dagger_{\dot\gamma}\ob^{\dot\gamma}, &\qquad&
R^{a}_{b}&=&\oc^\dagger_{b}\oc^{a}
-\frac{1}{4} \delta^{a}_{b}\oc^\dagger_{c}\oc^{c},
\end{array}
\eeq
\beqa
C=
1-\frac{1}{2} \oa^\dagger_{\gamma}\oa^{\gamma}
+\frac{1}{2} \ob^\dagger_{\dot\gamma}\ob^{\dot\gamma}
-\frac{1}{2} \oc^\dagger_{c}\oc^{c}, \qquad B= 1+\frac{1}{2} \oa^\dagger_{\gamma}\oa^{\gamma}
-\frac{1}{2} \ob^\dagger_{\dot\gamma}\ob^{\dot\gamma}\,,
\eeqa
\beq
\begin{array}{cccccccccccc}
Q^a_{\alpha}&=&\oa^\dagger_\alpha \oc^{a}\,, &\qquad&
\bar{Q}_{\dot\alpha a}&=& \ob^\dagger_{\dot\alpha} \oc^\dagger_{a}\,, &\qquad&
P_{\alpha \dot \beta}&=&\oa^\dagger_{\alpha}\ob^\dagger_{\dot \beta}\,,\\
S^{\alpha}_a &=& \oc^\dagger_{a} \oa^\alpha\,, &\qquad&
\bar{S}^{\dot\alpha a} &=& \ob^{\dot\alpha} \oc^{a}\,, &\qquad&
K^{\alpha \dot \beta} &=& \oa^{\alpha}\ob^{\dot \beta}\,.
\end{array}
\eeq
Using formulas \eqref{abckr} one can easily show that the above realization of the generators obeys the  commutation relations \eqref{erstekr}-\eqref{letztekr}. The set of physical excitations on the vacuum state (which is annihilated by all ``undaggered'' oscillators) is spanned by states for which the central charge 
\beq
C=
1-\frac{1}{2} \oa^\dagger_{\gamma}\oa^{\gamma}
+\frac{1}{2} \ob^\dagger_{\dot\gamma}\ob^{\dot\gamma}
-\frac{1}{2} \oc^\dagger_{c}\oc^{c}
\eeq
vanishes. One can easily check that this set coincides with the set of the irreducible fields \eqref{Feldermenge}
\beqa \label{ODdF}
&&\cder^k \fldF \simeq
  (\oa^\dagger)^{k+2}\,
  (\ob^\dagger)^{k}\,
  (\oc^\dagger)^0\,
  \vert 0 \rangle, \nonumber \\
&&\cder^k \Psi \simeq
  (\oa^\dagger)^{k+1}\,
  (\ob^\dagger)^{k}\,
  (\oc^\dagger)^1\,
  \vert 0 \rangle, \nonumber \\
&&\cder^k \Phi \simeq
  (\oa^\dagger)^{k}\,
  (\ob^\dagger)^{k}\,
  (\oc^\dagger)^2\,
  \vert 0 \rangle, \nonumber \\
&&\cder^k \dot\Psi \simeq
  (\oa^\dagger)^{k}\,
  (\ob^\dagger)^{k+1}\,
  (\oc^\dagger)^3\,
  \vert 0 \rangle, \nonumber \\
&&\cder^k \dot \fldF \simeq
  (\oa^\dagger)^{k}\,
  (\ob^\dagger)^{k+2}\,
  (\oc^\dagger)^4\,
  \vert 0 \rangle .
\eeqa
Unfortunately, a very disadvantageous feature of this definition is that the vacuum state itself does not belong to this class
\beq
C \vert 0 \rangle=1\,.
\eeq
One way to overcome this difficulty is to replace the $\oc^3$ and $\oc^4$  oscillators by $\od_{\dot a}\ , \dot a=1,2$
\beq
\od^\dagger _1=\oc^4, \qquad \od^\dagger _2 =\oc^3 \,.
\eeq
In this notation the $\fldZ$ field is annihilated by all oscillators and can therefore serve as a vacuum state. The consequence of such redefinition is the breakdown of the $\mathfrak{su}(4)$ symmetry to $\mathfrak{su}(2)\times \mathfrak{su}(2)$. On the other hand, the central charge can be rewritten in a more transparent form
\beq \label{Cmitd}
C=\frac{1}{2}(N_{\rm \textbf{b}}+N_{\rm \textbf{d}})-\frac{1}{2}(N_{\rm \textbf{a}}+N_{\rm \textbf{c}}) \,,
\eeq
where $N$ stands for the counting operators  (e.g. $N_{\rm \textbf{a}}=\oa^{\dagger}_{\gamma} \oa^{\gamma} $). It follows from \eqref{Cmitd} that the excitations on the vacuum state $\vert \fldZ^L \rangle:=\underbrace{\vert \fldZ \rangle \otimes \ldots \otimes \vert \fldZ \rangle}_{L}$ can only be created pairwise
\beq \label{AAbar}
(A^{\dagger})^M (\bar{A}^{\dagger})^{M} \vert Z^L \rangle\,,
\eeq
with $A=\{\oa_1,\oa_2 ,\oc_1, \oc_2\}$ and $\bar{A}=\{\ob_1,\ob_2, \od_1,\od_2\}$. There are consequently $4\times4=16$ fundamental excitations on each lattice site $\vert \fldZ \rangle$. Each field in \eqref{ODdF} is either a fundamental excitation or can be represented by a multiple excitation (a composition of the fundamental excitations) of a lattice site.
\begin{center}
\begin{figure}
\setlength{\unitlength}{1pt}%
\small\thicklines%
\begin{center}
\begin{picture}(260,30)(0,-10)
\put(  0,00){\circle{25}}%
\put( 12,00){\line(1,0){26}}%
\put( 50,00){\circle{25}}%
\put( 62,00){\line(1,0){26}}%
\put( 100,00){\circle{25}}%
\put( 112,00){\line(1,0){26}}%
\put(150,00){\circle{25}}%
\put(162,00){\line(1,0){26}}%
\put(200,00){\circle{25}}%
\put(212,00){\line(1,0){26}}%
\put(250,00){\circle{25}}%
\put(262,00){\line(1,0){26}}%
\put(300,00){\circle{25}}%
\put( 42,-8){\line(1, 1){16}}%
\put( 42, 8){\line(1,-1){16}}%
\put(242,-8){\line(1, 1){16}}%
\put(242, 8){\line(1,-1){16}}%
\end{picture}
\end{center}
\caption{Beauty diagram} \label{fig:beautydiagram}
\end{figure}
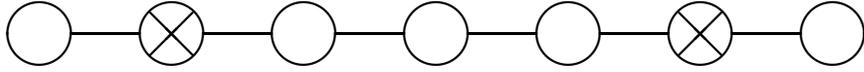
\end{center}

\subsection{One-loop Bethe Equations} \label{sec:einschleifenbethegleichungen}
It was shown in \cite{Beisert:2003yb} that the one-loop dilatation operator $D_2$  (see \eqref{Genentwicklung}) may be identified with the Hamiltonian of an integrable $\psua$ super spin chain.
The Bethe Ansatz techniques for integrable spin chains with simple Lie algebras as the symmetry algebras have been developed in \cite{Reshetikhin:1986vd}, \cite{Ogievetsky:1986hu} and subsequently generalized to the case of superalgebras  \cite{Saleur:1999cx}. This allowed Beisert and Staudacher \cite{Beisert:2003yb} to write down the one-loop Bethe equations of the planar $\N=4$ SYM theory.

As it was explained in section \ref{sec:physop}, in the planar limit it is sufficient to consider only the single trace operators. The eigenstates of the one-loop dilatation operator $\hat{D}_2 \Psi = \gamma^{\Psi}_2(g) \Psi$
are linear combinations of the basis states \eqref{Feldermenge}
\beq
\Psi=a_1 \Tr (\Omega_1 \ldots \Omega_L)+a_2 \Tr (\Omega'_1 \ldots \Omega'_L)+\ldots\,.
\eeq
Since the length operator commutes at the one-loop order with the dilatation operator, only basis states of equal length need to be taken into account. The complete one-loop dilatation operator was found in \cite{Beisert:2003jj}. Remarkably, it acts only on neighboring fields in each trace\footnote{The explicit form of $\hat{H}_{i,i+1}$ can be found in \cite{Beisert:2003jj}.}
\beq
\hat{D}_2=\sum^{L}_{i=1} \hat{H}_{i,i+1}\,,
\eeq
which allows to identify single trace operators with states of a closed spin chain
\beq
\Tr (\Omega_1 \ldots \Omega_L) \mapsto \vert \Omega_1 \ldots \Omega_L \rangle:=\vert \Omega_1 \rangle \otimes \ldots \otimes \vert \Omega_L \rangle \,.
\eeq
It follows from the cyclicity of the trace that
\beq \label{szykl}
\vert \Omega_1 \ldots \Omega_k \Omega_{k+1} \ldots \Omega_L \rangle=(-1)^{(\Omega_{k+1}\ldots \Omega_L \Omega_1 \ldots \Omega_k)}\vert \Omega_{k+1} \ldots \Omega_L \Omega_{1} \ldots \Omega_k \rangle\,,
\eeq
where the prefactor $(-1)^{(\Omega_{k+1}\ldots \Omega_L \Omega_1 \ldots \Omega_k)}$ corresponds to the overall sign of the permutations of fermionic fields.  The integrability of  $\hat{D}_2$ was shown in \cite{Beisert:2003yb} by an explicit construction of the $\psua$ R-Matrix.

Let $\mathcal{H}$ be a Hamiltonian of an integrable super spin chain with $\mathfrak{sl}(K\vert N)$ symmetry algebra. The diagonalisation of this operator can be performed using the techniques of nested Bethe Ansatz, as discussed in section \ref{sec:isk}. For arbitrary values of $K$ and $N$ and arbitrary spin representation it is, however, more convenient to use transfer matrices. The transfer matrices for such spin chains were constructed in \cite{Reshetikhin:1986vd} and \cite{Ogievetsky:1986hu}. The corresponding Bethe equations can be written in the following form \cite{Saleur:1999cx}

\beq \label{BGAA}
\left(\frac{u_j+\sfrac{i}{2}V_{K_j}}{u_j-\sfrac{i}{2}V_{K_j}} \right)^L=\prod^M _{l=1,j\neq j} \frac{u_j-u_l+\sfrac{i}{2} M_{K_j\, K_l}}{u_j-u_l-\sfrac{i}{2} M_{K_j\,K_l}}\,,
\eeq
where $M_{K_j\,K_l}$ denotes the symmetric Cartan matrix and $V_{K_j}$ is the spin representation vector. There are all together $M$ excitations (with respect to the vacuum state), among which $K_1$ of the type 1, $K_2$ of the type 2, etc. With the help of the Bethe roots $u_j$ one can parametrize each conserved quantity:
\begin{enumerate}
\item the momentum (the eigenvalue of the translation operator on the lattice, see \eqref{Verschiebungsoperator})
\beq
e^{i P}=e^{i (p_1+\ldots+p_M)}=\prod^{M}_{j=1} \frac{u_j+\sfrac{i}{2} V_{K_j}}{u_j-\sfrac{i}{2} V_{K_j}}\,,
\eeq
\item the energy
\beq \label{Energie}
E=c L \pm \sum_{j=1}^M\lr{\frac{i}{u_j+\sfrac{i}{2}V_{K_j}}-\frac{i}{u_j-\sfrac{i}{2}V_{K_j}}}\, \qquad \mbox{and}
\eeq
\item the higher conserved charges
\beq
Q_r=c_r L \pm \sum_{j=1}^M\lr{\frac{i}{(u_j+\sfrac{i}{2}V_{K_j})^{r-1}}-\frac{i}{(u_j-\sfrac{i}{2}V_{K_j})^{r-1}}}\,.
\eeq
\end{enumerate}
The constants $c$ and $c_r$ depend on the choice of the Dynkin diagram. Physically  they correspond to the values of the energy and the conserved higher charges of the chosen vacuum state.

The $\psua$ superalgebra is a real form of the complex $\mathfrak{sl}(4\vert 4)$ superalgebra and therefore the spectral equations of an integrable spin chain with the $\psua$ as the underlying symmetry algebra should also be contained in \eqref{BGAA}. The highest weight state corresponding to the Dynkin diagram presented in the previous section is a $\frac{1}{2}$-BPS field and as such is a suitable choice of the vacuum. The set of all excitations on this vacuum is a closed sector of the theory. Moreover, every highest weight state can be represented in the form \eqref{AAbar}, which implies that this set coincides with the excitation sector of the full planar $\N=4$ SYM theory. The choice of the vacuum, however, breaks the original $\psua$ symmetry. The residual symmetry is
\beq
\suzz \otimes \suzz
\eeq
and transforms the sets $\{\oa^\dagger_1,\oa^\dagger_2 ,\oc^\dagger_1, \oc^\dagger_2\}$ and $\{\ob^{\dagger}_1,\ob^{\dagger}_2, \od^\dagger_1,\od^\dagger_2\}$ in \eqref{AAbar} respectively. The central charge $\hat{H}$ of both superalgebras should be identified with the anomalous dilatation operator $\delta \hat{D}=\hat{D}-D_0$ in the following way (see \cite{Beisert:2004ry})
\beq \label{suzzzentraleladung}
\hat{H}=\frac{M}{2}+\frac{1}{2} \delta \hat{D}\,.
\eeq
It should be stressed that the overall central charge of the residual symmetry algebra is $2 \hat{H}$. The first level S-matrix can be decomposed into a product of two $\suzz$ S-matrices
\beq \label{SpsuazuSpsuzz}
\hat{S}_{\psua}(p_1,p_2)=S_0(p_1,p_2) \cdot \bigg(\hat{S}_{\suzz}(p_1,p_2) \otimes \hat{\bar{S}}_{\suzz}(p_1,p_2)\bigg)\,,
\eeq
where $S_0(p_1,p_2)$ is a scalar function. 

The form of the $\hat{S}_{\suzz}$ S-matrices in the asymptotic regime have been derived recently in \cite{Beisert:2005tm}.  The derivation makes use of the central extension of the $\suzz$, where the additional central elements are related to some braiding element which modifies the coalgebra structure, see \cite{Plefka:2006ze}. This S-matrix is also invariant under a Yangian \cite{Beisert:2007ds}, see also \cite{Spill:2008tp}.

The one-loop Bethe equations can be found directly form \eqref{BGAA}. The Cartan matrix corresponding to the Dynkin diagram presented in the previous section reads
\beq
M_{K_j,K_l}=\left(\begin{array}{c|c|c|c|c|c|c}
-2&+1&0&0&0&0&0   \\\hline
+1&0&-1&0&0&0&0   \\\hline
0&-1&+2&-1&0&0&0   \\\hline
0&0&-1&+2&-1&0&0   \\\hline
0&0&0&-1&+2&-1&0   \\\hline
0&0&0&0&-1&0&+1 \\\hline
0&0&0&0&0&+1&-2
\end{array}\right) \,,
\eeq
whereas the representation vector takes the following form
\beq \label{darstellungsvektor}
V=(0,0,0,1,0,0,0)\,.
\eeq
Thus, the one-loop spectral equations of the planar $\N=4$ SYM are given by
\beqa \nonumber
1&=&
\prod_{j=1,j\neq k}^{K_1}
\frac{u_{1,k}-u_{1,j}-i}{u_{1,k}-u_{1,j}+i} \prod_{j=1}^{K_2}
\frac{u_{1,k}-u_{2,j}+\sfrac{i}{2}}{u_{1,k}-u_{2,j}-\sfrac{i}{2}}
\nonumber \\
1&=&
\prod_{j=1}^{K_1}
\frac{u_{2,k}-u_{1,j}+\sfrac{i}{2}}{u_{2,k}-u_{1,j}-\sfrac{i}{2}}
\prod_{j=1}^{K_3}
\frac{u_{2,k}-u_{3,j}-\sfrac{i}{2}}{u_{2,k}-u_{3,j}+\frac{i}{2}}\,,
\nonumber \\
1&=&
\prod_{j=1}^{K_2}
\frac{u_{3,k}-u_{2,j}-\sfrac{i}{2}}{u_{3,k}-u_{2,j}+\sfrac{i}{2}}
\prod_{j=1, j \neq k}^{K_3}
\frac{u_{3,k}-u_{3,j}+i}{u_{3,k}-u_{3,j}-i}
\prod_{j=1}^{K_4}
\frac{u_{3,k}-u_{4,j}-\sfrac{i}{2}}{u_{3,k}-u_{4,j}+\sfrac{i}{2}}
\nonumber \\
\lr{\frac{u_{4,j}+\sfrac{i}{2}}{u_{4,j}-\sfrac{i}{2}}}^L&=&
\prod_{ j=1}^{K_3}
\frac{u_{4,k}-u_{3,j}-\sfrac{i}{2}}{u_{4,k}-u_{3,j}+\sfrac{i}{2}}
\prod_{j=1,j \neq k}^{K_4}
\frac{u_{4,k}-u_{4,j}+i}{u_{4,k}-u_{4,j}-i}
\prod_{j=1}^{K_5}
\frac{u_{4,k}-u_{5,j}-\sfrac{i}{2}}{u_{4,k}-u_{5,j}+\sfrac{i}{2}}
\nonumber \\
1&=&
\prod_{j=1}^{K_6}
\frac{u_{5,k}-u_{6,j}-\sfrac{i}{2}}{u_{5,k}-u_{6,j}+\sfrac{i}{2}}
\prod_{j=1, j \neq k}^{K_5}
\frac{u_{5,k}-u_{5,j}+i}{u_{5,k}-u_{5,j}-i}
\prod_{j=1}^{K_4}
\frac{u_{5,k}-u_{4,j}-\sfrac{i}{2}}{u_{5,k}-u_{4,j}+\sfrac{i}{2}}
\nonumber \\
1&=&
\prod_{j=1}^{K_7}
\frac{u_{6,k}-u_{7,j}+\sfrac{i}{2}}{u_{6,k}-u_{7,j}-\sfrac{i}{2}}
\prod_{j=1}^{K_5}
\frac{u_{6,k}-u_{5,j}-\sfrac{i}{2}}{u_{6,k}-u_{5,j}+\frac{i}{2}}\,,
\nonumber \\
1&=&
\prod_{j=1,j\neq k}^{K_7}
\frac{u_{7,k}-u_{7,j}-i}{u_{7,k}-u_{7,j}+i} \prod_{j=1}^{K_2}
\frac{u_{7,k}-u_{6,j}+\sfrac{i}{2}}{u_{7,k}-u_{6,j}-\sfrac{i}{2}} \,. \label{BGeinschleifen}
\eeqa
One notes that the above system of equations is symmetric with respect to the equation for the momentum-carrying roots $u_4$. This reflects the decomposition of the S-matrix \eqref{SpsuazuSpsuzz}.

Since the trace is cyclic, the total momentum must be a multiplicity of $2 \pi$ and consequently the Bethe equations must be supplied with the condition
\beq \label{impulsbedingung}
1=e^{iP}=
\prod_{j=1}^{K_4}
\frac{u_{4,j}+\sfrac{i}{2}}{u_{4,j}-\sfrac{i}{2}}\,.
\eeq
The excitation numbers $\{K_i\}, i=1,\ldots,7$ appearing in the above equations are uniquely determined through the Dynkin labels of a state\footnote{These relations can be derived from the weights of the corresponding oscillators.}
\beq
\left(\begin{array}{l} K_1\\K_2\\K_3\\K_4\\K_5\\K_6\\K_7 \end{array}\right)=\left(\begin{array}{l}
\frac{1}{2}\Delta_0-\sfrac{1}{2}(L-B)-\frac{1}{2} s_1\\
\phantom{\frac{1}{2}}\Delta_0-\phantom{\frac{1}{2}}(L-B)\\
\phantom{\frac{1}{2}}\Delta_0-\frac{1}{2}(L-B)-\frac{1}{2} p-\sfrac{3}{4}q_1-\sfrac{1}{4}q_2\\
\phantom{\frac{1}{2}}\Delta_0\phantom{\mathord{}-\sfrac{0}{2}(L-B)}-\phantom{\frac{1}{2}}p-\frac{1}{2} q_1-\frac{1}{2} q_2\\
\phantom{\frac{1}{2}}\Delta_0-\frac{1}{2}(L+B)-\frac{1}{2} p-\sfrac{1}{4}q_1-\sfrac{3}{4}q_2\\
\phantom{\frac{1}{2}}\Delta_0-\phantom{\frac{1}{2}}(L+B)\\
\frac{1}{2}\Delta_0-\frac{1}{2}(L+B)-\frac{1}{2} s_2
\end{array}\right).
\eeq
As mentioned before, the anomalous part of the dilatation operator can be identified with the Hamiltonian of the $\psua$ super spin chain and thus the energies of the states of this spin chain should be proportional to the eigenvalues of the dilatation operator. Explicitly, after solving the Bethe equations\footnote{The roots $u_j$ are in general complex. Since the Bethe equations are invariant under complex conjugation, the complex solutions form pairs $(u_j,u^{*}_j)$ rendering \eqref{gammaeinschleifen} real.} the anomalous dimension can be determined from 
\beq \label{gammaeinschleifen}
\gamma=2\,g^2\,\sum_{j=1}^{K_4}\lr{\frac{i}{u_j+\sfrac{i}{2}}-\frac{i}{u_j-\sfrac{i}{2}}}+\Op(g^4)\,,
\eeq
as follows from \eqref{Energie}. For the choice of the representation vector made in \eqref{darstellungsvektor} only the excitations (magnons) of type 4 carry the momentum. This allows for the following interpretation. A magnon is created by exciting the fourth node of the Dynkin diagram, which corresponds to the action of the fourth positive root on the vacuum state $\Tr \lr{\fldZ^L}$, see figure \ref{fig:beauty2}. Exciting subsequently the neighboring nodes changes the flavor of the magnon, but does not create any new ``particles''. In a similar manner, one can excite further nodes (though always those adjacent to the already excited ones) and as a result the following inequalities must be satisfied
\beq
K_1 \leq K_2 \leq K_3 \leq K_4 \geq K_5 \geq K_6 \geq K_7 \,.
\eeq
A detailed study of the $\psua$ representations \cite{Dobrev:1985qv} shows that no adjacent excitation numbers, $K_i$ and $K_{i+1}$, may be equal.
\begin{center}
\begin{figure}
\setlength{\unitlength}{1pt}%
\small\thicklines%
\begin{center}
\begin{picture}(260,50)(0,-20)
\put(  0,00){\circle{25}}%
\put(  0,15){\makebox(0,0)[b]{$K_1$}}%
\put(  0,-15){\makebox(0,0)[t]{$\oa^\dagger_2 \oa^1$}}
\put( 12,00){\line(1,0){26}}%
\put( 50,00){\circle{25}}%
\put( 50,15){\makebox(0,0)[b]{$K_2$}}%
\put( 50,-15){\makebox(0,0)[t]{$\oa^\dagger_1 \oc^1$}}%
\put( 62,00){\line(1,0){26}}%
\put( 100,00){\circle{25}}%
\put( 100,15){\makebox(0,0)[b]{$K_3$}}%
\put( 100,-15){\makebox(0,0)[t]{$\oc^\dagger_1 \oc^2$}}%
\put( 112,00){\line(1,0){26}}%
\put(150,00){\circle{25}}%
\put(150,15){\makebox(0,0)[b]{$K_4$}}%
\put(150,-15){\makebox(0,0)[t]{$\oc^\dagger_2 \od^\dagger_1$}}%
\put(162,00){\line(1,0){26}}%
\put(200,00){\circle{25}}%
\put(200,15){\makebox(0,0)[b]{$K_5$}}%
\put(200,-15){\makebox(0,0)[t]{$\od^\dagger_2 \od^1$}}%
\put(212,00){\line(1,0){26}}%
\put(250,00){\circle{25}}%
\put(250,15){\makebox(0,0)[b]{$K_6$}}%
\put(250,-15){\makebox(0,0)[t]{$\ob^\dagger_1 \od^2$}}%
\put(262,00){\line(1,0){26}}%
\put(300,00){\circle{25}}%
\put(300,15){\makebox(0,0)[b]{$K_7$}}%
\put(300,-15){\makebox(0,0)[t]{$\ob^\dagger_2 \ob^1$}}%
\put( 42,-8){\line(1, 1){16}}%
\put( 42, 8){\line(1,-1){16}}%
\put(242,-8){\line(1, 1){16}}%
\put(242, 8){\line(1,-1){16}}%
\end{picture}
\end{center}
\caption{Beauty diagram with the positive simple roots and the corresponding excitation numbers.} \label{fig:beauty2}
\end{figure}
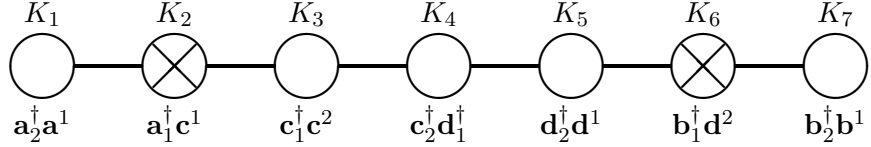
\end{center}
\section{Wrapping Interactions} \label{sec:wicklungswirkungen}
The anomalous dimensions at higher orders of perturbation theory can be determined in a two different manners. The first method amounts to evaluating the correlation functions \eqref{zweipunkt} or \eqref{dreipunkt}  of the corresponding operators to the desired order. Alternatively, as shown in \cite{Beisert:2003tq}, one can derive the dilatation operator using some further constraints (BMN scaling \cite{Berenstein:2002jq}, integrability, the closure of the symmetry algebra, etc.). For the full planar $\N=4$ SYM theory this is a very tedious task already at the two-loop order and this method was only applied in some closed subsectors \cite{Beisert:2003ys}-\cite{Zwiebel:2005er}. 

As it was explained in section \ref{sec:tHooft}, Feynman diagrams of a $U(N)$ gauge theory consist of the ``fat'' propagators and can be classified according to the genus of the associated surface. In each genus class one can furthermore distinguish the so called wrapping diagrams. In what follows, we will discuss these diagrams in the planar limit. 

Let $\hat{\Op}(x)$ be a single trace operator, which for the sake of simplicity we will assume to be built out of two scalar fields
\beq
\hat{\Op}=\Tr(\hat{\Phi}_1 \hat{\Phi}_2)\,.
\eeq
The corresponding two-point function can be determined using the well-know formula
\beq \label{zweipunktsr}
\langle 0 \vert \, \hat{\Op}(x) \hat{\Op}(y) \, \vert 0 \rangle=\frac{\langle 0 \vert \, \hat{\Op}_0(x) \hat{\Op}_0 (y)\, e^{i \int d^4 z \hat{\mathcal{L}}_{int} (z)} \, \vert 0 \rangle}{\langle 0 \vert \, e^{i \int d^4 z \hat{\mathcal{L}}_{int} (z)} \,\vert 0 \rangle}
\eeq
and the Wick theorem. In the above expression $\Op_0$ denotes the asymptotic field and $\mathcal{L}_{int}$ the interaction part of the Lagrangian. In order to define the wrapping interactions it is convenient to introduce the so called spectator fields $\psi, \bar{\psi}$, see \cite{Sieg:2005kd}. The fields $\psi$ are inserted into the trace $\Op(x)$ in the following way
\beq
\Tr\lr{\hat{\Phi}_1(x) \hat{\Phi}_2(x)} \mapsto \Tr \lr{\psi(x) \hat{\Phi}_1(x) \psi (x) \hat{\Phi}_2(x) \psi(x)}\,.
\eeq
Similarly $\bar{\psi}$ are inserted into $\hat{\Op}(y)$. The both auxiliary fields $\psi$ und $\bar{\psi}$ should be contracted while calculating \eqref{zweipunktsr} (it should be noted, however, that $\psi$ and $\bar{\psi}$ are free fields) and their contraction will be graphically represented by
\begin{center}
\begin{picture}(100,20)(0,-7)
\put(0,0){\makebox{$\psi(x)$}}
\put(30,0){\makebox{$\bar{\psi}(y)$}}
\put(10,-10){\line(0,1){5}}
\put(10,-10){\line(1,0){30}}
\put(40,-10){\line(0,1){5}}
\put(60,0){\makebox{$=$}}
\put(83,3){\line(1,0){3}}
\put(90,3){\line(1,0){3}}
\put(97,3){\line(1,0){3}}
\put(104,3){\line(1,0){3}}
\put(111,3){\line(1,0){3}}
\put(118,3){\line(1,0){3}}
\put(125,3){\line(1,0){3}}
\put(132,3){\line(1,0){3}}
\put(139,3){\line(1,0){3}}
\end{picture}
\end{center}
We exclude contractions for which two lines of the spectator fields cross each other. Also two adjacent parallel lines are considered to be equivalent.

\begin{figure}
\begin{center}
\includegraphics[scale=0.5]{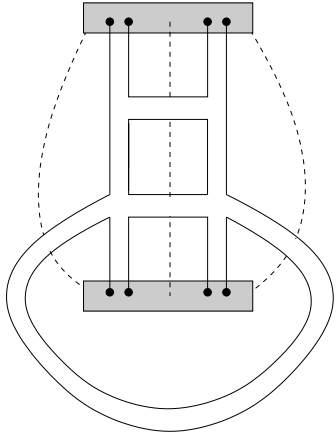}
\caption{An example of a planar wrapping diagram that contributes to the two-point correlation function. The grey rectangle represents a single trace operator, which is composed from two fundamental fields.} \label{fig:wrapping}
\end{center}
\end{figure}

A wrapping diagram is defined to be a Feynman diagram such that for every contraction of the spectator fields all the lines of the spectator fields cross the lines of the other fields. An example of such diagram is shown in figure \ref{fig:wrapping}. Because of their topological definition, wrapping diagrams can contribute starting from the order $\Op(g^{2L})$ only since at lower orders at least one contraction line of the spectator fields does not cross any contraction line of the other fields. In some special cases, the wrapping interactions may be delayed even beyond this order, as for example occurs for operators that can be identified within different closed subsectors and with different corresponding lengths.

As mentioned before, it was possible to determine the dilatation operator to the first few orders in some simple closed subsectors \cite{Beisert:2003ys}-\cite{Zwiebel:2005er}. Thereby it was assumed that the loop order $\ell$ is smaller than the lenght $L$ of a state on which the dilatation operator acts. The reason for this are precisely the wrapping interactions, which are highly non-local and cannot be determined through the symmetry algebra. Moreover, the form of the wrapping interactions is different for different lengths.
\section{Asymptotic Bethe Ansatz} \label{sec:ABA}
At higher-loop orders it is currently unknown with Hamiltonian of which integrable spin chain one should identify the dilatation operator. Despite this fact, based on the analysis of the closed subsectors \cite{Beisert:2003ys}, \cite{Beisert:2004hm}, a conjecture on the form of the asymptotic all-loop Bethe equations for the planar $\N=4$ SYM theory was made in \cite{Beisert:2005fw}. These were subsequently proven in \cite{Beisert:2005tm} under the assumption of factorizability of the S-matrix.

The Dynkin diagram (called ``Beauty'') that we have considered so far is, however, unsuitable for the purpose of formulating the asymptotic all-loop spectral equations. The reason being the fact that the $\mathfrak{so}(6)$ subsector, which can be consistently obtained by truncating all except the three middle nodes of the diagram, is not a closed subsector beyond the one-loop level. Let $\fldX$, $\fldY$ and $\fldZ$ denote the fields of the $\mathfrak{so}(6)$ subsector. The following mixing process
\beq \label{mischungsprozess}
\fldX \fldY \fldZ \longleftrightarrow \fldU \fldV\,,
\eeq
where $\fldU$, $\fldV$ denote fermionic fields, must be taken into account already at the two-loop order (this process also lowers the length of the state, see discussion below). One concludes therefore that $\mathfrak{so}(6)$ Dynkin diagram should not appear as a subdiagram of the $\psua$ at higher-loop order. 

A suitable all-loop choice of the $\psua$ Dynkin diagram is presented in figure \ref{fig:mulatadiagramm}. Contrary to the Beauty diagram it contains four rather then two simple positive fermionic roots. Another pleasant feature of this choice is that the corresponding vacuum state remains unchanged
\beq
\vert \fldZ \rangle^L = \Tr \left(\fldZ^L \right) \,.
\eeq
Similarly to the previous case, there are all together sixteen fundamental excitations. We present them in the figure \ref{fig:mulatadiagramm}, which should be understood as follows. The field $\fldX$ corresponds to a single excitation of the main (momentum-carrying) node and generally a state in the $i$-th column and $j$-th row is obtained by exciting once $(i-1)$ nodes left to the central node and $(j-1)$ nodes to the right.

As in the one-loop case, one associates to each single trace operator a state of a closed spin chain. The all-loop system exhibits, in contradistinction to the usual spin chains, many novel features. First of all, the length of the spin chain must be smaller then the loop-order, else the non-local wrapping interactions need to be taken into account and their influence on the dynamics of the system is currently unknown. Generically, the $\ell$-loop Hamiltonian acts simultaneously on $\ell+1$ adjacent lattice sites. A feature that rarely occurs for integrable systems. In what follows, the regime $L>\ell$ will be called the asymptotic region and all quantities defined therein will be marked as asymptotic. Probably the most striking feature of this asymptotic all-loop spin chain are the fluctuations of length due to the flavor mixing processes, e.g.~\eqref{mischungsprozess}. This may seem to invalidate the above definition of the asymptotic region, however the supersymmetry delays the wrapping interactions for the operators with lower length mixing with the higher length operators.  Despite all these unusual properties, the asymptotic all-loop spin chain seems to be integrable, though this has not been proven rigorously yet. On the other hand, the perturbative asymptotic integrability for the first few orders of perturbation theory have been confirmed in some subsectors of the full theory, see \cite{Staudacher:2004tk} and \cite{Beisert:2005fw}. Under the assumption that this holds to all-loop order in the asymptotic region, Beisert have derived \cite{Beisert:2005tm}  the corresponding S-matrix \eqref{SpsuazuSpsuzz} up to an overall scalar factor. The asymptotic scattering matrix, due to the assumed integrability, remains local, in contradistinction to the Hamiltonian, also at higher loop orders. Because of the decomposition \eqref{SpsuazuSpsuzz}, it suffices to construct one copy of  the $\suzz$ S-matrix. It turns out, however, that each $\suzz$ algebra must be extended with two additional local charges in order to overcome the very restrictive particle representations of $\suzz$. These additional charges vanish on the physical states, which in both cases is equivalent to the momentum constraint $1=e^{iP}=e^{i(p_1+\ldots+p_M)}$. It is interesting to note that after introduction of these charges the corresponding S-matrix is uniquely determined from the invariance condition
\beq
\forall \hat{J} \in \suzz \ltimes \mathfrak{u}(1)^2: \qquad [\hat{J} \otimes \mathbb{I}+\mathbb{I} \otimes \hat{J},\hat{S}_{\suzz} (p_1,p_2)]=0\,.
\eeq
As was explained in section \ref{sec:isk}, the Bethe equations can be obtained from the periodicity conditions of the $k$-th level spin chain, where in this case $k=1,2,3,4$ (the total number of excitations is four).

To formulate these equations it is necessary to introduce, in addition to the rapidity $u$, the deformation variables
\beq \label{uparametrisierung}
x(u)=\frac{1}{2} u \lr{1+\sqrt{1-\frac{4g^2}{u^2}}}\,, \qquad x^\pm (u)=x(u \pm \frac{i}{2})\,.
\eeq
The asymptotic all-loop Bethe equations \cite{Beisert:2005fw}, \cite{Beisert:2005tm} can then be written as
\begin{center}
\begin{figure}[t]\label{magic square}
\begin{minipage}{360pt}
\setlength{\unitlength}{1pt}
\small\thicklines
\begin{center}
\begin{picture}(200,200)(30,-180)
\put( 55,-144){\makebox(0,0)[t]{$\oc^\dagger_1 \oa^{2}$}}
\put( 80,-152){\circle{15}}
\put( 80,-119){\line(0,-1){26}}
\put( 55,-104){\makebox(0,0)[t]{$\oa^\dagger_2 \oa^{1}$}}
\put( 80,-112){\circle{15}}
\put( 80,-79){\line(0,-1){26}}
\put( 55,-64){\makebox(0,0)[t]{$\oa^\dagger_1 \oc^{2}$}}
\put( 80,-72){\circle{15}}
\put( 80,-07){\line(0,-1){58}}
\put( 80,00){\circle{15}}
\put( 80,25){\makebox(0,0)[t]{$\oc^\dagger_2 \od^\dagger_2$}}
\put( 87,00){\line(1,0){66}}
\put(120,-35){\makebox(0,0)[b]{$\fldX$}}
\put(120,-75){\makebox(0,0)[b]{$\dot{\fldU}$}}
\put(120,-115){\makebox(0,0)[b]{$\dot{\fldV}$}}
\put(120,-155){\makebox(0,0)[b]{$\bar{\fldY}$}}
\put(160,00){\circle{15}}
\put(160,25){\makebox(0,0)[t]{$\od^2 \ob^\dagger_1$}}
\put(160,-35){\makebox(0,0)[b]{$\fldU$}}
\put(160,-75){\makebox(0,0)[b]{$\cder$}}
\put(160,-115){\makebox(0,0)[b]{$\dot{\cder}$}}
\put(160,-155){\makebox(0,0)[b]{$\bar{\fldV}$}}
\put(167,00){\line(1,0){26}}
\put(200,00){\circle{15}}
\put(200,25){\makebox(0,0)[t]{$\ob^1 \ob^\dagger_2$}}
\put(200,-35){\makebox(0,0)[b]{$\fldV$}}
\put(200,-75){\makebox(0,0)[b]{$\dot{\bar{\cder}}$}}
\put(200,-115){\makebox(0,0)[b]{$\bar{\cder}$}}
\put(200,-155){\makebox(0,0)[b]{$\bar{\fldU}$}}
\put(207,00){\line(1,0){26}}
\put(240,00){\circle{15}}
\put(240,25){\makebox(0,0)[t]{$\ob^2 \od^\dagger_1$}}
\put(240,-35){\makebox(0,0)[b]{$\fldY$}}
\put(240,-75){\makebox(0,0)[b]{$\dot{\bar{\fldV}}$}}
\put(240,-115){\makebox(0,0)[b]{$\dot{\bar{\fldU}}$}}
\put(240,-155){\makebox(0,0)[b]{$\bar{\fldX}$}}
\put( 75,-157){\line(1,1){10}}
\put( 75,-147){\line(1,-1){10}}
\put( 75,-77){\line(1,1){10}}
\put( 75,-67){\line(1,-1){10}}
\put(155,-5){\line(1,1){10}}
\put(155,5){\line(1,-1){10}}
\put(235,-5){\line(1,1){10}}
\put(235,5){\line(1,-1){10}}
\linethickness{0.75pt}
\multiput(100,-15)(40,0){5}{\line(0,-1){160}}
\multiput(100,-15)(0,-40){5}{\line(1,0){160}}
\end{picture}
\end{center}
\end{minipage}
\caption{All-loop Dynkin diagram of $\psua$ together with the sixteen fundamental excitations} \label{fig:mulatadiagramm}
\end{figure}
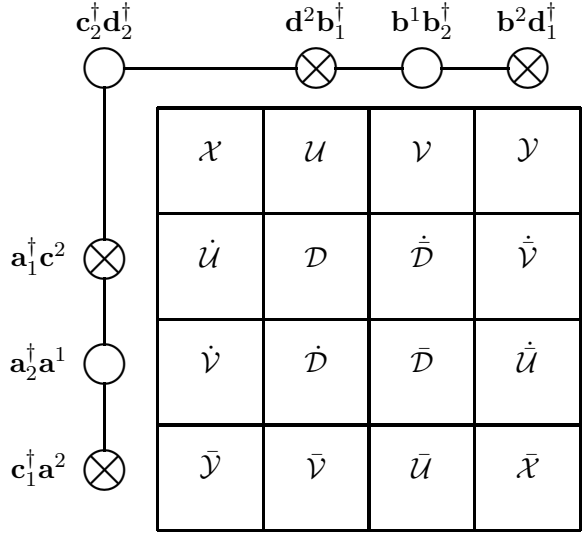
\end{center}
\beqa \label{allschleifenbg}
\nonumber
1&=&e^{iP}=e^{i(p_1+\ldots+p_{K_4})}=
\prod_{j=1}^{K_4}
\frac{x_{4,j}^+}{x_{4,j}^-}\,,
\nonumber\\
1&=&
\prod_{\textstyle}^{K_2}
\frac{u_{1,k}-u_{2,j}+\sfrac{i}{2}}{u_{1,k}-u_{2,j}-\sfrac{i}{2}}
\prod_{j=1}^{K_4}
\frac{1-g^2/x_{1,k}x_{4,j}^{+}}{1-g^2/x_{1,k}x_{4,j}^{-}}\,,
\nonumber \\
1&=&
\prod_{j=1, j\neq k}^{K_2}
\frac{u_{2,k}-u_{2,j}-i}{u_{2,k}-u_{2,j}+i}
\prod_{j=1}^{K_3}
\frac{u_{2,k}-u_{3,j}+\sfrac{i}{2}}{u_{2,k}-u_{3,j}-\frac{i}{2}}
\prod_{j=1}^{K_1}
\frac{u_{2,k}-u_{1,j}+\sfrac{i}{2}}{u_{2,k}-u_{1,j}-\frac{i}{2}}\,,
\nonumber \\
1&=&
\prod_{\textstyle}^{K_2}
\frac{u_{3,k}-u_{2,j}+\sfrac{i}{2}}{u_{3,k}-u_{2,j}-\sfrac{i}{2}}
\prod_{j=1}^{K_4}
\frac{x_{3,k}-x_{4,j}^{+}}{x_{3,k}-x_{4,j}^{-}}\,,
\nonumber \\
1&=&
\lr{\frac{x^-_{4,k}}{x^+_{4,k}}}^L
\prod_{ j=1,j\neq k}^{K_4}
\lr{
\frac{x_{4,k}^{+}-x_{4,j}^{-}}{x_{4,k}^{-}-x_{4,j}^{+}}\,
\frac{1-g^2/x_{4,k}^+ x_{4,j}^-}{1-g^2/x_{4,k}^- x_{4,j}^+}\,
\sigma^2(x_{4,k},x_{4,j})}
\nonumber \\ \qquad
&&\times
\prod_{}^{K_1}
\frac{1-g^2/x_{4,k}^{-} x_{1,j}}{1-g^2/x_{4,k}^{+}x_{1,j}}
\prod_{j=1}^{K_3}
\frac{x_{4,k}^{-}-x_{3,j}}{x_{4,k}^{+}-x_{3,j}}
\prod_{j=1}^{K_5}
\frac{x_{4,k}^{-}-x_{5,j}}{x_{4,k}^{+}-x_{5,j}}
\prod_{j=1}^{K_7}
\frac{1-g^2/x_{4,k}^{-}x_{7,j}}{1-g^2/x_{4,k}^{+}x_{7,j}}\,,
\nonumber \\
1&=&
\prod_{}^{K_6}
\frac{u_{5,k}-u_{6,j}+\sfrac{i}{2}}{u_{5,k}-u_{6,j}-\sfrac{i}{2}}
\prod_{j=1}^{K_4}
\frac{x_{5,k}-x_{4,j}^{+}}{x_{5,k}-x_{4,j}^{-}}\,,
\nonumber \\
1&=&
\prod_{ j=1,j\neq k}^{K_6}
\frac{u_{6,k}-u_{6,j}-i}{u_{6,k}-u_{6,j}+i}
\prod_{j=1}^{K_5}
\frac{u_{6,k}-u_{5,j}+\sfrac{i}{2}}{u_{6,k}-u_{5,j}-\frac{i}{2}}
\prod_{j=1}^{K_7}
\frac{u_{6,k}-u_{7,j}+\sfrac{i}{2}}{u_{6,k}-u_{7,j}-\frac{i}{2}}\,,
\nonumber \\
1&=&
\prod_{ j=1}^{K_6}
\frac{u_{7,k}-u_{6,j}+\sfrac{i}{2}}{u_{7,k}-u_{6,j}-\sfrac{i}{2}}
\prod_{j=1}^{K_4}
\frac{1-g^2/x_{7,k}x_{4,j}^{+}}{1-g^2/x_{7,k}x_{4,j}^{-}}\,,
\eeqa
The scalar factor $\sigma^2(u,v)$ appearing on the right hand side of the fourth eqution is the so called dressing factor and is closely related to $S_0(p_1,p_2)$ in \eqref{SpsuazuSpsuzz}
\beq
S_0\lr{u,v}=\frac{x^{-}(u)-x^{+}(v)}{x^{+}(u)-x^{-}(v)} \, \frac{1-\frac{g^2}{x^{+}(u) x^{-}(v)}}{1-\frac{g^2}{x^{-}(u) x^{+}(v)}}\,\sigma^2 \lr{u,v}\,.
\eeq
In \cite{Beisert:2005wv} it was advocated that its most general form is
\beq \label{dpdefinition}
\sigma^2(u,v)=\exp(2\,i\,\theta(u,v))\,,
\eeq
where the phase $\theta(u,v)$ is given through
\beq \label{thetaphase}
\theta(u_k,u_j)=\sum^{\infty}_{r=2} \sum^{\infty}_{\nu=0} \beta_{r,r+1+2\nu}(g)\lr{q_r(u_k)\,q_{r+1+2\nu}(u_j)-q_r(u_j)\,q_{r+1+2\nu}(u_k)}\,.
\eeq
In \cite{Beisert:2006ez}, based on the transcendental properties of the scaling function and homogeneity of its power expansion in $g$, the explicit form of the coefficients $\beta_{r,r+1+2\nu}(g)$ was proposed 
\beq
\beta_{r,r+1+2\nu}(g)=\sum^{\infty}_{\mu=\nu} g^{2r+2\nu+2\mu} \beta^{(r+\nu+\mu)}_{r,r+1+2\nu}\,,
\eeq
where
\beq
\beta^{(r+\nu+\mu)}_{r,r+1+2\nu}=0 \qquad \mbox{for} \qquad \mu < r+\nu-1 \,,
\eeq
and elsewhere
\beq
\beta^{(r+\nu+\mu)}_{r,r+1+2\nu}=2(-1)^{r+\nu+\mu} \frac{(r-1)(r+2\nu)}{2\mu+1}\,{2\mu+1 \choose \mu-r-\nu+1}\,{2\mu+1 \choose \mu-\nu}\,\zeta(2\mu+1) \,.
\eeq
It turns out that the dressing factor contributes at weak coupling starting from the four-loop order only, see \cite{Bern:2006ew}-\cite{Beisert:2007hz}. The excitation numbers, similarly to the one-loop case, are uniquely determined through the labels of the state in question. The explicit relation for the all-loop Dynkin diagram reads
\beq \label{allschleifenaz}
\left(\begin{array}{l} K_1\\K_2\\K_3\\K_4\\K_5\\K_6\\K_7 \end{array}\right)=\left(\begin{array}{l}
\sfrac{1}{2}(L-B)-\sfrac{1}{2}p-\sfrac{1}{4}(3q_1+q_2)\\
\frac{1}{2}\Delta_0-\frac{1}{2}(p+s_1)-\sfrac{1}{4}(3q_1+q_2)\\
\Delta_0-\frac{1}{2}(L-B)-\frac{1}{2} p-\sfrac{1}{4}(3q_1+q_2)\\
\Delta_0-p-\sfrac{1}{2}(q_1+q_2)\\
\Delta_0-\frac{1}{2}(L+B)-\frac{1}{2} p-\sfrac{1}{4}(q_1+3q_2)\\
\frac{1}{2}\Delta_0-\sfrac{1}{2}(p+s_2)-\sfrac{1}{4}(q_1+3q_2)\\
\frac{1}{2}(L+B)-\sfrac{1}{2}p-\sfrac{1}{4}(q_1+3q_2)
\end{array}\right).
\eeq
The expresion for the anomalous dimension \eqref{gammaeinschleifen} is generalized as follows
\beq \label{allschleifenad}
\gaba=g^2 \gaba_1+g^4 \gaba_2+\ldots=2\,g^2\,\sum^{K_4}_{j=1} \lr{\frac{1}{(x^+ (u_j))}-\frac{1}{(x^- (u_j))}} \,.
\eeq
As it was discussed in the preceding section, starting from the order $\Op(g^{2L})$ one needs to take the wrapping interactions into account and consequently equations \eqref{allschleifenbg} are generically not valid at and beyond this order. In some subsectors, as for example in the case of the $\mathfrak{sl}(2)$ subsector discussed in the next section, the wrapping corrections are delayed by supersymmetry and the asymptotic Bethe equations \eqref{allschleifenbg} remain valid up to the order $\Op \lr{g^{2L+2}}$. In this article all quantities that have been calculated with help of the asymptotic Bethe ansatz will be marked with the label ``ABA''. 

The higher conserved charges are given by 
\beq \label{hLadungen}
{Q_r\inddowns{ABA}}=\sum^{K_4}_{j=1} q_r \inddowns{ABA} (u_j)=2\,g^2\,\sum^{K_4}_{j=1} \lr{\frac{1}{(x^+ (u_j))^{r-1}}-\frac{1}{(x^- (u_j))^{r-1}}}\,.
\eeq
For symmetric root distribution all odd charges \eqref{hLadungen} vanish.

Physical solutions of the system \eqref{allschleifenbg}, that is solutions that correspond to physical states of the theory, must have different values of the rapidities at each nesting level
\beq
\forall j\neq k: \qquad u_{n,k} \neq u_{n,j} \qquad n=1,\ldots,7
\eeq
since in the opposite case the corresponding wave function vanishes.
\chapter{Analytical Properties of Twist Operators}
Twist operators have played a major role in performing tests of the AdS/CFT correspondence since they are conjectured to be dual to the so called spinning-string on the $AdS_5 \times S^5$, which in some limit may be treated semiclassically. In this section we will review the analytic properties of twist operators.
\section{The $\mathfrak{sl}(2)$ Subsector} \label{sec:sl2untersektor}
The $\mathfrak{sl}(2)$ subsector is a closed subsector with operators being composed from scalar fields $\fldZ$ and covariant derivatives $\cder$
\beq \label{twistop}
\Tr \left(\cder^\M\, \fldZ^L \,\right) + \ldots\, .
\eeq
The dots in  \eqref{twistop}  stand for all possible permutations of the derivatives over the $\fldZ$ fields with suitable coefficients in front. The length of these operators is equal to the number of the $\fldZ$ fields (see subsection \ref{sec:psualgebra}) and the covariant derivatives $\cder$ should be interpreted as excitations on the $\Tr \lr{\fldZ^L}$ vacuum. The number of excitations is unbounded $M=0,1,2,...\ldots$ and may exceed the length of the operator. This is due to the fact that the $\mathfrak{sl}(2)$ representation is infinite-dimensional. The simplest operators in this sector are composed of two scalar fields und arbitrary number of covariant derivatives
\beq \label{twistzwei}
\Tr \left(\,\fldZ \, \cder^\M\, \fldZ \,\right) + \ldots\, .
\eeq
 
The weights of the $\mathfrak{sl}(2)$ operators can be immediately read off from their field content \eqref{twistop}. The labels of the primary state must be, however, slightly shifted in order to comply with the unitarity \cite{Dobrev:1985qv}
\beq \label{sl2gewichte}
\{\Delta_0,s_1,s_2,q_1,p,q_2,B,L\}=\{L+M-1,M-1,M-1,1,L-2,1,0,L\}\,.
\eeq
This corresponds to the action of some lowering operators $J^-$ in \eqref{senkoperatoren}. A twist of an operator is defined as 
\beq
T=\Delta_0-\frac{1}{2} \lr{s_1+s_2}\,.
\eeq
According to this definition the twist of these operator is equal $L$.

The closure of the $\mathfrak{sl}(2)$ subsector is also reflected in the compact form of the corresponding Bethe equations. It follows from \eqref{sl2gewichte} together with \eqref{allschleifenaz} that the excitation numbers on the all-loop diagram (figure \ref{fig:mulatadiagramm}) are given by
\beq
\{K_1,K_2,K_3,K_4,K_5,K_6,K_7\}=\{0,0,M-1,M,M-1,0,0\}\,.
\eeq
Since the third and the fifth equations can be written as
\beq
P(x_{3,k})=0, \qquad P(x_{5,k})=0 \,,
\eeq
with
\beq \label{Pdef}
P(x)= \prod^M_{j=1} \lr{x-x^+ _{4,j}}-\prod^M_{j=1} \lr{x-x^- _{4,j}}=c\,\prod^{M-1}_{j=1} \lr{x-x_{3,j}}=c\,\prod^{M-1}_{j=1} \lr{x-x_{5,j}}\,,
\eeq
one concludes that
\beq
x_{3,k}=x_{5,k} \qquad \mbox{for} \qquad k=1,\ldots, M-1\,.
\eeq
Moreover, it follows from \eqref{Pdef} that
\beq
\frac{P(x^+ _{4,k})}{P(x^- _{4,k})}=\prod^M _{j=1,j\neq k} \frac{x^+ _{4,k}-x^- _{4,j}}{x^- _{4,k}-x^+ _{4,j}}=\prod^M _{j=1} \frac{x^+ _{4,k}-x_{3,k}}{x^- _{4,k}-x_{3,k}}=\prod^M _{j=1} \frac{x^+ _{4,k}-x_{5,k}}{x^- _{4,k}-x_{5,k}} \,.
\eeq
This identity allows to reduce the system of three equations to the equation for the main roots $u_{4,k}$ only
\beq \label{allschleifensl2}
\left(\frac{x^+_k}{x^-_k}\right)^L=
\prod_{\substack{j=1\\j \neq k}}^\M
\frac{x_k^- -x_j^+}{x_k^+-x_j^-}\,
\frac{1-g^2/x_k^+x_j^-}{1-g^2/x_k^-x_j^+}\,
\exp\left(2\,i\,\theta(u_k,u_j)\right)\,.
\eeq
On the solutions of this equation one needs to impose the momentum constraint
\beq
\prod_{k=1}^M \frac{x^+_k}{x^-_k}=1\, .
\eeq
At the one-loop order equation \eqref{allschleifensl2} reduces to the Bethe equations of the non-compact  $XXX_{-\frac{1}{2}}$ spin chain
\beq \label{einschleifensl2}
\bigg(\frac{u_k+\frac{i}{2}}{u_k-\frac{i}{2}}\bigg)^L =
\prod^{M}_{\substack{j=1 \\j\neq k}}\frac{u_k-u_j-i}{u_k-u_j+i}\, .
\eeq
The non-compactness is due to the covariant derivatives, which can occur in arbitrary number at each lattice site.  

The Baxter equation, \textit{cf.} \eqref{BG}, corresponding to \eqref{einschleifensl2}  takes the following form
\beq \label{baxtergleichung}
\left(u+\frac{i}{2} \right)^L \,Q(u+i)+\left(u-\frac{i}{2} \right)^L \,Q(u-i)\,=\,t(u)\, Q(u)\, ,
\eeq
where $t(u)=T^{(a)}(u)$ is the eigenvalue of the auxiliary transfer matrix
\beq \label{kleinetm}
t(u)=2 u^L + q_{L-2}\, u^{L-2}+...+q_{0} \, .
\eeq
The coefficient  $q_{L-1}$ is always zero and $q_{L-2}$ corresponds to the eigenvalue of the Casimir opertor
\beq \label{qcasimir}
q_{L-2}=-J_0 (J_0+1)-\frac{1}{4}L, \qquad \qquad J_0=M+\frac{1}{2}L \,.
\eeq
The remaining charges $q_r$, $r=0,\ldots,L-3$ consitute a complete set of quantum numbers that describe uniquely any given state. Equation \eqref{baxtergleichung}, upon fixing the charges, is a difference equation of the second order and thus has two algebraically independent solutions. One of the solutions is however non-polynomial and we exclude it by writting
\beq \label{Qansatz}
Q(u)=\prod^M_{j=1} (u-u_j)\,.
\eeq
Given a solution,  the one-loop anomalous dimension  \eqref{gammaeinschleifen} may be found from
\beq \label{gammadurchQ}
\gaba=2\,g^2\,\lr{Q'(\frac{i}{2})-Q'(-\frac{i}{2})} + \Op(g^4)\,,
\eeq
as can be easily checked using \eqref{Qansatz}.

A very interesting feature of twist operators is their scaling behavior for large values of the spin. This was first observed in \cite{Belitsky:2006en} at the one-loop order and was confirmed to hold to all orders in \cite{Eden:2006rx} and \cite{Beisert:2006ez}. More precisely, the limit is defined through $M\to \infty$, with $L$ growing slower then logarithmically with $M$. The anomalous dimension to the leading order is hence given by 
\beq \label{skalenlimes}
\Delta -\Delta_0  = \gamma(g)=f(g)\; \log M +\ldots \qquad L\to\infty, M \to \infty \qquad L \ll \log M\,.
\eeq
The universal scaling function $f(g)$ depends only on the coupling constant $g$ and conicides, at least up to three-loop order (see \cite{Eden:2006rx}), with the $L=2$ scaling function. We will discuss this limit in detail in chapter \ref{sec:tests}.
\section{Twist-Two Operators}\label{sec:t2resultate}
In this section we will discuss a special case of $L=2$, in which the anomalous dimension may be found in a closed form as a function of $M$.  
We start by noting that in this case the transfer matrix takes a particularly simple form
\beq
t(u)= 2 u^2-(M^2+M+\frac{1}{2}) \, ,
\eeq
and the difference eqution \eqref{baxtergleichung}
\beq
(u+\frac{i}{2})^2 \,Q(u+i)+(u-\frac{i}{2})^2 \, Q(u-i)\,=\,(2u^2-(M^2+M+\frac{1}{2})) \, Q(u)
\eeq
is solved by the continuos Hahn polynomials
\beq \label{Hahnpolynome}
Q(u)\, =\, {_3 F_{2}} (-M,M+1,\frac{1}{2}-i u;1,1;1)\,.
\eeq
The Bethe roots are simply zeros of this polynomial, see \eqref{Qansatz}. It is clear that for twist-two operators $M$ must take even values since the roots of \eqref{Hahnpolynome} do not obey the momentum constraint \eqref{impulsbedingung} for odd $M$. 

Using the formula \eqref{gammadurchQ}, one finds
\beq \label{t2gamma}
\gaba (M)\, = \, 8\,g^2 S_{1} (M) +\Op \lr{g^4}\, ,
\eeq
where $S_1$ is the harmonic sum
\beq
S_1 (M)= \sum^M _{j=1} \frac{1}{j} \,.
\eeq
The formula \eqref{t2gamma} determines the anomalous dimension of twist-two operators as a function of the spin $M$; a rare occurrence even at the one-loop level.

At higher loops it is also possible to reformulate the Bethe equations  \eqref{allschleifensl2} in the form of the Baxter equation, see \cite{Belitsky:2006wg}. This allows to expand and subsequently solve the Baxter equation order by order in perturbation theory, see \cite{Kotikov:2008pv}. In this article, however, we will not present these solutions, but rather concentrate on the corresponding expressions for the anomalous dimension. Remarkably, one can guess them in a fairly simple manner using the so called transcendentality principle. This method was introduced in \cite{Kotikov:2007cy} and is based on the previous observations made in \cite{Kotikov:2004ss}. It assumes that at each order of the perturbation theory $\ell$ the anomalous dimension is expressed through the generalized hamonic sums of the order $(2\ell-1)$, or through the products of zeta functions and harmonic sums for which the sum of the arguments of the zeta functions and the orders of the harmonic sums is equal to $(2\ell-1)$. The generalized harmonic sums are defined by the following recursive procedure \cite{Vermaseren:1998uu}
\beq \label{vhs}
S_a (M)=\sum^{M}_{j=1} \frac{(\mbox{sgn}(a))^{j}}{j^a}\, , \qquad
S_{a_1,\ldots,a_n}(M)=\sum^{M}_{j=1} \frac{(\mbox{sgn}(a_1))^{j}}{j^{a_1}}
\,S_{a_2,\ldots,a_n}(j)\, .
\eeq
The order $\ell$ of each sum $S_{a_1,\ldots,a_n}$ is given by the sum of the absolute values of its indices
\beq
\ell=\vert a_1 \vert +\ldots \vert a_n \vert\,,
\eeq
and the order of a product of harmonic sums is equal to the sum of orders of its constituents. The canonical basis of the harmonic sums of $\ell$-th order is spanned by
\beqa \label{basis}
&&\left\{S_{a_{11}}, S_{a_{21},a_{22}},\, \ldots,\, S_{a_{\ell1},a_{\ell2},\ldots,a_{\ell\ell}}: \right.\nonumber \\
&&\left.\qquad \qquad \ell=\vert a_{11} \vert=\vert a_{21} \vert+\vert a_{22} \vert=\ldots=\vert a_{\ell 1} \vert+\vert a_{22} \vert+\ldots+\vert a_{\ell\ell} \vert \right\}\,,
\eeqa
where $M$ dependence of the sums is implicit. Each $\ell$-th order product of harmonic sums can be decomposed in this basis. 

We will discuss the aforementioned method of determining the higher order corrections to the anomalous dimension of twist-two operators by taking the two-loop order as an example. In this case, according to the transcendentality principle, the order of the allowed harmonic sums is three (the products of zeta functions and harmonic sums do not contribute at this order) and thus the basis  \eqref{basis} takes the following form
\beqa \label{basiszweischleifen}
&&\{S_3,S_{-3},S_{2,1},S_{1,2},S_{-2,1},S_{1,-2},S_{2,-1},S_{-1,2},S_{-2,-1},S_{-1,-2},\nonumber\\ &&\ \, S_{1,1,1},S_{-1,1,1},S_{1,-1,1},S_{1,1,-1},S_{-1,-1,1},S_{-1,1,-1},S_{1,-1,-1},S_{-1,-1,-1}\} \,.
\eeqa
It was conjectured in \cite{Kotikov:2002ab} that the index $-1$ does not appear in the harmonic sums contributing to the anomalous dimension. Although some Feynman diagrams may lead to harmonic sums with index $-1$, their total contribution should cancel. Moreover, each sum for which its first $k$ indices are equal $1$ scales in the limit $\M \to \infty$ as 
\beq
S_{1,\ldots,1, a_{k+1},...,a_n}(M)\simeq \#\log^k(M) \qquad \mbox{for} \qquad M\gg 1 \,.
\eeq
One thus concludes that, in view of the scaling properties of the anomalous dimension \eqref{skalenlimes}, also $S_{1,1,1}$ may not contribute to this order. The physical basis of the harmonic sums at the two-loop order is consequently a small subset of  \eqref{basiszweischleifen}
\beq
\{S_3,S_{-3},S_{2,1},S_{1,2},S_{-2,1},S_{1,-2}\} \,.
\eeq
The two-loop $\gaba_4$ anomalous dimension can be expanded in this basis
\beqa
\gaba_4(M) &=& c_1\,S_3(M) + c_2 \,S_{-3}(M)+c_3\,S_{2,1}(M)+c_4\,S_{1,2}(M) \nonumber \\
&&+c_5\,S_{-2,1}(M)+c_6\,S_{1,-2}(M)\,,
\eeqa
with $c_1,\ldots,c_6$ being coefficients one still needs to determine. However, it is possible   to find the two-loop correction to the anomalous dimension for the first few values of $M$ from \eqref{allschleifensl2} together with \eqref{allschleifenad}. This is further simplified due to the fact that the one-loop roots are known, see \eqref{Hahnpolynome}. Expanding the roots to the two-loop order in \eqref{allschleifensl2}
\beq
u_k= u^{(0)}_k+g^2\,u^{(1)}_k+\Op(g^4)\,,
\eeq
one finds using \eqref{einschleifensl2} a linear system of equations for $u^{(1)}_k$, which can be easily solved with help of any algebra program (e.g. \texttt{Mathematica}). The coefficients $c_1,\ldots,c_6$ may thus be easily determined
\beq \label{gamma4t2}
\frac{\gaba_4(M)}{16}= S_3 +S_{-3}-2 \bigg(S_{1,2}+2 S_{2,1}\bigg)-2 S_{1,-2} \,.
\eeq
It is only a bit more involved to repeat this procedure for the three-loop correction, with the result being
\beqa \label{gamma6t2}
\frac{\gaba_6(M)}{64}&=&2 S_{5}+2 S_{-5}-4 \bigg(S_{1,4}+S_{4,1}\bigg)-5 \bigg(S_{2,3}+S_{3,2}\bigg)\nonumber \\
&&+4 \bigg(S_{1,2,2}+S_{2,1,2}+ S_{2,2,1}\bigg)+4 \bigg(S_{3,1,1}+S_{1,3,1}+S_{1,1,3}\bigg) \nonumber\\
&&-4 S_{-4,1}-2 S_{-3,-2}-S_{-3,2}-2 S_{-2,-3}-8 S_{1,-4}\nonumber\\
&&-2 S_{3,-2}-9 S_{2,-3}+2 S_{-2,-2,1}+2 S_{-2,1,-2}+8 S_{1,-3,1}\nonumber\\
&&+2 S_{1,-2,-2}+2 S_{1,-2,2}+12 S_{1,1,-3}+4 S_{1,2,-2}+6 S_{2,-2,1}\nonumber\\
&&+4 S_{2,1,-2}-8 S_{1,1,-2,1} \,.
\eeqa
The both expressions \eqref{gamma4t2} and \eqref{gamma6t2} coincide with the field theory computations, see \cite{Kotikov:2002ab} and  \cite{Kotikov:2004ss}. The effectiveness of this method can be however appreciated starting from the four-loop order, where the usual perturbative calculations become very involved. In \cite{Kotikov:2007cy} the four-loop contribution to anomalous dimension was found.  We reproduce this result in table  \ref{sec:gamma8t2}, where in the last row the contribution of the dressing phase was printed in boldface. In the next chapter we will use this result in order to test the veracity of the equations \eqref{allschleifensl2} since one expects the wrapping interactions to contribute at this order.

One can also use the above procedure in order to determine the higher conserved charges. It turns out that the $\ell$-th loop correction to the $2r$-th charge $Q_{2r}$\footnote{All odd charges vanish due to the $u \to -u$ symmetry of the root distribution.} is composed of harmonic sums of the $\lr{2\ell+2r-3}$-th order. Since harmonic sums of different orders are algebraically independent, this confirms the algebraic independence of the higher charges and furnishes an indirect proof of the asymptotic integrability. 
\begin{table}[hp!]
\beqa
& &\bm{4\, \HS_{-7}+6\, \HS_{7}}+ 2\,( \HS_{-3,1,3} + \HS_{-3,2,2} +
      \HS_{-3,3,1} + \HS_{-2,4,1} )
  + 3\,( -\HS_{-2,5}\nonumber\\
  &+& \HS_{-2,3,-2} ) +4\,( \HS_{-2,1,4}
- \HS_{-2,-2,-2,1} -
      \HS_{-2,1,2,-2} - \HS_{-2,2,1,-2} -
      \HS_{1,-2,1,3} \nonumber \\
&-& \HS_{1,-2,2,2} -
      \HS_{1,-2,3,1} )
  + 5\,( -\HS_{-3,4}
+ \HS_{-2,-2,-3} )
  + 6\,(- \HS_{5,-2} \nonumber\\
&+& \HS_{1,-2,4} -
      \HS_{-2,-2,1,-2} - \HS_{1,-2,-2,-2} )
  + 7\,( -\HS_{-2,-5}
+ \HS_{-3,-2,-2} \nonumber\\
&+&
      \HS_{-2,-3,-2} + \HS_{-2,-2,3} )
  + 8\,( \HS_{-4,1,2} + \HS_{-4,2,1} - \HS_{-5,-2} - \HS_{-4,3}\nonumber \\  &-&
      \HS_{-2,1,-2,-2}
+ \HS_{1,-2,1,1,-2} )
  + 9\,\HS_{3,-2,-2}
  -10\,\HS_{1,-2,2,-2}
  + 11\,\HS_{-3,2,-2}\nonumber\\
 &+& 12\,( -\HS_{-6,1} + \HS_{-2,2,-3}
+ \HS_{1,4,-2} + \!\HS_{4,-2,1} + \!
      \HS_{4,1,-2} - \!\HS_{-3,1,1,-2} -\!
      \HS_{-2,2,-2,1}\nonumber\\
&-& \!\HS_{1,1,2,3} - \!
      \HS_{1,1,3,-2} - \!\HS_{1,1,3,2} - \!
      \HS_{1,2,1,3}
- \HS_{1,2,2,-2} -
      \HS_{1,2,2,2} - \HS_{1,2,3,1} -
      \HS_{1,3,1,-2} \nonumber \\
&-& \HS_{1,3,1,2} -
      \HS_{1,3,2,1} - \HS_{2,-2,1,2} -
      \HS_{2,-2,2,1} - \HS_{2,1,1,3}
-
      \HS_{2,1,2,-2} - \HS_{2,1,2,2}\nonumber\\
&-&
      \HS_{2,1,3,1} - \HS_{2,2,1,-2} -
      \HS_{2,2,1,2} - \HS_{2,2,2,1} -
      \HS_{2,3,1,1} - \HS_{3,1,1,-2} -
      \HS_{3,1,1,2}
- \HS_{3,1,2,1}\nonumber\\
&-& \HS_{3,2,1,1} )
+ 13\,\HS_{2,-2,3}
  -14\,\HS_{2,-2,1,-2}
  + 15\,( \HS_{2,3,-2} + \HS_{3,2,-2} )
 \nonumber \\
&+& 16\,( \HS_{-4,1,-2}
+ \HS_{-2,1,-4} - \!
      \HS_{-2,-2,1,2} - \!\HS_{-2,-2,2,1} - \!
      \HS_{-2,1,-2,2} - \!\HS_{-2,1,1,-3} \nonumber\\
&-& \!
      \HS_{1,-3,1,2} - \!\HS_{1,-3,2,1} - \!
      \HS_{1,-2,-2,2}
- \HS_{2,-2,-2,1} +
      \HS_{-2,1,1,-2,1} + \HS_{1,1,-2,1,-2}\nonumber \\
 &+&
      \HS_{1,1,-2,1,2} + \HS_{1,1,-2,2,1} )
  -17\,\HS_{-5,2}
  + 18\,( -\HS_{4,-3}
- \HS_{6,1} + \HS_{1,-3,3} )\nonumber\\
  &+& 20\,( -\HS_{1,-6} - \HS_{1,6} -
      \HS_{4,3} + \HS_{-5,1,1} +
      \HS_{-4,-2,1} + \HS_{-3,-2,2} +
      \HS_{-2,-4,1} \nonumber\\
&+& \HS_{-2,-3,2} +
      \HS_{1,3,3} + \HS_{3,1,3} +
      \HS_{3,3,1} - \HS_{1,1,-2,3} -
      \HS_{1,2,-2,-2} - \HS_{2,1,-2,-2} )\nonumber \\
 &-&21\,\HS_{3,4}
+ 22\,( \HS_{1,-2,-4} + \HS_{2,2,3} +
      \HS_{2,3,2} + \HS_{3,-2,2} + \HS_{3,2,2})
+ 23\,( -\HS_{-3,-4} \nonumber \\
&-& \HS_{5,2} +
      \HS_{2,-2,-3} )
+ 24\,( -\HS_{-4,-3} + \HS_{1,-4,-2} -
      \HS_{1,-3,1,-2} - \HS_{1,1,1,4} -
      \HS_{1,1,4,1}\nonumber\\
&-& \HS_{1,3,-2,1} -
      \HS_{1,4,1,1} - \HS_{3,-2,1,1}
- \HS_{3,1,-2,1} - \HS_{4,1,1,1} +
      \HS_{-2,-2,1,1,1} + \HS_{-2,1,-2,1,1}\nonumber\\
&+&
      \HS_{1,-2,-2,1,1} + \HS_{1,-2,1,-2,1} +
      \HS_{1,1,-2,-2,1}
+ \HS_{1,1,1,-2,-2} +
      \HS_{1,1,2,-2,1} + \HS_{1,2,1,-2,1}\nonumber\\
&+&
      \HS_{2,1,1,-2,1} )
+ 25\,\HS_{2,-3,-2}
  + 26\,( -\HS_{2,5} + \HS_{1,4,2}
+ \HS_{2,4,1} + \HS_{4,1,2} + \HS_{4,2,1})\nonumber\\
&+& 28\,( \HS_{1,2,4} + \HS_{2,1,4} -
      \HS_{-3,1,-2,1} - \HS_{-2,1,-3,1} -
      \HS_{1,-2,1,-3} )
+ 30\,\HS_{-3,1,-3} \nonumber\\
  &+& 32\,( \HS_{1,5,1} + \HS_{5,1,1} -
      \HS_{-3,-2,1,1} - \HS_{-2,-3,1,1} -
      \HS_{1,-3,-2,1} - \HS_{1,-2,-3,1} \nonumber\\
&-&\HS_{2,2,-2,1} + \HS_{1,2,-2,1,1} +
      \HS_{2,1,-2,1,1} - \HS_{1,1,1,-2,1,1} )
  + 36\,( \HS_{1,1,5} + \HS_{1,3,-3} \nonumber\\
&+&      \HS_{3,1,-3}
- \HS_{1,1,-3,-2} - \!
      \HS_{1,1,-2,-3} - \!\HS_{1,1,2,-3} - \!
      \HS_{1,2,-2,2} - \!\HS_{1,2,1,-3} - \!
      \HS_{2,1,-2,2} \nonumber \\
&-& \!\HS_{2,1,1,-3} )
  + \!38\,\HS_{-3,-3,1}
+ 40\,( -\HS_{1,-4,1,1} - \HS_{2,-3,1,1} +
      \HS_{1,1,1,-2,2} )\nonumber\\
&-&41\,\HS_{3,-4}
  + 42\,( -\HS_{2,-5} + \HS_{1,-4,2} +
      \HS_{1,-3,-3} )
+ 44\,( \HS_{1,-5,1} + \HS_{2,-3,2}
+      \HS_{3,-3,1} )\nonumber \\
&+& 46\,\HS_{2,2,-3}
  + 48\,\HS_{1,1,-3,1,1}
  + 60\,( \HS_{1,1,-5} - \HS_{1,1,-3,2} )
+ 62\,\HS_{2,-4,1}
+ 64\,\HS_{1,1,1,-3,1}\nonumber\\
 &+& 68\,( \HS_{1,2,-4} + \HS_{2,1,-4} -
      \HS_{1,2,-3,1}- \HS_{2,1,-3,1} )
-72\,\HS_{1,1,1,-4}
-80\,\HS_{1,1,-4,1}\nonumber\\
&-&\bm{\mathbf{\zeta(3)}\HS_1(\HS_3-\HS_{-3}+2\,\HS_{-2,1})}.
\nonumber
\eeqa
\caption{The result for $\frac{\gaba_8}{256}$, see \cite{Kotikov:2007cy}.} \label{sec:gamma8t2}
\end{table}
\section{The One-Loop Non-Linear Integral Equation}
In this section we will introduce the concept of the holes which are dual excitations to the magnons. With help of the counting function we will rewrite the one-loop Bethe equations for the $\mathfrak{sl}(2)$ subsector in the form of a non-linear integral equation. This new representation allows to understand the integrability from the ``dual'' side and to investigate analytically many limiting cases.

\subsection{Magnons and Holes} \label{sec:mundl}
Using the ansatz \eqref{Qansatz} the left hand side of the Baxter equation \eqref{baxtergleichung} becomes a polynomial of the  $(L+M)$-th order and according to the fundamental theorem of algebra it must posses $(L+M)$ roots. Putting
\beq
u=u_k \qquad k=1,2,\ldots,M\,,
\eeq
the right hand side of \eqref{baxtergleichung} vanishes and one obtains the Bethe equations \eqref{einschleifensl2}. The remaining  $L$ solutions are roots of the transfer matrix eigenvalue $t(u)$
\beq \label{kleintmprodukt}
t(u)=2\,\prod_{k=1}^L(u-u_{\rm h}^{(k)})\, .
\eeq
These additional roots will be called holes in what follows. For $u=u^{(k)} _{\rm k}$ the Baxter equation \eqref{baxtergleichung} gives
\beq \label{einschleifenlg}
\bigg(\frac{u^{(k)} _{\rm h}+\frac{i}{2}}{u^{(k)}_{\rm h}-\frac{i}{2}}\bigg)^L =
\prod^{M}_{\substack{j=1 \\ j\neq k}}\frac{u^{(k)}_{\rm h}-u_j-i}{u^{(k)} _{\rm h}-u_j+i}  \qquad k=1,2,\ldots,L \,.
\eeq
The right hand side of this equation contains the product over the usual Bethe roots and consequently the hole roots may be found only after the Bethe roots are determined. This is related to the fact that the charges $q_j$ discussed in the previous section may be expressed through the Bethe roots $u_k$, as can be seen from \eqref{baxtergleichung} together with \eqref{Qansatz}. By comparing  \eqref{kleinetm} and \eqref{kleintmprodukt} one concludes that
\beq
\sum^L_{j=1} u^{(j)}_{\rm h}=0\,,
\eeq
which corresponds to the momentum constraint \eqref{impulsbedingung} for the holes. Intuitively, one can consider the hole roots as rapidities of the  $\fldZ$ fields. We will discuss this point below. 

In what follows we will confine ourselves to the ground states (states with the lowest anomalous dimension in the sector) and to even values of the Lorentz spin $M$, in which case both the magnon roots as well as the hole roots are real and symmetrically distributed around the origin. We will further assume that $L \ll M$. The charges $q_k$ in \eqref{kleinetm} depend in general on $M$, however for the assumed values of parameters it was shown in \cite{Belitsky:2006en} that the second charge $q_2$ is strongly dominating. This implies that two hole roots are much bigger then the other
\beq \label{u1u2}
u^{(1)}_{\rm h}=-u^{(2)}_{\rm h}\simeq \sqrt{\frac{q_2}{2}}\approx \frac{M}{\sqrt{2}}\,.
\eeq
In order to derive an integral equation we rewrite the one-loop equations in the logarithmic form 
\beq
2L\,\arctan \lr{2\,u_k}=2\,\pi\,n_k-2\,\sum^{\frac{M-1}{2}}_{j=-\frac{M-1}{2}} \arctan \lr{u_k-u_j}\,,
\eeq
where we have chosen $\rm Arctan$ branch of the logarithms and grouped the roots as follows
\beq
u_{-k}=-u_{k} \qquad k=\frac{1}{2},\,\frac{3}{2},\ldots, \frac{M-1}{2}\,.
\eeq
In \cite{Eden:2006rx} it was conjectured that the mode numbers  $n_k$ of the ground states are given by
\beq \label{modenzahlen}
n_k=k+\frac{L-2}{2}\, \mbox{sgn}(k) \qquad \mbox{for} \qquad k \,=\,\pm \frac{1}{2} \pm \frac{3}{2},...,\pm \frac{M-1}{2}\, .
\eeq
The value of the roots grow monotonically with $n_k$. The mode numbers of the hole roots, as we will show below using the counting function, can be splitted into two groups. The two universal holes have large mode numbers
\beq \label{uniloecher}
n^{u,1}_{\rm h}=\frac{L+M-1}{2}\,, \qquad n^{u,2}_{\rm h}=-\frac{L+M-1}{2}\, .
\eeq
The rest of the hole roots fills the gap around zero opened by the distribution of the magnon roots 
\beq \label{restloecher}
n^{r}_{\rm h} \in \left\{-\frac{L-3}{2}, \ldots, \frac{L-3}{2} \right\}\,.
\eeq
Hence, the following inequlity between magnon and hole roots is satisfied 
\beq
\vert u_{\rm h}^{(1,2)}\vert > \vert u_k \vert > u_{\rm h}^{(j)} \quad (j\neq1,2)\,.
\eeq

\subsection{The Counting Function and the Non-Linear Integral Equation} \label{sec:zfplusie}
A particularly suitable quantity to describe simultaneously magnons and holes is the scaling function, see \cite{Feverati:2006tg}. In the case of the  $\mathfrak{sl}(2)$ operators it is convenient to adopt the following definition (see \cite{Freyhult:2007pz}) 

\beq \label{zaehlenfunktion}
Z(u)=L\,\phi(u,\frac{1}{2})+\sum_{k=-\frac{M-1}{2}}^{\frac{M-1}{2}}\phi(u-u_k,1),\quad \mbox{where}\quad
\phi(u,\xi)=i\,\log\left(\frac{i\xi+u}{i\xi-u}\right)\,.
\eeq
The origin of its name is due to the relation
\beq \label{sfbu}
Z(\pm\infty)=\pm \,\pi\,(L+M)
\eeq
in conjunction with
\beqa \label{ZBW}
Z(u_j)&=&\pi\,(2\,n_j+\delta-1)\,, \qquad j=\pm \frac{1}{2},\ldots, \pm \frac{M-1}{2}\,,  \\
\label{ZLW}
Z(u_{\rm h}^{(k)})&=&\pi\,(2\,n_{\rm h}^{(k)}+\delta-1)\,, \qquad k=1,\ldots, L\,,
\eeqa
as can be easily confirmed using \eqref{zaehlenfunktion} and \eqref{einschleifensl2}. Here, we have introduced
\beq
\delta=L \quad\mbox{mod}\quad 2\,.
\eeq
Therefore $Z(u)$ is a continuos function which, whenever $u$ is equal to the magnon or the hole root, returns the corresponding mode number. With the help of the counting function it is straightforward to determine the expression for the mode numbers of the holes. Directly from \eqref{modenzahlen} together with \eqref{sfbu}-\eqref{ZLW} one confirms  the validity of \eqref{uniloecher} and \eqref{restloecher}.

The derivation of the non-linear integral equation is based on the following fundamental identity (see \cite{Feverati:2006tg} and references therein)
\beqa \label{wesentlicheid}
\sum_{k=1}^M
f(u_k)+\sum_{j=1}^L\,f(u_{\rm h}^{(j)})&=&-\int_{-\infty}^\infty\frac{du}{2\pi}\,f'(u)\,Z(u) \nonumber \\
&& +\int_{-\infty}^\infty
\frac{du}{\pi}\,f'(u)\,\mbox{Im}\log\left[1+(-1)^\delta\,e^{i\,Z(u+i\,0)}\right]\,,
\eeqa
which can be proven by contour integration methods. Applying this identity to the sum in  \eqref{zaehlenfunktion}, one finds after performing Fourier transformation \cite{Freyhult:2007pz} the integral equation for the scaling function\footnote{Due to superficial divergencies, it is more transparent to apply \eqref{wesentlicheid} to the second derivative of the corresponding sum subsequently integrating twice. The integration constants are fixed by antisymmetry of $Z(u)$ and
\begin{equation}\label{cond1}
\lim_{u \to \infty} Z'(u)=0\, .
\end{equation}
}

\beqa\label{Zuraum}
\nonumber Z(u)&=&i\,L\,\log\frac{\Gamma\left(1/2+i\,u\right)}{\Gamma\left(1/2-i\,u\right)}+\sum_{j=1}^L\,i\,\log\frac{\Gamma\left(-i\,(u-u_{\rm h}^{(j)})\right)}{\Gamma\left(i\,(u-u_{\rm h}^{(j)})\right)}\\
&&+\lim_{\alpha \to \infty}\int_{-\alpha}^\alpha \frac{dv}{\pi}\,i\,\frac{d}{du}\log\frac{\Gamma(-i\,(u-v))}{\Gamma(i\,(u-v))}\,\mbox{Im}\log\left[1+(-1)^\delta\, e^{i\,Z(v+i\,0)}\right].
\eeqa
The non-linear term must be regularised due to the asymptotic behavior of the integrand. It should be stressed that the non-linear term may not be integrated by parts, else the equation would become linear what contradicts the non-linearity of the Bethe equations. The reason for this are discontinuities of the term $\mbox{Im}\log\left[1+(-1)^\delta\, e^{i\,Z(v+i\,0)}\right]$, which need to be taken into account while integrating. 

The identity \eqref{wesentlicheid} allows to express all higher charges \eqref{hLadungen} through the counting function $Z(u)$
\beq \label{Qpeinschleifen}
Q_p=-\int\frac{du}{2\pi}q_p'(u)Z(u)-\sum_{j=1}^L\,q_p(u_h^{(j)})+\int\frac{dv}{\pi}q_p'(v)\textrm{Im}\log\left[1+(-1)^\delta e^{iZ(v+i0)}\right] \,.
\eeq
Here, $q_p(u)$ denote the corresponding charge densities. The first charge also needs to be regularized
\beqa
P&=&\lim_{\alpha \to \infty}\left(-\int_{-\alpha}^\alpha \frac{du}{2\pi}\,p'(u)\,Z(u)-\sum_{j=1}^L
p(u_{\rm h}^{(j)})\right. \nonumber \\
&&+\left.\int_{-\alpha}^\alpha \frac{du}{\pi}\,p'(u)\,\textrm{Im}\log\left[1+(-1)^\delta\,e^{i\,Z(u+i\,0)}\right]\right)\,,
\eeqa
with $p(u)\equiv q_1(u)$ being the momentum density of a magnon
\beq
p(u)=\frac{1}{i}\,\log\frac{u+i/2}{u-i/2}.
\eeq
Using the antisymmetry of $Z(u)$ and $p(u)$ one finds the usual momentum constraint \eqref{impulsbedingung}
\beq
P=0.
\eeq
In a similar manner, the one-loop anomalous dimension $\gaba_2$ is given by
\beqa \label{einschleifengammadurchZ}
\nonumber
\gaba_2&=&
4\,\gamma_{\rm E}\, L+2\,\sum_{j=1}^L\left\{\psi(1/2+i\,u^{(j)}_{\rm h})+\psi(1/2-i\,u^{(j)}_{\rm h})\right\}\\
&&+2\,\int_{-\infty}^\infty\frac{dv}{\pi}\,i\,\frac{d^2}{dv^2}\,\left(\log\frac{\Gamma\left(1/2+i\,v\right)}{\Gamma\left(1/2-i\,v\right)}\right)\mbox{Im}\log\left[1+(-1)^\delta\,e^{i\,Z(v+i\,0)}\right]\,,
\eeqa
where $\gamma_{\rm E}$ denotes the Euler-Mascheroni constant. 

Equation \eqref{Zuraum} together with the equation for the holes \eqref{ZLW} and the mode  numbers  \eqref{uniloecher}-\eqref{restloecher} are completely equivalent to the Bethe equations \eqref{einschleifensl2} for the ground states\footnote{The generalization to all operators in the sector is fairly simple, though we will not consider it in this article.}.

As it was shown in \cite{Freyhult:2007pz}, the last term in  \eqref{Zuraum}, which we will subsequently call the  $\rm{Im log}$-term, at large values of $M$ and for $u \ll M$ can be approximated by
\beq \label{imlogterm}
\lim_{\alpha \to \infty}\int_{-\alpha}^\alpha \frac{dv}{\pi}\,i\,\frac{d}{du}\log\frac{\Gamma(-i\,(u-v))}{\Gamma(i\,(u-v))}\,\mbox{Im}\log\left[1+(-1)^\delta\, e^{i\,Z(v+i\,0)}\right]\simeq 2\log{2}\, u \,.
\eeq
Therefore, the equation for the ``small'' holes \eqref{ZLW} $u^{(j)}_{\rm h}, j=3,4,\ldots,L$ takes the following explicit form
\beq \label{klg}
2^{-2\,i\,u^{(k)}_{\rm h}}\lr{\frac{\Gamma(\frac{1}{2}+i\,u^{(k)}_{\rm h})}{\Gamma (\frac{1}{2}-i\,u^{(k)}_{\rm h})}}^L=\prod^L_{j=1,j\neq k} \frac{\Gamma\lr{i(u_k-u_j)}}{\Gamma\lr{-i(u_k-u_j)}}\,.
\eeq
The product on the right hand side runs over all hole roots, while the equation itself is valid for
 $k=3,4,\ldots,L$. This equation was first derived with the help of the Baxter equation in \cite{Belitsky:2006en}, where also its solutions have been studied in detail. In particular, it was shown that the small hole roots $u^{(j)}_{\rm h}\, (j=3,4,\ldots L)$ scale like
\beq \label{klskalierung}
\max \left\{u^{(3)}_{\rm h},u^{(4)}_{\rm h},\ldots,u^{(L)}_{\rm h}\right\} \sim \tilde{c}\,\frac{L}{\log M}+\Op\lr{\frac{L^2}{\log^2 M}}
\eeq
when $M \gg 1$.

All physical quantities can be expressed through hole rapidities and therefore holes can be considered as dual excitations of the spin chain. Since the number of holes is equal to the number of $\fldZ$ fields, this suggests to identify them with the  $\fldZ$ fields. In the dual description the $\fldZ$ fields scatter on the (unphysical) $\Tr(\cder^M)$ vacuum. Thereby two of the holes move very fast and scatter with itself and the remaining slow holes $u^{(j)}_{\rm h},\ j=3,4,\ldots,L$. It is instructive to note that for the twist-two operators there are no slow holes and the scattering of the fast holes can be seen as the two-body problem. This clarifies the exact solvability in this case. 

\subsection{The Thermodynamic Limit} \label{sec:thermolimes}
In the case of compact spin chains one usually defines the thermodynamic limit by taking the length of the spin chain to be infinite. Non-compact spin chains, in addition, offer the possibility 
to take the number of excitations very large as compared to the length. In the context of the AdS/CFT correspondence this is often much more interesting than the usual thermodynamic limit. In this section we will assume $L \to \infty$ and $M \to \infty$ such that
\beq \label{skalenlimes2}
L \ll M \,.
\eeq
For these values of $L$ and $M$ the mode numbers are explicitly known, see \eqref{modenzahlen}. 

It was conjectured in \cite{Korchemsky:1995be} that in this limit the magnon roots densly cover   the interval $(-\infty,-a)\cup (a,\infty)$\footnote{Accordingly, one may show using the properties of orthogonal polynomials that the roots of \eqref{Hahnpolynome} condense on the whole real axis when $M\to \infty$.}. Therefore it is appropriate to introduce the density of roots. We define it through
\beq \label{dichtedef}
\rho_0(u)=\frac{1}{M} \sum^{\frac{M-1}{2}}_{j=-\frac{M-1}{2}} \delta\lr{u-u_j}\,,
\eeq
from which the normalization condition follows
\beq \label{rhonormierung}
\lr{\int^{-a}_{-\infty}+\int^{\infty}_{a}}dv \rho_0(v)=1\,.
\eeq
Each sum over the Bethe roots can be expressed through the density
\beq \label{fdurchrho}
\sum^{\frac{M-1}{2}}_{j=-\frac{M-1}{2}} f(u_j)=M \, \lr{\int^{-a}_{-u(M)}+\int^{u(M)}_{a}}dv \rho_0(v) f(v) \,,
\eeq
where $u(M)$ denotes the boundary of the magnon distribution
\beq
\lim_{M \to \infty} u(M)=\infty \,.
\eeq
Some expressions will need to be regularized with $u(M)$.

The mode numbers \eqref{modenzahlen} can be also obtained from the function\footnote{This continuation is, contrary to the counting function, not smooth for finite $M$.}
\beq \label{nvonu}
n(u)=-\frac{M}{2}+\sum^{\frac{M-1}{2}}_{j=-\frac{M-1}{2}} \theta \lr{u-u_j}+\frac{L-2}{2}\, \mbox{sgn}(u), \quad \qquad n(u_k)=n_k\,.
\eeq
One then notes the following relation between $n(u)$ and the density
\beq
\rho_0(u)=\frac{1}{M} \frac{d}{du} \lr{n(u)-\frac{L-2}{2}\mbox{sgn}(u)}\,.
\eeq
Thus in the limit $M\to \infty$, \textit{cf.} \eqref{ZBW},
\beq \label{Zzurho}
\frac{1}{M} \frac{d}{du} Z(u)=2\,\pi \rho_0(u)+2\,\pi\,\frac{(L-2)}{M}\delta(u)\,.
\eeq
Upon differentiating \eqref{zaehlenfunktion}, one obtains with help of  \eqref{fdurchrho} and \eqref{Zzurho} an integral equation for the density
\beq \label{rhointgleich}
2\,\pi\,\rho_0(u)+2\,\pi\,\frac{L-2}{M}
\delta(u)
-\frac{L}{M} \frac{1}{u^2+\frac{1}{4}}
-2
\left(\int_{-u(M)}^{-a} dv + \int_a^{u(M)} dv\right)
\frac{\rho_0(v)}{(u-v)^2+1}
=0\,.
\eeq
This equation must be supplemented with the normalization condition \eqref{rhonormierung}. We will discuss equation \eqref{rhointgleich} in detail in section \ref{sec:tests}.

It follows from \eqref{klskalierung} that in the limit \eqref{skalenlimes} three cases should be distinguished. In the first one $L$ is kept fixed or goes to infinity slower then logarithmically  with respect to $M$; then $a=0$. In the second case $L=j\,\log M$ and according to \eqref{klskalierung} the boundary parameter $a$ is constant $a=a(j)$. The last case applies to $L$ diverging faster then logarithmically, when $a$ is non-constant anymore and grows with $M$. We will not discuss the last case in this review.

\section{The Non-Linear Integral Equation at Higher Orders} \label{sec:allschleifenig}
In this section we will generalize the non-linear integral equation to higher orders of the perturbation theory.

Similarly to the one-loop case, the counting function is defined as the logarithm of the corresponding Bethe equations \eqref{allschleifensl2}
\beqa \label{allschleifenzf}
\nonumber Z(u)&=&i\,L\,\log\frac{x(i/2+u)}{x(i/2-u)}+i\,\sum^{M}_{k=1}
\log\frac{i+u-u_k}{i-(u-u_k)}\\
&&-2\,i\,\sum^{M}_{k=1}
\log\frac{1+\frac{g^2}{x(i/2+u)x(i/2-u_k)}}{1+\frac{g^2}{x(i/2-u)x(i/2+u_k)}}+2\sum^{M}_{k=1}
\theta (u,u_k).
\eeqa
With help of \eqref{wesentlicheid} one obtains the desired result
\beqa \label{allschleifennig}
\nonumber Z(u)&=&i\,L\,\log\frac{x(i/2+u)}{x(i/2-u)}+\int_{-\infty}^\infty\frac{dv}{2\pi}\,\phi'(u-v,1)\,Z(v)\\
\nonumber&&-\sum_{j=1}^L\,\phi(u-u_{\rm h}^{(j)},1)-\int_{-\infty}^\infty\frac{dv}{\pi}\,\phi'(u-v,1)\,\mbox{Im}\log\left[1+(-1)^\delta \,e^{i\,Z(v+i\,0)}\right]\\
\nonumber&&+\int_{-\infty}^\infty\frac{dv}{2\pi}\,\left(2\,i\,\frac{d}{dv}\log\frac{1+\frac{g^2}{x(i/2+u)x(i/2-v)}}{1+\frac{g^2}{x(i/2-u)x_(i/2+v)}}-\theta(u,v)\right)\,Z(v) \\ \nonumber
&&+\sum_{j=1}^L \left(2\,i\,\log\frac{1+\frac{g^2}{x(i/2+u)x(i/2-u_{\rm h}^{(j)})}}{1+\frac{g^2}{x(i/2-u)x_(i/2+u_{\rm h}^{(j)})}}-\theta(u,u_{\rm h}^{(j)})\right)\\
&&-\nonumber\int_{-\infty}^\infty\frac{dv}{\pi}\,\left(2\,i\,\frac{d}{dv}\,\log\frac{1+\frac{g^2}{x(i/2+u)x(i/2-v)}}{1+\frac{g^2}{x(i/2-u)x_(i/2+v)}}-\theta(u,v)\right)\times\\
&&\qquad \qquad \quad \mbox{Im}\log\left[1+(-1)^\delta \,e^{i\,Z(v+i\,0)}\right] \, .
\eeqa
Contrary to the previous case \eqref{Zuraum}, linear terms in $Z$ occur on the right hand side of this equation. This is due to the fact that the integral kernels are not of a difference form anymore and cannot be diagonalized simultaneously with the Fourier transform.

The mode numbers are not influenced by quantum corrections as long as the coupling constant is small and, thus, the relations \eqref{ZLW} remain valid beyond the one-loop level. Equation \eqref{allschleifennig} together with \eqref{ZLW} is fully equivalent to the asymptotic Bethe equations \eqref{allschleifensl2} for the ground states.

The all-loop counting function \eqref{allschleifenzf} exhibits similar properties to its one-loop counterpart. In particular
\beq \label{Zzurhoallschleifen}
\frac{1}{M} \frac{d}{du} Z(u)=2\,\pi \rho(u)+2\,\pi\,\frac{(L-2)}{M}\delta(u)\,,
\eeq
with $\rho(u)$ being the all-loop density.

Equation \eqref{allschleifennig} can be further simplified by performing a Fourier transformation. Defining
\beq \label{fouriertransformation}
\hat{F}(t)=\int_{-\infty}^{\infty} du\,e^{-i\,t\,u}\,F(u) \,,
\eeq
one finds after partial diagonalization
\begin{eqnarray}\label{Ztraum}
 \hat{Z}(t)&=&\frac{2 \,\pi \,L
e^{\frac{t}{2}}}{i\,t\,(e^{t}-1)} J_{0}(2gt)-\sum^{L}_{j=1}
\frac{2\,\pi\,\cos \left(t\,u^{(j)}_{\rm h} \right)}{i\,t\,(e^t-1)} -\frac{2}{e^t-1} \hat{\cal L}(t)\nonumber \\
 &+&8\,g^2 \frac{e^{\frac{t}{2}}}{e^t-1} \int^{\infty}_{0}\,dt'\,
e^{-\frac{t'}{2}}\,\hat{K}(2gt,2gt')\,\bigg(t'\,\hat{\cal L}(t')
\nonumber\\
&+&\frac{\pi}{i}
\sum^{L}_{j=1} \cos\left(t'\,u^{(j)}_{\rm h}\right) \bigg) \nonumber \\
&-&4
\,g^2\,\frac{e^{\frac{t}{2}}}{e^{t}-1}\int^{\infty}_{0}\,dt'\,e^{-\frac{t'}{2}}\,t'\,
\hat{K}(2gt,2gt')\,\hat{Z}(t'),
\end{eqnarray}
where $\hat{{\cal L}}(t)$ is the Fourier transform of
\beq
{\cal L}(u)=\mbox{Im}\log\left[1+(-1)^\delta\, e^{i\,Z(u+i\,0)}\right]\,.
\eeq
The kernel $\hat{K}(t,t')$ is defined through (see \cite{Beisert:2006ez})
\beq \label{Kern}
\hat{K}(t,t')=\hat{K}_0(t,t')+\hat{K}_1(t,t')+\hat{K}_{\rm d}(t,t')\,,
\eeq
with
\beq \label{Kern0}
\hat{K}_0(t,t')=\frac{t\,J_1(t)\,J_0(t')-t'\,J_0(t)\,J_1(t')}{t^2-t'^2}\,,
\eeq
and
\beq \label{Kern1}
\hat{K}_1(t,t')=\frac{t'\,J_1(t)\,J_0(t')-t\,J_0(t)\,J_1(t')}{t^2-t'^2}\,.
\eeq
The dressing kernel $\hat{K}_{\rm d} (t,t')$ corresponds to the dressing phase \eqref{thetaphase} and is a convolution of the both previous kernels
\beq \label{KernD}
\hat{K}_{\rm d}(t,t')=8\,g^2\,\int^{\infty}_{0} dt''\, \hat{K}_1(t,2gt'')\,\frac{t''}{e^{t''}-1}\,\hat{K}_0(2gt'',t')\, .
\eeq
The asymptotic conserved charges can be expressed through the counting function in a similar manner to the one-loop case 
\beq \label{Qpallschleifen}
Q_p \inddowns{ABA}=-\int\frac{dv}{2\pi} q_p  \inddowns{ABA}\phantom{}' (v)Z(v)-\sum_{j=1}^L\,q_p \inddowns{ABA} (u_h^{(j)})+\int\frac{dv}{\pi}q_p  \inddowns{ABA}\phantom{}'(v) \mbox{Im}\log\left[1+(-1)^\delta e^{iZ(v+i0)}\right] \,,
\eeq
where $q_p(u)$ is the all-loop charge density \eqref{hLadungen}.

According to the equation \eqref{Ztraum} the function $\hat{Z}(t)$ exhibits a first-order pole at  $t=0$. The reason for this are the asymptotic properties of the counting function \eqref{allschleifenzf}, which is Fourier transformable only in the principle value sense, similarly to e.g. $\arctan(u)$. Equation \eqref{Ztraum} is particularly suitable to investigate different limits in $L$ and $M$.

\chapter{Dynamical Tests of the Asymptotic Bethe Ansatz}
In this chapter we will carry out dynamical tests of the asymptotic Bethe equations in the $\mathfrak{sl}(2)$ subsector. For this purpose we will apply the BFKL equation originating from the high energy QCD. This equation, which was derived by  analyzing leading contributions to the hadronic scattering amplitudes, predicts the pole structure of the anomalous dimension of twist-two operators. With the help of the four-loop result presented in the previous section we will demonstrate that the asymptotic Bethe ansatz is invalid at the wrapping order. This supports the hypothesis that if the system remains integrable the wrapping interactions must modify the structure of the equations \eqref{allschleifenbg}.

\section{The BFKL Equation and the Double Logarithmic Constraints} \label{sec:bfklunddle}

An important problem in the theory of hadrons is to determine the behavior of the scattering amplitudes in the so called Regge limit, when the invariant mass $s$ is much bigger then $t$
\beq \label{reggelimes}
s \gg -t \sim M^2 \,,
\eeq
where $M$ denotes the mass scale of hadrons. Regge discovered that in this limit the high-energy scattering in quantum mechanics is governed by singularities in angular momentum of the partial waves.
Later on, this phenomenon was understood to apply to quantum field theories as well and in particular to the scattering theory of hadrons. Let $A(s,t)$ be a scattering amplitude. The decomposition into partial waves can be performed with help of the Mellin transformation, see \cite{Forshaw:1997dc} and \cite{Korchemsky:1994um} for pedagogical reviews,
\beq
A(s,t)=is \int^{\delta+i\infty}_{\delta-i\infty} \frac{d \omega}{2\pi i} \lr{\frac{s}{M^2}}^{\omega} \bar{A}(\omega,t)\,.
\eeq
Every pole $\omega_0$ of the partial wave $\bar{A}(\omega,t)$ contributes to the amplitude $A(s,t)$ the   term
\beq
f(t)\,s^{\omega_0}\,,
\eeq 
where $f(t)$ is a function of $t$ only. Therefore, knowing all the poles one can reproduce the asymptotic behavior of the scattering amplitude. The nearest (dominating) pole is often called the Pomeranchuk pole.

In the limit \eqref{reggelimes} it was shown, see \cite{Korchemsky:1994um} and references therein, that any $2\to 2$ scattering in the Regge limit \eqref{reggelimes} can be represented in the following form
\beq
A(s,t)=i \sum^{\infty}_{m=0} \alpha s \bigg((\alpha \log s)^m\,f_{m,m}(t)+\alpha(\alpha \log s)^{m-1}\,f_{m,m-1}(t)+\ldots+\alpha^m f_{m,0}(t)\bigg)+\Op\lr{s^0}\,,
\eeq
where $\alpha$ denotes the coupling constant. The analytic structure in the $s$ plane is related to the convergence properties of this series and can be determined only after resummation. Generally, for the QCD or the $\N=4$ SYM, the function $f_{m,n}(t)$ cannot be found explicitly and one is forced to make certain approximations. To the leading order one may neglect all coefficients except for $f_{m,m}$. This is the so called leading logarithmic approximation (LLA). In next-to-leading logarithmic approximation (NNLA) one takes additionally $f_{m,m-1}$ into account, etc. It turns out that in the LLA approximation the hadronic scattering amplitudes are dominated by interacting gluons propagating in the $t$ channel. These gluons, due to their mutual interactions, form collective excitations (see \cite{Fadin:1975cb}-\cite{Kuraev:1977fs}), the so called reggeons. To each reggeon corresponds a sum of infinitely many Feynman diagramms. The scattering of two such excitations is equivalent to taking the LLA appoximation and more generally one can refine the approximation scheme by taking further reggeons into account. It was found in \cite{Fadin:1975cb}-\cite{Kuraev:1977fs} that the gluon-gluon partial waves obey an integral equation (the so called LO BFKL equation), which upon iteration describes the contribution to the scattering amplitude coming from a pair of regeeized gluons. Subsequently, the complete basis of the homogeneous BFKL equation was deterimined \cite{Balitsky:1978ic}, which allowed to derive a functional equation for the eigenvalues. Due to the fact that solely gluons determine the leading behavior of the scattering amplitudes, a similar equation may be found in other gauge theories.

Although $\N=4$ SYM theory does not have hadrons in its spectrum, it is still possible to formulate mathematically the BFKL equation, see \cite{Kotikov:2002ab}. Roughly speaking, the pomeron of the $\N=4$ gauge theory is described by the non-local gauge-invariant operator
\beq
\label{pomeron}
{\rm pomeron} = \Tr\left(  \fldZ\, \cder^{-1+\omega}\, \fldZ\,\right) \,,
\eeq
where the parameter $\omega$ is assumed to be small. It can be shown \cite{Kotikov:2002ab} that the anomalous dimension of this operator (for $\omega \to 0$) in the LLA approximation can be found from the LO BFKL equation
\beq \label{BFKLLLAt2}
\frac{\omega}{- 4\,g^2}=
\Psi\left(-\frac{\gamma}{2}\right)+\Psi\left(1+\frac{\gamma}{2}\right)-
2\,\Psi\left(1\right)\,,
\eeq
where $\Psi(x)$ is the psi function. Due to the resummation of infinitely many Feynman diagramms, the BFKL equation determines \textit{non-perturbatively} the leading singularities. The perturbative solution can be obtained by expanding in $\gamma$ 
\beq \label{bfkl}
\frac{\omega}{- 4\,g^2}=
\frac{2}{\gamma}
-2\sum_{k=1}^{\infty}\left(\frac{\gamma}{2}\right)^{2k}\zeta(2k+1)\,,
\eeq
and subsequently substituting the perturbative expansion of the anomalous dimension
\beq \label{gammaent}
\gamma(g)=g^2 \gamma_2+g^4 \gamma_4+g^6 \gamma_6+\ldots \,.
\eeq
To the first few orders one finds the following perturbative expansion around $\omega=0$
\beq \label{bfklvoraussage}
\gamma=2\,\left(\frac{-4\,g^2}{\omega}\right)
-0\,\left(\frac{- 4\,g^2}{\omega}\right)^2
+0\,\left(\frac{-4\,g^2}{\omega}\right)^3
-4\,\zeta(3)\,\left(\frac{-4\,g^2}{\omega}\right)^4
\pm \ldots \,.
\eeq
It should be noted that the LO BFKL equation predicts the order and the residue of the leading pole at each order of the perturbation theory. In order to find the sub-leading poles one would need to go beyond the leading approximation scheme. 

The quantum numbers of  \eqref{pomeron} can be formally obtained through analytic continuation in $M$ of the corresponding quantum numbers of twist-two operators
\beq
\label{omega}
\{2,3,\ldots\} \ni M\to M=-1+\omega\, .
\eeq
In this sense one can also interpret  the state  \eqref{pomeron}.  This is supported by the well-defined analytical properties of the anomalous dimension and higher conserved charges of twist-two operators for which closed formulas in $M$ are known. Harmonic sums occurring in these expressions, as shown in \cite{Kotikov:2005gr}, may be continued analytically to the whole complex plain.  For example, the one-loop contribution \eqref{t2gamma} can be expressed through the psi function $\Psi(x)$
\beq
\gamma_2=8\,S_1(M)=8 \lr{\Psi(M+1)-\Psi(1)} \,,
\eeq
which is defined for $M \in \mathbb{C}$. It was shown in \cite{Kotikov:2005gr} that all nested harmonic sums \eqref{vhs} can be consistently continued to complex values of $M$ by means of psi functions and higher transcendental functions. After analytic continuation every harmonic sum of the order $\ell$ exhibits poles at negative integer values of $M$ and the order of the highest pole is always smaller or equal $\ell$. In particular, the anomalous dimension of twist-two operators is singular for $M=-1$. This allows to compare the order and the residue of the highest pole with the prediction coming from the BFKL equation. We will use this possibility to verify the veracity of the four-loop result derived in section  \ref{sec:t2resultate} from the asymptotic Bethe equations \eqref{allschleifensl2}. 

Apart from the BFKL equation, predicting the structure of the poles at $M=-1$, there also exist constraints following from the double logarithmic behavior of the scattering amplitudes, which enable to predict the leading singularities at negative \textit{even} vaues of $M$.  Scattering amplitudes in this limit were studied for QED and QCD in \cite{GGLF}, \cite{KirLi} and \cite{BER}. According to the hypothesis formulated in \cite{Kotikov:2002ab,Kotikov:2000pm}, despite the fact that they originate from the double logarithmic behavior, the singularities at $M=j-2=-r$ ($r=2,3,...$) may be predicted from the generalized BFKL equation
\beq \label{VBFKL}
\frac{\omega}{- 4\,g^2}=
\Psi\left(-\frac{\gamma}{2}\right)+\Psi\left(1+
\frac{\gamma}{2}+|n|\right)-
2\,\Psi\left(1\right)\,.
\eeq
This equation generalizes \eqref{bfkl} to the case of higher twist operators due to the relation $\vert n \vert =T-2$. It should be stressed, however, that these higher twist operators are not embedded in the $\mathfrak{sl}(2)$ sector and they still have not been fully identified in the $\N=4$ SYM theory.

Allowing for $|n|$ to become negative $|n|=-r+1$, $r=2,4,\ldots\ $ such that
\beq
\omega =M+r \to 0 \qquad \mbox{and} \qquad |n|+r-1=C_1(r)\,\omega^2+{\cal{O}}(\omega^3)\,,
\eeq
one can, after replacing the argument of the second psi function by
\beq
1+
\frac{\gamma}{2}+|n| \mapsto 1+
\frac{\gamma}{2}+|n|+\omega
\eeq
as explained in \cite{Kotikov:2002ab,Kotikov:2000pm}, derive the following relation from \eqref{VBFKL}
\beq \label{DLE}
\gamma\,(2\,\omega+\gamma)=-16 g^2\, .
\eeq
Physically, this corresponds to taking into account the double logarithmic contributions to scattering amplitudes $\sim (\alpha \mbox{ln}^2 s)^n\,s^{-r+2}$. 

Equation \eqref{DLE} is a polynomial equation of the second order. One of the solutions is unphysical since the anomalous dimension must vanish when $g \to 0$ and thus
\beqa \label{dlvoraussage}
\gamma&=&-\omega+\omega\, \sqrt{1-\frac{16 g^2}{\omega^2}}
\nonumber \\
&=&
2\,\frac{(-4\, g^2)}{\omega}
-2\,\frac{(-4\, g^2)^2}{\omega^3}
+4\,\frac{(-4\, g^2)^3}{\omega^5}
-10\,\frac{(-4\, g^2)^4}{\omega^7}-\ldots\, .
\eeqa

\section{Dynamical Test of the Asymptotic Bethe Equations}

The predictions \eqref{bfklvoraussage} and \eqref{dlvoraussage} can be compared with the analytic continuations of the one-, two- and three-loop corrections \eqref{t2gamma}, \eqref{gamma4t2}, \eqref{gamma6t2} and most importantly with the analytic continuation the four-loop result. For $M=-1+\omega$ one finds
\beq \label{zusammenbruch}
\gaba=2\,\left(\frac{-4\,g^2}{\omega}\right)
-0\,\left(\frac{-4\,g^2}{\omega}\right)^2
+0\,\left(\frac{-4\,g^2}{\omega}\right)^3
-2\,\frac{(-4\,g^2)^4}{\omega^7}
\pm \ldots \,.
\eeq
Around the negative even values $M=-2 +\omega,-4 +\omega,\ldots$ one derives
\beq
\gaba=2\,\frac{(-4\, g^2)}{\omega}
-2\,\frac{(-4\, g^2)^2}{\omega^3}
+4\,\frac{(-4\, g^2)^3}{\omega^5}
-10\,\frac{(-4\, g^2)^4}{\omega^7}-\ldots\,.
\eeq
One thus infers that the one-, two- and three-loop leading singularities, as derived from the asymptotic $\mathfrak{sl}(2)$ Bethe equations \eqref{allschleifensl2}, coincide with the LO BFKL and the double logarithmic predictions. On the other hand, the four-loop correction violates strongly the pole structure predicted by the BFKL equation. After analytic continuation some of the harmonic sums of the seventh order exhibit poles at  $M=-1$ of the order higher then four. According to BFKL equation, however, these poles should cancel each other. This contradiction proves unambiguously that the four-loop result is \textit{incorrect} and consequently that the asymptotic Bethe ansatz \eqref{allschleifensl2}  must fail for twist-two operators at four-loop order. Moreover, this also confirms that the wrapping interactions must be taken into account at this order.

It turns out that this maximal violation of the BFKL prediction can be easily traced back in table \ref{sec:gamma8t2} since after analytic continuation only the first two sums exhibit poles of seventh order
\beq
S_7 (-r+\omega)=-\frac{1}{\omega^7}+\Op\lr{\frac{1}{\omega^6}},\qquad \ S_{-7} (-r+\omega)=-\frac{(-1)^r}{\omega^7}+\Op\lr{\frac{1}{\omega^6}}\,,
\eeq
with $r=-1,-2,\ldots$. The expansion \eqref{bfklvoraussage} necessitates that coefficients in front of the both sums should be equal, while consistency with \eqref{dlvoraussage} fixes these coefficients to $5$. On the other hand, the coefficients following from the ABA are $4$ and $6$ respectively (see table \ref{sec:gamma8t2}).

\section{The NLO BFKL Equation and the Generalized Double Logarithmic Constraints}

In this subsection we will briefly discuss refinements of the BFKL and the double logarithmic predictions. These provide additional constraints on the form of the perturbative anomalous dimension of twist-two operators and may therefore be used in order to check the validity of any yet-to-be-found system of spectral equations of the $\N=4$ gauge theory.

The leading order BFKL equation discussed in subsection \ref{sec:bfklunddle} has recently been generalized by including sub-leading effects in the Regge kinematics. This was first calculated for the QCD  in \cite{Fadin:1998py} and later the $\N=4$ part of the QCD answer was extracted in  \cite{Kotikov:2002ab,Kotikov:2000pm}. Using the dimensional reduction scheme the NLO BFKL equation may be written as
\beq
\frac{\omega}{-4\,g^2} = \chi (\gamma )-g^2\,\delta (\gamma )\,,
\eeq
where
\beqa
\chi (\gamma ) &=&
\Psi\left(-\frac{\gamma}{2}\right)+\Psi\left(1+\frac{\gamma}{2}\right)-
2\,\Psi\left(1\right)\, ,\\[4mm]
\delta (\gamma ) &=&4\,\chi ^{\,\prime \prime } (\gamma )
+6\,\zeta(3)+2\,\zeta(2)\,\chi (\gamma )+4\,\chi (\gamma )\,\chi ^{\,\prime} (\gamma )  \nonumber \\[2mm]
& & -\frac{\pi^3}{\sin \frac{\pi \gamma}{2}}- 4\,\Phi \left(-\frac{\gamma}{2}
\right) -4\,\Phi \left(1+\frac{\gamma}{2} \right)\,.
\eeqa

The function $\Phi (\gamma )$ is given by the following expression
\beqa
\Phi (\gamma ) =~\sum_{k=0}^{\infty }\frac{(-1)^{k}} {(k+\gamma)^2 }\biggl[\Psi
\left(k+\gamma +1\right)-\Psi (1)\biggr]. \label{9}
\eeqa

The perturbative pole structure of the anomalous dimension may be found similarly as in the previous subsection. Since the NLO BFKL equation contains information about the next-to-leading effects, it predicts the residue of the next-to-leading pole at each order of perturbation theory.  Thus, expression \eqref{bfklvoraussage} is refined to
\beqa \label{nbfklvoraussage}
\gamma&=&\left(2+0\,\omega\right)
\left(\frac{-4\,g^2}{\omega}\right) -\left(0+0\,\omega
\right)\,\left(\frac{-4\,g^2}{\omega}\right)^2
+\left(0+\,\zeta(3)\,\omega\right)\,\left(\frac{-4\,g^2}{\omega}\right)^3
\\
&&-\left(4\,\zeta(3)+\frac{5}{4}\,\zeta(4)\,\omega\right)\,\left(\frac{-4\,g^2}{\omega}\right)^4
\pm \ldots \,. \nonumber
\eeqa
This expansion is reproduced through the one-, two- and three-loop results \eqref{t2gamma}, \eqref{gamma4t2} and \eqref{gamma6t2}. At the four-loop order one obtains from \eqref{nbfklvoraussage}  all together four constraints. The first three leading poles must be absent (though, exceptionally, the vanishing of the seventh-order pole implies the vanishing of the pole of the sixth-order) and the residues of two further poles must coincide with \eqref{nbfklvoraussage}.

In the publication \cite{Kotikov:2004er} a conjecture has been put forward, according to which the inclusion of the NLO and the NNLO corrections to the doube lograithmic scaling amounts to modifying equation \eqref{DLE}  in the following way\beqa \label{VDLE}
\gamma\,(2\,\omega+\gamma)\ &=&-16\, g^2\left(1-\HS_1\,
\omega-(\HS_2+\zeta _2)\,\omega ^2\right)
-64\,g^4(\HS_2+\zeta _2-\HS_1^2)\nonumber\\
&&-4\,g^2\,
(\HS_2+\HS_{-2})\,\gamma^2\,.
\eeqa
Harmonic sums in this formula are all functions of $(r-1)$. It is straightforward to find the corresponding perturbative pole structure
\beq \label{VDLEV1}
\gamma =2\,\sum _{\ell=1}^{\infty}c_\ell(\omega)\,
\left(-4\,g^2\right)^\ell\,,
\eeq
with the coefficients $c_i$ given by
\beqa \label{VDLEV2}
c_1(\omega)&=&\frac{1}{\omega}-\HS_1-\omega(\zeta_2+\HS_2)+...\,,\nonumber\\
c_2(\omega)&=&-\frac{1}{\omega^3}+\frac{2\,\HS_1}{\omega^2}+
\frac{\zeta_2+\HS_2}{\omega}+...\,,\nonumber\\
c_3(\omega)&=&\frac{2}{\omega^5}-\frac{6\,\HS_1}{\omega^4}+
\frac{-4\,(\zeta_2+\HS_2)+4\,\HS_1^2+(\HS_2+\HS_{-2})}{\omega ^3}
+...\,,\nonumber\\
c_4(\omega)&=&-\frac{5}{\omega ^7}+\frac{20\,\HS_1}{\omega ^6}+
\frac{14\,(\zeta _2+\HS_2)-24\,\HS_1^2-4\,(\HS_2+\HS_{-2})}{\omega ^5}+\ldots \,.
\eeqa
It should be stressed that the expression \eqref{VDLEV1} together with \eqref{VDLEV2} determines residues of the poles at each negative even value of $M=-2,-4,\ldots$, thus allowing to compare the residue functions and not single numbers only.

Yet another possibility of testing the validity of the spectral equations offer the reciprocity relations \cite{Dokshitzer:2005bf}-\cite{Basso:2006nk}, which originate from some curious hidden symmetry of the anomalous dimension of twist-two operators. Surprisingly, the same relations may be found from the string theory side \cite{Beccaria:2008tg} suggesting their non-perturbative validity.

Very recently, the wrapping correction to the four-loop result, \textit{cf.} table \ref{sec:gamma8t2}, has been derived in \cite{Bajnok:2008qj} by evaluating the first finite-size corrections to the string sigma model on the $\ads$ at weak coupling. This novel procedure was shown to reproduce correctly the four-loop anomalous dimension of the Konishi operator calculated in \cite{Fiamberti:2007rj} and \cite{Velizhanin:2008jd} with the usual perturbative methods, and the corrected anomalous dimension of twist-two operators has been found to pass positively all tests discussed in this chapter! 
\chapter{Dynamical Test of the Planar AdS/CFT Correspondence}\label{sec:tests}
In this chapter we will discuss interpolating observables, which allow to dynamically verify the AdS/CFT duality.
 
\section{Interpolating Functions}\label{sec:interpol}
The difficulty in proving the AdS/CFT correspondence is mainly caused by the fact that the non-perturbative quantization of the $\N=4$ SYM theory and that of the string theory on $\ads$ space is currently not understood. Luckily, in special limits we will discuss, one can use the asymptotic Bethe equations to define non-perturbatively gauge theory observables, which can be then continued to the results of the string theory obtained from the semiclassical quantization.

One of the simplest dynamical tests of the AdS/CFT correspondence can be performed with help of the $\frac{1}{2}$-BPS operators, e.g. $\Tr \lr{\mathcal{Z}^L}$, see section \ref{sec:psualgebra}. These operators are primary fields with respect to the superconformal algebra and are annihilated additionally by half of the supersymmetry generators $Q$. It is thus straightforward to derive from \eqref{SQKR} and \eqref{SBQBKR} the all-loop relation
\beq \label{12BPS}
\Delta(g)=p=L\,.
\eeq
Since the Cartan labels of the $\mathfrak{su}(4)$ symmetry algebra do not receive quantum corrections, it follows from \eqref{12BPS} that the scaling dimension is protected and non-perturbatively equal to its classical counterpart. On the other hand, the string theory on the $\ads$ space has massless excitations which should be identified with the aforementioned BPS operators. Such an identification was done in \cite{Witten:1998qj}, confirming the validity of the correspondence in this simple case.

The asymptotic integrability discussed in chapter \ref{sec:integrabilitaet} is the first step to the non-perturbative definition of the $\N=4$ gauge theory. With help of the asymptotic Bethe equations \eqref{allschleifenbg} it is possible to calculate perturbatively anomalous dimensions of the operators only up to the order $\Op(g^{2L})$. However, for $L \to \infty$ one should be able to reproduce the full perturbative expansion, which could be then continued beyond its convergence radius. Below we will discuss this possibility and argue that the results can be compared with the predictions of string theory. 

A particularly suitable subsector for this purpose is again the $\mathfrak{sl}(2)$ subsector. As already mentioned in \eqref{skalenlimes}, states of this subsector with the minimal anomalous dimension scale for $M \to \infty$ as\footnote{This logarithmic scaling is a special case of the so called Sudakov scaling, see \cite{Collins:1989bt}.}
\beq \label{skalenlimesnu}
\Delta -\Delta_0  = \gamma(g)=f^{(L)}(g)\; \log M +\ldots \,.
\eeq
The universal scaling function\footnote{It should be stressed that the scaling function of twist-two operators $f^{(2)}(g)$ does not need to coincide with the scaling function of twist-three operators $f^{(3)}(g)$, etc. It would be interesting to prove whether the universal scaling function $f(g)$ coincides with its finite-length counterparts
\beq
f(g)=^?f^{(2)}(g)=^? f^{(3)}(g)=^?\ldots\,.
\eeq
Recently, the above equality was confirmed up to the fourth-loop order for $f^{(2)}(g)$ and $f^{(3)}(g)$, see \cite{Eden:2006rx}, \cite{Kotikov:2007cy} and \cite{Bajnok:2008qj}.} $f(g)$ is defined in the limit $L\to \infty$, $L \ll \log M$, and thus in the region of validity of the asymptotic Bethe equations. This scaling function was defined at the one-loop order in \cite{Belitsky:2006en} and subsequently investigated at higher-loops in \cite{Eden:2006rx} and \cite{Beisert:2006ez} by means of the asymptotic Bethe equations \eqref{allschleifensl2}. It turns out that the logarithmic behavior \eqref{skalenlimesnu} is reproduced by the leading magnon fluctuation density and that this density can be determined from an integral equation, the BES equation, allowing to compute the perturbative expansion of the scaling function to arbitrary order

\beqa \label{skalenfunktionpe}
f_{\rm ABA}(g)&=&8\, g^2
-\frac{8}{3}\, \pi^2 g^4
+\frac{88}{45}\, \pi^4  g^6
- 16\left(\frac{73}{630}\,\pi^6 + 4\, \zeta(3)^2 \right)g^8 \nonumber \\
&&+32 \lr{\frac{887}{14175} \pi^8+\frac{4}{3} \pi^2 \zeta(3)^2+40 \zeta(3) \zeta(5)}g^{10}\pm\ldots\, .
\eeqa
At each order of perturbation theory $\ell$  only zeta functions or their combinations of the order $(2\ell-1)$ contribute, moreover all with the overall sign $(-1)^{\ell+1}$. The convergence properties of this oscillating series were studied in \cite{Beisert:2006ez}. It was shown that the convergence radius is equal $\frac{1}{4}$ and that the series admits a natural analytical continuation to the complex plane.

On the string theory side this limit corresponds to the so called ``spinning-string'', with spin $M$ on the $AdS_5$ space and angular momentum $L$ on $S^5$. The classical equations of motion in this limit can be exactly solved, as shown in \cite{Gubser:2002tv,Frolov:2002av} and \cite{Roiban:2007jf,Roiban:2007dq}, and the corresponding solution can serve as the starting point for the semiclassical quantization. The latter was performed to the two-loop order in perturbation theory (in $g \gg 1$), resulting in the following expansion of the scaling function

\beq \label{sfstringvorhersage}
f(g)=4\, g
-\frac{3\,\log 2}{\pi}
-\frac{\rm{K}}{4\,\pi^2} \frac{1}{g}
-\ldots \,.
\eeq
Here, $K=\beta(2)$ denotes the Catalan constant. This expansion may be confronted with the strong coupling expansion of the BES equation. We will discuss this further in section \ref{sec:starkekopplung}.

The function $f(g)$ is thus the first interpolating observable of the AdS/CFT correspondence. It is natural to pose the question whether it is possible to find its generalizations, such that asymptotic Bethe equations \eqref{allschleifenbg} may once again be used to define it non-perturbatively. A detail analysis of the semiclassical quantization of string theory suggests that there exists a natural generalization of \eqref{skalenlimesnu}, see \cite{Frolov:2002av,Frolov:2006qe,Alday:2007mf}, when the length is assumed to grow logarithmically with $M$
\beq \label{vskalenlimes}
M \rightarrow \infty\, ,
L \rightarrow \infty\, ,
\qquad L=j\,\log M \,.
\eeq
In the framework of semiclassical quantization $g \gg 1$ and it is also convenient to introduce the following parameter
\beq
z:= \frac{4g \log M}{L}=\frac{4g}{\LoverlogM}\,.
\eeq
The classical energy of the spinning string in the limit \eqref{vskalenlimes} is then given by \cite{Frolov:2002av}
\beq \label{vsfklassisch}
E_0=M+L\,\sqrt{1+z^2}\, + \ldots\, ,
\eeq
and the one-loop correction found in \cite{Frolov:2006qe} takes the following form
\beqa \label{vsfeinschleifen}
\lefteqn{E_1
=\frac{L}{\sqrt{\lambda}}
\frac{1}{\sqrt{1+z^2}} \left\{ z\sqrt{1+z^2}-(1+2z^2)
\log\left[z+\sqrt{1+z^2}\right]\right. }\nonumber \\
&&\hspace{1.5cm}\left.-z^2+2(1+z^2)\log(1+z^2)
-(1+2z^2)\log\left[\sqrt{1+2z^2}\right]
 \right\}. \label{E1}
\eeqa
This expression may be also derived from the asymptotic Bethe ansatz at strong coupling, see \cite{Casteill:2007ct}.
Upon identifying the energy with the scaling dimension of twist operators and expanding in small $j \ll 1$, the following refinement of the scaling behavior \eqref{skalenlimesnu} can be found from \eqref{vsfklassisch} and \eqref{vsfeinschleifen}
\beq \label{vsfstringvorhersage}
\Delta(g)-M-L=
\Bigg(4\,g-\frac{3\,\log 2}{\pi}-\LoverlogM -\frac{\rm{K}}{4\,\pi^2} \frac{1}{g}
+\frac{\LoverlogM^2}{8\,g}+\ldots \Bigg)
\, \log M\,
+ \ldots\, .
\eeq
This suggests that the limit \eqref{vskalenlimes} could also be defined on the gauge theory side. In sections \ref{sec:ftlimes} and \ref{sec:allschleifenvsl} we will show that this is indeed the case. In the generalized scaling limit \eqref{vskalenlimes} the anomalous dimension exhibits again logarithmic scaling
\beq \label{vsf}
\Delta -M -L = \gamma(g)=f(g,  \LoverlogM) \log M
+\ldots \, ,
\eeq
with the new scaling function $f(g, \LoverlogM)$ depending now on two parameters. The ordinary scaling limit \eqref{skalenlimesnu} is recovered for $j \to 0$
\beq
f(g,j) \bigg\vert_{j=0} = f(g)\,.
\eeq
In section \ref{sec:allschleifenvsl} we will derive a closed integral equation for the leading density of the roots and argue that the function $f(g,j)$ is analytic in $g$ and $j$.
\section{Scaling Limit and the BES Equation} \label{sec:BESgleichung}
In this section we will discuss extensively the scaling limit \eqref{skalenlimesnu}, in which the number of the excitations diverges $M \to \infty$, while the length $L$ remains finite or grows slower then logarithmically with respect to $M$. 
\subsection{The Leading One-Loop Density}
It was shown in  \cite{Eden:2006rx} that the roots in this limit cover densely the interval $(-\frac{M}{2}, \frac{M}{2})$. Since the number of the small holes grows slower then logarithmically with $M$, it follows from \eqref{klskalierung} that they do not form a gap in the magnon root distribution. The integral equation for the density \eqref{rhointgleich} may thus be written as 
\beq \label{ES}
2\,\pi\,\rho_0(u)+2\,\pi\,\frac{L-2}{M}
\delta(u)
-\frac{L}{M} \frac{1}{u^2+\frac{1}{4}}
-2
\int_{-\frac{M}{2}}^{\frac{M}{2}} dv \frac{\rho_0(v)}{(u-v)^2+1}
=0\,.
\eeq
It should be supplemented with the proper normalization condition
\beq
\int^{\infty}_{-\infty} \rho_0 (u)=1 \,.
\eeq
The equation \eqref{ES} was derived in \cite{Eden:2006rx} and thoroughly analyzed therein. Contrary to naive expectations, even though $u(M)\simeq \frac{M}{2}$ diverges for $M\to \infty$, one cannot use Fourier transformation for the sake of solving this equation. Rather then that, one defines the rescaled density
\beq \label{umskalierung}
\bar{\rho}_0(\bar{u})=M \, \rho_0 (u), \qquad \mbox{where} \qquad  u=M\,\bar{u}.
\eeq
This redefinition preserves the measure of the integration $d\bar{u} \, \bar{\rho}_0 (\bar{u})=du\,\rho_0 (u)$. Rescaling equation \eqref{ES} and using $\bar{\rho}_0 (\bar{u})$ together
with the relations
\beqa
\frac{1}{2\,M} \frac{1}{\bar{u}^2+\frac{1}{4 M^2}}&=&\pi \,\delta(\bar{u}) + \Op\lr{\frac{1}{M}}\,, \\[3pt]
\frac{1}{M} \frac{1}{(\bar{u}-\bar{u}')^2+\frac{1}{M^2}}&=&\pi \,\delta(\bar{u}-\bar{u}') + \frac{1}{M} \frac{\mathcal{P}}{(\bar{u}-\bar{u}')^2} + \Op\lr{\frac{1}{M^2}}\,,
\eeqa
where $\mathcal{P}$ denotes the Cauchy principle value, one derives
\beq \label{singintgleich}
0=4\,\pi\,\delta(\bar{u})+2\, \mathcal{P} \int^{\frac{1}{2}}_{-\frac{1}{2}} \bar{u}' \frac{\bar{\rho}_0 (\bar{u}')}{(\bar{u}-\bar{u}')^2} \,.
\eeq
It was shown in \cite{Eden:2006rx} that this singular equation is solved by the Korchemsky density\footnote{The density \eqref{korchemskydichte} was first found by G.~Korchemsky in \cite{Korchemsky:1995be} while analyzing certain Baxter equation. }
\beq \label{korchemskydichte}
\bar{\rho}_K (\bar{u}) = \frac{1}{\pi} \log \frac{1+\sqrt{1-4\,\bar{u}^2}}{1-\sqrt{1-4\,\bar{u}^2}}=\frac{2}{\pi} \rm{arctanh} \lr{\sqrt{1-4\,\bar{u}^2}}\,.
\eeq
Since this solution is singular at $\bar{u}=0$, the original density $\rho_0(u)=\frac{1}{M} \bar{\rho}_0 (M \bar{u})$ should be considered as a distribution rather then a function. The one-loop anomalous dimension to the leading order in $M$ can be found directly from \eqref{fdurchrho} and \eqref{korchemskydichte} 
\beq \label{8logM}
\gamma^{\rm ABA}_2 = \frac{2}{M} \int^{\frac{1}{2}}_{-\frac{1}{2}}dv \frac{\bar{\rho}_K (\bar{v})}{\bar{v}^2+\frac{1}{4 M^2}}= -\frac{4}{M\,\pi} \int^{\frac{1}{2}}_{-\frac{1}{2}}dv \frac{\log{|\bar{v}|}}{\bar{v}^2+\frac{1}{4 M^2}} + \Op \lr{M^0}= 8 \log M + \Op \lr{M^0}\,.
\eeq
It should be noted that the leading result does not depend on $L$, which confirms the universal scaling behavior at the one-loop order.

The solution \eqref{korchemskydichte} can be also partially recovered from \eqref{Zuraum}. Upon rescaling $u$ as in \eqref{umskalierung}, and using \eqref{Zzurho} together with \eqref{u1u2}, \eqref{imlogterm} and \eqref{Zuraum}, one finds
\beq \label{korchemskyauszraum}
\bar{\rho}_0 (\bar{u})= \frac{1}{\pi} \log {\frac{1-2 \bar{u}^2}{\bar{u}^2}}\,.
\eeq
Comparing \eqref{korchemskyauszraum} with \eqref{korchemskydichte}, one concludes that the result \eqref{korchemskyauszraum} approximates well the exact solution in the interval $\bar{u} \in (-\frac{1}{2},\frac{1}{2})$ and only at the boundaries deviates significantly from \eqref{korchemskydichte}. The discrepancy between \eqref{korchemskyauszraum} and \eqref{korchemskydichte} is caused by the $\rm{Im log}$ term in \eqref{Zuraum} since in the vicinity of $\bar{u}= \frac{1}{2}$ und $\bar{u}=-\frac{1}{2}$ the approximation \eqref{imlogterm} ceases to be valid. It should be stressed, however, that the difference between \eqref{korchemskydichte} and \eqref{korchemskyauszraum} is negligible at the leading order.
\subsection{The BES Equation}
To the leading order in $M$ it is possible to recast equation \eqref{Ztraum} in the form of a linear integral equation for the density. For this purpose the large $M$ expansion of the following type of integrals needs to be determined
\beq \label{grossesMintegrale}
f(M)=\int^{\infty}_{0}dx\, h(x)\,\sin \left(u(M)\, x \right)\,.
\eeq
Here, $\lim_{M\to \infty} u(M)=\infty$ and $h(x)$ is a sufficiently smooth function on the interval $[0,\infty)$. We start by noting that
\beq
\lim_{M \to \infty} f(M)=0\,,
\eeq
and therefore the following asymptotic expansion of \eqref{grossesMintegrale} may be assumed
\beq \label{fasymptotisch}
f(M)=\sum^{\infty}_{j=0} \frac{c_j}{u(M)^{1+j}}\, .
\eeq
To determine the first coefficient $c_0$ it is enough to integrate \eqref{grossesMintegrale} by parts
\beq \label{c0koeffizient}
c_0=\lim_{M \to \infty} u(M) \, f(M)=\lim_{M \to \infty}
\int^{\infty}_{0} dx \,h(x)\left(-\frac{d}{dx} \cos \left(u(M)\,
x\right)\right)=h(0) 
\eeq
since the boundary terms vanish. Integrating by parts $(n+1)$ times we find
\beq
c_n=\lim_{M \to \infty} u(M)^{n+1} \, \left(f(M)-\sum^{n-1}_{j=1}
\frac{c_j}{u(M)^{1+j}}\right)=(-1)^{\frac{n}{2}} \, h^{(n)}(0)
\qquad \mbox{for even $n$}\, .
\eeq
All odd $c_n$ vanish because of the $u(M)\mapsto -u(M)$ symmetry of \eqref{grossesMintegrale}. 

The derivation of the leading equation is based on the observation that in the scaling limit the non-linear term in \eqref{Ztraum} may be neglected and equation \eqref{Ztraum} becomes a Fredholm integral equation of the second kind. However, not all inhomogeneous terms contribute to the leading order. Indeed, one can drop all terms except for the one containing the fast (universal) holes  $u^{(1,2)}_{\rm h}$,
\beq
\frac{2 \,\pi \,
e^{\frac{t}{2}}}{i\,t\,(e^{t}-1)}-
\frac{2\,\pi\,\cos \left(t\,u^{(1,2)}_{\rm h} \right)}{i\,t\,
(e^t-1)}\,,
\eeq
since the rapidities of the remaining holes vanish in this limit (see \eqref{klskalierung})
\beq \label{fplusendlich}
u_{\rm h}^{(j)}\simeq 0 \qquad j=3,\ldots, L \,.
\eeq
Therefore the equation for the leading scaling function reads
\beqa \label{Ztraumf}
\nonumber \hat{Z}(t)&=&\frac{4 \,\pi \,
e^{\frac{t}{2}}}{i\,t\,(e^{t}-1)}-
\frac{4\,\pi\,\cos \left(t\,u^{(1)}_{\rm h} \right)}{i\,t\,(e^t-1)}\\
&&-4
\,g^2\,\frac{e^{\frac{t}{2}}}{e^{t}-1}\,\int^{\infty}_{0}\,dt'\,e^{-\frac{t'}{2}}\,t'\,
\hat{K}(2gt,2gt')\,\hat{Z}(t')\,.
\eeqa
Despite the fact that the universal holes receive quantum corrections, these are sub-leading at large $M$ and the one-loop asymptotic behavior \eqref{u1u2} is still valid. One way to motivate this is to note that the eigenvalue of the Casimir operator \eqref{qcasimir} is additively renormalized (see \cite{Belitsky:2006wg})
\beq
J_0 \mapsto J=M+\frac{1}{2}L + \frac{\gamma(g)}{2}\,.
\eeq
Subtracting the one-loop part from the counting function
\beq \label{ZinZ0unddeltaZ}
\hat{Z}(t)=\hat{Z}_0(t)+\delta \hat{Z}_{\rm BES}(t) \,,
\eeq
and upon identifying, in accordance with \eqref{Zzurhoallschleifen}, $\delta \hat{Z}_{\rm \tiny BES}(t)$ with the fluctuation density\\ (see \cite{Eden:2006rx})
\beq \label{Zzufluktuationsdichte}
\delta\hat{Z}_{\rm \tiny BES}(t)=16\,\pi\,i\,g^2\,e^{\frac{t}{2}}\,
\frac{\oldtildesigma_{\rm \tiny BES}(t)}{t}\, \log(M)\,,
\eeq
one derives
\beq \label{BESgleichung}
\oldtildesigma_{\rm \tiny BES}(t)=\frac{t}{e^t-1}\,\left(\hat{K}(2gt, 0)-4\,g^2\,\int^{\infty}_{0}\,dt'\,\hat{K}(2gt,2gt')\,
\oldtildesigma_{\rm \tiny BES}(t')\right)\,,
\eeq
where we have used the formula \eqref{c0koeffizient} to evaluate the integral
\beqa
\int^{\infty}_{0}\,dt'\,e^{-\frac{t'}{2}}\,t'\,
\hat{K}(2gt,2gt')\,\hat{Z}_0(t')&=& \int d u(M) \int^{\infty}_{0}\,dt'\,e^{-\frac{t'}{2}}\,t'\,
\hat{K}(2gt,2gt')\,\frac{\partial}{\partial u(M)} \hat{Z}_0(t') \nonumber \\
&=&-4\,\pi\,i \,\hat{K}(2gt,0) \, \log M + \Op(M^0)\,.
\eeqa

Equation \eqref{BESgleichung} is the celebrated BES equation and was first found in \cite{Beisert:2006ez}.  Let us note that the above derivation  differs significantly from the original one and confirms that the decomposition into the one-loop density and fluctuation density is mathematically well defined even non-perturbatively.
\subsection{The Strong Coupling Limit} \label{sec:starkekopplung}
Directly from the definition \eqref{Zzurhoallschleifen} and \eqref{8logM} together with \eqref{Zzufluktuationsdichte} it follows that the anomalous dimension to the leading order is given by
\beq
\gamma^{\rm ABA}(g)=\lr{8\,g^2-64\,g^2\,\int^{\infty}_0 dt' \hat{K}(2gt,2gt')\oldtildesigma_{\rm BES}(t')}\log M + \Op(M^0)\,.
\eeq
One therefore identifies
\beq
f^{\rm ABA}(g)=\lr{8\,g^2-64\,g^2\,\int^{\infty}_0 dt' \hat{K}(2gt,2gt')\,\oldtildesigma_{\rm BES}(t')}\,.
\eeq
Consequently, the scaling function $f(g)$ is completely determined by the solution of the BES equation. This allows to compare with the string theory result \eqref{sfstringvorhersage}. A systematic method\footnote{The BES equation at strong coupling was a subject of extensive studies in many different publications, see \cite{Kostov:2008ax} and references therein.} of expansion of the BES equation at large values of the coupling constant was proposed in \cite{Basso:2007wd}. To the first few orders one finds
\beq
f^{\rm ABA}(g)=4\, g
-\frac{3\,\log 2}{\pi}
-\frac{\rm{K}}{4\,\pi^2} \frac{1}{g}
-\ldots
\eeq
in complete agreement with \eqref{sfstringvorhersage}. This constitutes one of the first dynamical tests of the planar AdS/CFT correspondence. It should be stressed that this would not have been possible without the asymptotic integrability. 
\subsection{Finite Size Corrections}
\label{sec:subleading}
Beyond the leading order the large $M$ expansion of the anomalous dimension consists of the finite part $f_{\rm finite} (g,L)$ and terms that vanish with $M\to \infty$

\beq
\gamma=f(g)\,\log{M}+f_{{\rm finite}}(g,L)+\mathcal{O}\left(\frac{1}{\log^2{M}}\right)\,.
\eeq
The somewhat extraordinary order of the first of these terms $\Op\lr{\frac{1}{ \log^2 M}}$ comes from the quadratic contribution of the small holes \eqref{klskalierung}\footnote{These corrections are present already at the one-loop level as may be seen by expanding \eqref{einschleifengammadurchZ} in $u^{(j)}_{\rm h}$.}. For the finite part $\Op\lr{M^0}$, however, the approximation \eqref{fplusendlich} is sufficient and the perturbative expansion of $f_{{\rm finite}}(g,L)$ may be found directly from \eqref{Ztraum}
\beqa \label{fendlich}
\nonumber f^{\rm ABA}_{\rm finite}(g,L)&=&\left(\gamma-(L-2)\,\log 2\right)\,f(g)-8\,(7-2\,L)\,\zeta(3)\,g^4\\\nonumber
&&+8\,\left(\frac{4-L}{3}\,\pi^2\,\zeta(3)+(62-21\,L)\,\zeta(5)\right)\,g^6\\
\nonumber&-&\frac{8}{15}\,\bigg((13-3\,L)\,\pi^4\,\zeta(3)+5\,(32-11\,L)\,\pi^2\,\zeta(5) \\ \nonumber
&&+75\,(127-46\,L)\,\zeta(7)\bigg)\,g^8 \nonumber \\
&\pm&\ldots \,.
\eeqa
The function $f^{\rm ABA}_{\rm finite}(g,L)$ exhibits transcendentality properties similar to the scaling function, however in contradistinction to the latter, it explicitly depends on $L$. Thus, one cannot trust the perturbative expansion beyond the wrapping order. Despite this fact though, the part strictly proportional to $L$ should not be influenced by the wrapping interactions suggesting that also $f^{\rm ABA}_{\rm finite}(g,L)$ remains unaffected.

\section{The Generalized Scaling Limit} \label{sec:ftlimes}
In this section we will define the generalized scaling limit \eqref{vskalenlimes} at the first order of perturbation theory.

In this limit the length scales as $L=j\,\log M$ and one concludes from \eqref{klskalierung} that the small holes occupy the interval $(-\holegap,\holegap)$. Numerical analysis suggests that the largest of the magnon roots is again of the order $\pm \frac{M}{2}$ and consequently the Bethe roots condense on the interval $(-\frac{M}{2},-a)\cup (a, \frac{M}{2})$. We will prove later that $a=\holegap$.

It is convenient to decompose the corresponding density\footnote{We will denote the density in this limit by $\magnonrho(u)$.} into the singular part
$\korrho(u)$ and the fluctuation density $\tilde \sigma(u)$
\beq
\magnonrho(u)=\korrho(u)+\tilde \sigma(u)\,,
\eeq
where $\korrho(u)=1/M\, \bar{\rho}_{K}(u/M)$, see \eqref{umskalierung}. Upon adding the following term to \eqref{rhointgleich}
\beq
2 \int_{-a}^{a} dv \frac{\korrho(v)}{(u-v)^2+1}
=\frac{4\,\log M}{\pi\, M}\left(\arctan(u+a)-\arctan(u-a)\right)+ \Op(M^0)\,,
\eeq
one finds that the fluctuation density $\tilde \sigma(u)$ must scale as $\log M/M$. This justifies the following redefinition

\beq \label{dichtezerlegung}
\magnonrho (u)=\korrho (u)-\frac{8\,\log M}{M}\,\sigma(u)\, .
\eeq
The equation for the density \eqref{rhointgleich} may now be transformed into an equation for the fluctuation density
\beqa \label{sigmaeinschleifen}
&&2\,\pi\,\sigma(u)
-\frac{1}{2\pi} \left(\arctan(u+a)-\arctan(u-a)\right)
+\frac{\LoverlogM}{8}\,\frac{1}{u^2+\frac{1}{4}}\\\nonumber
&&-2
\left(\int_{-\infty}^{-a} dv + \int_a^{\infty} dv\right)
\frac{\sigma(v)}{(u-v)^2+1}=0\, .
\eeqa
This should be supplemented with the normalization condition for the density 
\beq
\left(\int^{-a}_{-\infty}+\int^{\infty}_{a} \right)du \,\rho_m(u)=1\,,
\eeq
from which one derives
\beq \label{sigmaeinschleifennormierung}
\LoverlogM=\frac{4\,a}{\pi}-8\,\int^{a}_{-a} du \,\sigma(u)\,.
\eeq
Intuitively, the fluctuation density describes perturbations around the Korchemsky density, caused by the gap in the root distribution. The size of the gap $2a$ may be determined in the following way. One substitutes \eqref{sigmaeinschleifennormierung} in \eqref{sigmaeinschleifen} and subsequently solves the equation for $\sigma(u,a)$. Putting the solution into \eqref{sigmaeinschleifennormierung} results in the relation $j=j(a)$, which must be then inverted.

The one-loop anomalous dimension may be found immediately using \eqref{korchemskydichte} and \eqref{dichtezerlegung}
\beqa \label{einschleifenadvsl}
\gamma^{\rm ABA}_2 (\LoverlogM,M)&=&
\lr{8-\frac{16}{\pi}\,\arctan\lr{2a(j)}-16\,\left(\int_{-\infty}^{-a(j)}\,du + \int_{a(j)}^{\infty}\,du\right)
\frac{\sigma(u)}{u^2+\frac{1}{4}}} \log M \nonumber \\
&&+ \Op\lr{M^0}\, .
\eeqa
This proves that the validity of \eqref{vskalenlimes} at the one-loop order.
\subsection{Fluctuation Density in the Fourier Space}

Equation \eqref{sigmaeinschleifen} becomes particularly simple in the Fourier space. We rewrite the equation in the form 

\beqa \label{sigmaeinschleifenumgeschrieben}
\sigma(u)&=&\frac{1}{4\pi^2}\,\left(\arctan(u+a)-\arctan(u-a)\right)-\frac{\LoverlogM}{16\,\pi}\frac{1}{u^2+\frac{1}{4}} \\ \nonumber
&&+ \int^{\infty}_{-\infty} \frac{dv}{\pi}\frac{\sigma(v)}{1+(u-v)^2}-\int^{a}_{-a} \frac{dv}{\pi} \frac{\sigma(v)}{1+(u-v)^2}\,.
\eeqa
Sticking to the conventions in \cite{Eden:2006rx}, we define the Fourier transformation of the density to be\footnote{Every other quantity in the Fourier space is defined through \eqref{fouriertransformation}.}
\beq \label{fdfr}
\hat{\sigma}(t)=
e^{-\frac{t}{2}} \int_{-\infty}^{\infty} du\,e^{-i\,t\,u}\,\sigma(u)\,.
\eeq
It then follows from \eqref{sigmaeinschleifenumgeschrieben} that $\hat{\sigma}(t)$ satisfies the following integral equation
\beq \label{seft}
\hat{\sigma}(t)=\frac{t}{e^t-1}\left(\hat{K}_{\rm h}(t,0;a)-\frac{\LoverlogM}{8\,t}\,-4\,\int^{\infty}_0 dt' \,\hat{K}_{\rm h}(t,t';a)\,\hat{\sigma}(t')\right)\,,
\eeq
with the integral kernel $\hat{K}_{\rm h}(t,t';a)$ defined through
\beqa \label{fdfrkern}
\hat{K}_{\rm h}(t,t';a)
&=&\frac{e^{\frac{t'-t}{2}}}{4\,\pi\,t}\,\int^{a}_{-a}du\,\cos(t\,u)\,\cos(t'\,u) \nonumber \\
&=&\frac{1}{2\,\pi\,t}\,e^{-\frac{t}{2}}\,\frac{t\cos(at')\sin(at)-t'\cos(at)\sin(at')}{t^2-{t'}^2}\,e^{\frac{t'}{2}}\,.
\eeqa
The normalization condition \eqref{sigmaeinschleifennormierung} is equivalent to
\beq \label{sigmaeinschleifennormierungft}
j=\frac{4a}{\pi}-\frac{16}{\pi} \int^{\infty}_{0} dt \, \hat{\sigma}(t) e^{\frac{t}{2}} \, \frac{\sin at}{t}\,.
\eeq
The significant difference between  \eqref{sigmaeinschleifen} and  \eqref{seft} is the fact that the domain of the integration of the latter does not depend on the boundary parameter $a$ and the equation may be iterated more easily.  Once \eqref{seft} has been solved to the desired order, one can find the corresponding anomalous dimension from
\begin{eqnarray}
\frac{\gamma^{\rm ABA}_2(\LoverlogM,M)}{\log M}&=&8\,\left[1-\frac{2}{\pi}\,\arctan{2a} \right.\\
\nonumber &&\quad \ -\left. 4\,\int^{\infty}_{0}dt \,\left(\hat{\sigma}(t)-4\,t\,\int^{\infty}_{0} dt'\, \hat{K}_{\rm h}(t,t';a)\,\hat{\sigma}(t') \right) \right]+\Op\lr{\frac{1}{\log M}}\,.
\end{eqnarray}
%

\subsection{The Density For the Holes}
The above integral equation for the density in the generalized scaling limit was derived using magnon roots. However, as pointed out in the section \ref{sec:zfplusie}, the system may also be equivalently described by holes. Below we will take this point of view and show that the resulting equation for the hole density is equivalent to \eqref{seft}.

\begin{figure}
\begin{center}
\includegraphics[scale=1]{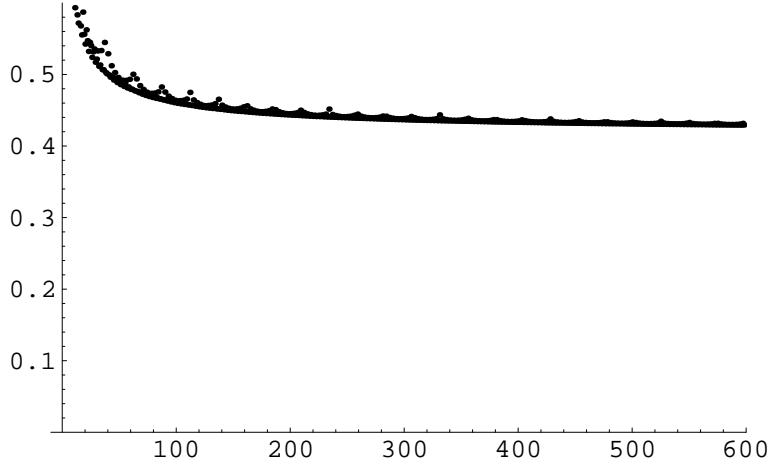}
\caption{The convergence function $r(k)$ for $k=1,\ldots,600$\,.}\label{fig:konvergenzfunktion}
\end{center}
\end{figure}

As discussed above, in the limit \eqref{vskalenlimes} the small hole roots cover densely the interval $(-\holegap,\holegap)$. Similarly to the relation \eqref{Zzurho}, the hole density can be be defined thorough\footnote{In what follows by hole density we understand the density of the small holes $u^{(j)}_{\rm h}, j=3,\ldots, L$ .}
\beq
\frac{1}{L}\,\frac{d}{du}\,Z(u)=2\,\pi\,\holerho(u)
+\Op\left(\frac{1}{L}\right)\,, \qquad \qquad u \in (-\holegap,\holegap)\,.
\eeq
It is also very natural to normalize the density to one
\beq \label{ldnormierung}
\int_{-c}^c du\,\holerho(u)=1\, .
\eeq
Approximating the non-linear term in \eqref{Zuraum} as in \eqref{imlogterm} we find
\beqa
\holerho(u)&=&
\frac{1}{L}\,
\big(\psi(i\,(u-u_{\rm h}^{(1)}))+\psi(-i\,(u-u_{\rm h}^{(1)})
+\psi(i\,(u+u_{\rm h}^{(1)}))+\psi(-i\,(u+u_{\rm h}^{(1)}))\big)
\nonumber \\
& &
+\frac{1}{L} \, 2\log 2
-\frac{1}{2\,\pi}\,\big(\psi(\frac{1}{2}+i\,u)+\psi(\frac{1}{2}-i\,u)\big)
\nonumber \\
& &
+\int_{-\holegap}^{\holegap}\frac{dv}{2\pi}\,\big(\psi(i\,(u-v))+\psi(-i\,(u-v))\big)\,
\holerho(v)\,.
\eeqa
Moreover, using \eqref{u1u2} one gets to the leading order
\beq \label{lig}
\holerho(u)=
\frac{2}{\pi \LoverlogM}
-\frac{1}{2\pi}\,\big(\psi(\frac{1}{2}+i\,u)+\psi(\frac{1}{2}-i\,u)\big)
+\int_{-\holegap}^{\holegap}\frac{dv}{2\pi}\,\big(\psi(i\,(u-v))+\psi(-i\,(u-v))\big)
\holerho(v)\, .
\eeq
Upon solving this equation, the one-loop generalized scaling function may be found from \eqref{einschleifengammadurchZ}
\beq \label{eaddld}
f_{(1)} (j)=\lim_{M \to \infty} \frac{\gamma^{\rm ABA}_2(\LoverlogM,M)}{\log M}=
8+2\,  \LoverlogM\,\int_{-\holegap}^{\holegap}du\,\holerho(u)\,
\big(\psi(\frac{1}{2}+i\,u)+\psi(\frac{1}{2}-i\,u)-2\,\psi(1)\big)\,.
\eeq
Even though \eqref{lig} together with \eqref{ldnormierung} constitute the sought-for system of equations, it is more convenient for the computational purposes to transform them in a way such that the boundary parameter $c$ does not specify the integration domain. Hence, we rescale the variable $u$ and the density itself as in \eqref{umskalierung},
\beq \label{lumskalierung}
\bar u = \frac{u}{\holegap}\,,
\qquad  \qquad
\holebarrho(\bar u)=  \LoverlogM\,\holegap\, \holerho(u)\,,
\eeq
and define the non-singular kernel
\beq
K(\bar u,\bar v)=\frac{\holegap}{2\pi}\,
\bigg(
\psi(i\,\holegap\,(\bar u-\bar v))+\psi(-i\,\holegap\,(\bar u-\bar v))
-\psi(\frac{1}{2}+i\,\holegap\,\bar u)-\psi(\frac{1}{2}-i\,\holegap\,\bar u)
\bigg)\, .
\eeq
Equation \eqref{lig} may now be rewritten as
\beq \label{slig}
\holebarrho(\bar u)=
\frac{2}{\pi}\,\holegap
+\int_{-1}^1 d \bar v\, K(\bar u,\bar v)\,
\holebarrho(\bar v)\, .
\eeq
One notes that the dependence on the boundary parameter $c$ is hidden in the integral kernel. It follows from \eqref{ldnormierung} and \eqref{lumskalierung} that the normalization condition for the rescaled density is given by the following expression
\beq \label{sldnormierung}
  \LoverlogM=\int_{-1}^1dv\, \holebarrho(\bar u)\,.
\eeq
Finally, the one-loop relation \eqref{eaddld} may be rewritten in the following form
\beq \label{eaddsld}
f_{(1)}(j)=
8+2\,\int_{-1}^1d \bar u\,\holebarrho(\bar u)\,
\big(\psi(\frac{1}{2}+i \holegap \bar u)+\psi(\frac{1}{2}-i \holegap \bar u)-2\,\psi(1)\big).
\eeq
The integral equation \eqref{slig} is particularly suitable for the iteration. To the first few orders one finds from \eqref{sldnormierung} and \eqref{eaddsld} the following expansion of the one-loop generalized scaling function
\beqa
f_{(1)}(j)=&=&
8-8\,\LoverlogM\,\log 2+\frac{7}{12}\,\LoverlogM^3\,\pi^2 \,\zeta(3)
-\frac{7}{6}\,\LoverlogM^4 \,
   \pi ^2 \, \log 2\, \zeta(3)
\nonumber \\ \label{evsl}
&&+
2 \,\LoverlogM^5 \, \left(\frac{7}{8} \, \pi ^2 \, \log^2 2\, \zeta(3)-\frac{31}{640} \, \pi ^4\,
   \zeta(5)\right)+
   \Op(\LoverlogM^6).
\eeqa
It turns out that the equation \eqref{slig} can be also very effectively analyzed numerically. In figure \ref{fig:konvergenzfunktion} we present the convergence function of the series $f_{(1)}(j)=\sum^{\infty}_{k=0} f_{1,k}\,j^k$,
\beq
r(k)=(f_{1,k})^{-\frac{1}{k}}\,,
\eeq
for the first six hundred terms. One concludes from this numerical analysis that the convergence radius
\beq
r=\limsup_{k\to \infty} r(k) \simeq 0.4
\eeq
exists and thus the function \eqref{evsl} is an analytic function in the region $\vert j \vert < r$.\\

It should be noted that due to \eqref{Zzurho} and \eqref{dichtezerlegung} the hole density should be considered as an analytic continuation of the fluctuation density $\sigma(u)$
\beq
\LoverlogM \,\holerho (u)=\frac{2}{\pi}-8\,\sigma(u) \qquad \qquad u\in (-c,c) \,.
\eeq
Using \eqref{fdfr} one can rewrite the above relation as
\beq \label{ldzufd}
\LoverlogM \,\holerho (u)=\frac{2}{\pi}-\frac{8}{\pi}\,\int^{\infty}_{0}dt\, \hat{\sigma}(t)\,e^{\frac{t}{2}} \,\cos{tu}\, .
\eeq
This relation may in turn be used to prove the equivalence of \eqref{seft} and \eqref{lig}. Multiplying \eqref{seft} with  $e^{\frac{t}{2}} \,\cos{tu}$ and integrating in $t$ over the positive real axis one finds, using \eqref{fdfrkern} together with \eqref{sigmaeinschleifennormierungft} and \eqref{ldzufd}, the equation \eqref{lig}. Thus, we conclude that $a=c$ must hold and that there is no gap between the hole and the magnon distributions.

\section{The Generalized Scaling Function To All Orders} \label{sec:allschleifenvsl}
In this section we will generalize equation \eqref{sigmaeinschleifen} to all orders in perturbation theory, thus proving the existence of the scaling function \eqref{vsf}.

\subsection{Derivation}
The asymptotic Bethe equations \eqref{allschleifensl2} in the novel limit \eqref{vskalenlimes} are particularly easy to study with the formalism developed in section \ref{sec:allschleifenig}. The non-linear term in \eqref{Ztraum} may be again neglected. On the other hand, the small holes cannot be disregarded anymore. This is due to the fact that their overall contribution is proportional to $L=\LoverlogM \, \log M$ and therefore must be taken into account.  Thus, equation \eqref{Ztraum} in the limit \eqref{vskalenlimes} takes the following form
\beqa \label{Ztraumvsl}
 \hat{Z}(t)&=&\frac{2 \,\pi \,L
e^{\frac{t}{2}}}{i\,t\,(e^{t}-1)}J_0(2gt)-\sum^{L}_{j=1}
\frac{2\,\pi\,\cos \left(t\,u^{(j)}_{\rm h} \right)}{i\,t\,(e^t-1)}\nonumber \\
 &&+8\pi\,\,g^2\, \frac{e^{\frac{t}{2}}}{i\,(e^t-1)} \sum^{L-2}_{j=1}\,\int^{\infty}_{0}\,dt'\,
e^{-\frac{t'}{2}}\,\hat{K}(2gt,2gt') \cos\left(t'\,u^{(j)}_{\rm h}\right) \nonumber \\
&&-4
\,g^2\,\frac{e^{\frac{t}{2}}}{e^{t}-1}\,\int^{\infty}_{0}\,dt'\,e^{-\frac{t'}{2}}\,t'\,d
\hat{K}(2gt,2gt')\,\hat{Z}(t').
\eeqa
In similarity to \eqref{ZinZ0unddeltaZ} we subtract the one-loop part

\beq \label{ZinZ0unddeltaZvsl}
\hat{Z}(t)=\hat{Z}_0(t)+\delta \hat{Z}(t)\,,
\eeq
and identify $\delta \hat{Z}(t)$ with the fluctuation density $\oldtildesigma(t)$ \footnote{The fluctuation density in the $u$-space $\sigma(u)$ for $u\in (-a,a)$ must be considered as an analytic continuation.}
\beq \label{allschleifensigmadef}
\delta \hat{Z}(t)=16\,\pi\, i\,e^{\frac{t}{2}}\,\frac{\oldtildesigma(t)}{t} \, \log {M}\, .
\eeq
This allows to rewrite the equation \eqref{Ztraumvsl} in the form
\beqa \label{sigmaallschleifenmitsummen}
\nonumber\oldtildesigma(t)&=&\frac{t}{e^t-1} \left[g^2\,\hat{K}(2gt,0)-\frac{\LoverlogM}{8}\frac{J_0(2gt)}{t}+\frac{1}{8\log M}\,\sum_{j=3}^{L}\frac{e^{-t/2}\cos(t\,u_{\rm h}^{(j)})}{t}\right.\\
&&\ -\left.
\frac{g^2}{2}\,\frac{1}{\log M}\,\sum_{j=3}^{L}\,\int_0^\infty\,dt'\,\hat{K}(2gt,2gt')\,e^{-t'/2}\,\cos(t'\,u_{\rm h}^{(j)})\right.\\ \nonumber
&&\ -\left.4\,g^2\,\int_0^\infty\,dt'\,\hat{K}(2gt,2gt')\,\oldtildesigma(t')\right].
\eeqa
It follows from \eqref{Qpallschleifen} that the anomalous dimension to the leading order in $M$ can be written as
\beqa \label{adallschleifenmitsummen}
\gamma=8\,g^2\,\log{M}&&\left(1-\frac{1}{\log M}\,\sum_{j=3}^{L}\,\int_0^\infty\,dt\,\frac{J_1(2gt)}{2gt}e^{-t/2}\,\cos(tu_{\rm h}^{(j)}) \right.\\ \nonumber
&&\ \ -\left. 8\,\int_0^\infty\,dt\,\frac{J_1(2gt)}{2gt}\,
\oldtildesigma(t)\right)+\Op(M^0).
\eeqa
As discussed in the section \ref{sec:allschleifenig}, the equations for the holes to all-loop order (but for perturbative values of $g$) are given by
\beq \label{allschleifenklfr}
Z(u_{\rm h}^{j})=\pi\,(2\,n_{\rm h} ^j+\delta-1)\,.
\eeq
Upon performing Fourier transformation of this equation, one finds the relation
\beq
\frac{i}{\pi} \,\int^{\infty}_{0}\, \sin \left(t\,u_{\rm h}^j \right )\,\hat{Z}(t) = \pi\,(2\,n_{\rm h} ^j+\delta-1)\,,
\eeq
which after substituting \eqref{ZinZ0unddeltaZvsl} in \eqref{allschleifenklfr} can be written in the following form
\beqa \label{klfrlogform}
2\,\pi\,n_{\rm h}^{(k)}&=&4\,F(u_{\rm h}^{(k)},u_{\rm h}^{(1)})-16\,\log M\,\int_0^\infty\,dt\,\frac{\oldtildesigma(t)}{t}\,e^{t/2}\,\sin(tu_{\rm h}^{(k)})\,,
\eeqa
with
\beqa
\nonumber F'(x,y)&\equiv&\int^{\infty}_0 dt\, \cos{t x} \, \frac{e^{\frac{t}{2}}-\cos{t\, y}}{e^t-1}\\
\nonumber &=& \frac{1}{4}\,\bigg(\psi \left(i\,(x-y)\right)+\psi \left(-i\,(x-y)\right)+\psi \left(i\,(x+y) \right)+\psi \left(-i\,(x+y) \right) \\
&&-2\,\psi\left(\frac{1}{2}-i\,x \right)-2\,\psi \left(\frac{1}{2}+i\,x \right) \bigg)\,.
\eeqa

For $L=j\,\log M \to \infty$ the hole roots, as shown before, occupy a finite interval around zero. Therefore one easily derives from \eqref{klfrlogform} the following integral relation\footnote{The magnon density is related to $\oldtildesigma(t)$ through
\beq
\rho_m(u)=\frac{2}{\pi}\,\frac{1}{M}\,F'(u,u_{\rm h}^{(1)})-\frac{8\,\log{M}}{\pi\,M}\,\int_0^\infty\,dt\,\oldtildesigma(t)\,e^{t/2}\,\cos(t\,u).
\nonumber
\eeq
} for the all-loop hole density $\holerho(u)$
\beq \label{ldzuallschleifensigma}
\LoverlogM\,\holerho(u)=\frac{2}{\pi \log M}\,F'(u,u_{\rm h}^{(1)})-\frac{8}{\pi}\,\int_0^\infty\,dt\,\oldtildesigma(t)\,e^{t/2}\,\cos(t\,u).
\eeq
The first term in this equation, $\frac{2}{\pi}\,\frac{1}{\log M}\,F'(u,u_{\rm h}^{(1)})$, is in the limit $M \to \infty$ identical with \eqref{korchemskyauszraum} when expressed in the original variable $u=M \, \bar{u}$. Therefore it follows from \eqref{korchemskydichte} that
\beq \label{Fprimf}
F'(u,u_{\rm h}^{(1)})=\log M + \Op(M^0) \qquad \qquad u \in (-a,a).
\eeq
The sums in \eqref{sigmaallschleifenmitsummen} to the leading order may be replaced by integrals over the hole density. Using \eqref{ldzuallschleifensigma} and \eqref{Fprimf} one finds the equation
\beqa \label{allschleifenigvsl}
\nonumber\oldtildesigma(t)&=&\frac{t}{e^t-1}\left[-\frac{\LoverlogM}{8\,t}\,J_0(2gt)+\hat{K}_{\rm h}(t,0;a)-4\,\int_0^\infty dt'\,\hat{K}_{\rm h}(t,t';a)\oldtildesigma(t')\right.\\ \nonumber
&&\left.\quad\,\,\qquad+g^2\,\hat{K}(2gt,0)-4\,g^2\,\int_0^\infty dt'\,\hat{K}(2gt,2gt')\,\oldtildesigma(t') \right.\\ \nonumber
\lefteqn{\left.~~~-4\,g^2\,\int_0^\infty dt'\,t'\,\hat{K}(2gt,2gt')\,\bigg(\hat{K}_{\rm h}(t',0;a)
-4\,\int_0^\infty dt''\,\hat{K}_{\rm h}(t',t'')\,\oldtildesigma(t'')\bigg)\right]}
\\
\eeqa
with $\hat{K}_{{\rm h}}(t,t';a)$ defined in \eqref{fdfrkern}. From the normalization condition for the hole density
\beq \label{allschleifenldnormierung}
\int_{-a}^{a} du\,\rho_{\rm h}(u)=1
\eeq
one derives the same relation as in the one-loop case \eqref{sigmaeinschleifennormierungft}
\beq \label{allschleifennormierung}
j=\frac{4a}{\pi}-\frac{16}{\pi} \int^{\infty}_{0} dt \, \hat{\sigma}(t) e^{\frac{t}{2}} \, \frac{\sin at}{t}\,.
\eeq

Finally, putting \eqref{allschleifennormierung} in \eqref{allschleifenigvsl} one finds the desired integral equation for the leading density in the generalized scaling limit \eqref{vskalenlimes}

\beq \label{endergebnis1}
\oldtildesigma(t)=\frac{t}{e^t-1}\left(\hat {\cal K}(t,0)-4\,\int^{\infty}_{0}dt'\,\hat {\cal K}(t,t')\,\oldtildesigma(t')\right)\,.
\eeq
The new kernel $\hat {\cal K}(t,t')$ is an intricate combination of the hole an magnon kernels
\beqa \label{endergebnis2}
\hat {\cal K}(t,t')&=&g^2\,\hat K(2gt,2gt')+\hat{K}_{\rm h}(t,t';a)-\frac{J_0(2gt)}{t}\,\frac{\sin{at'}}{2\,\pi\,t'}\,e^{\frac{t'}{2}}\nonumber\\
&&-4\,g^2\,\int_0^\infty dt''\,t''\,\hat K(2gt,2gt'')\,\hat{K}_{\rm h}(t'',t';a)\,.
\eeqa
It is interesting to note that equation \eqref{endergebnis1} takes structurally similar form to the BES equation \eqref{BESgleichung}. The leading anomalous dimension \eqref{adallschleifenmitsummen} may be expressed through $\oldtildesigma(t)$ with help of \eqref{ldzuallschleifensigma}, which in conjunction with \eqref{vsf} leads to the following expression
\beqa \label{vsfallschleifen}
\nonumber f^{\rm ABA}(g,j)&=&8\,g^2\,\left[1-8\,\int_0^\infty\,dt\,\frac{J_1(2gt)}{2gt}\,t\,\hat{K}_{\rm h}(t,0;a(j))\right.\\
\nonumber&-&8\,\left.\int_0^\infty\,dt\,\frac{J_1(2gt)}{2gt}\,\left(\oldtildesigma(t)-4\,t\,\int_0^\infty\,dt'\,\hat{K}_{\rm h}(t,t';a(j))\,\oldtildesigma(t')\right)\right]\\
&&=16\,\left(\oldtildesigma(0)+\frac{\LoverlogM}{16}\right).
\eeqa
Since the wrapping interactions may be neglected in the limit in question, the above formula proves the existence of the generalized scaling function \eqref{vsf} to all orders in perturbation theory. This novel function depends on two variables and thus it is possible to define two infinite families of scaling functions by expanding in $g$ or $j$ respectively
\beqa \label{finfnentwicklung}
f(g,\LoverlogM)=f(g)+\sum_{n=1}^\infty f^{(n)}(g)\,\LoverlogM^n=f_{(1)}(j)+\sum_{n=2}^\infty f_{(n)}(\LoverlogM)\,g^{2n-2}\,.
\eeqa
In particular, the function $f^{(1)}(g)$ coincides with \eqref{fendlich} after setting $L=j\,\log M$ and keeping the leading terms in the expansion only
\beqa
\nonumber f^{(1)}(g)&=&- f(g)\,\log 2+16\,g^4\,\zeta(3)-g^6\bigg(\frac{8}{3}\,\pi^2\,\zeta(3)+168\,\zeta(5)\bigg)\\
&&+g^8\left(\frac{8}{5}\,\pi^4\,\zeta(3)+\frac{88}{3}\,\pi^2\,\zeta(5)+1840\,\zeta(7)\right) +\ldots\,.
\eeqa
This suggests that the $L$ dependent part of \eqref{fendlich} is not influenced by the wrapping effects. Another curious observation is the vanishing of the $f^{(2)}(g)$ function to all orders in perturbation theory. An indication of such behavior on the string theory side was observed in \cite{Radu+Arkady}.

Interestingly, for $\LoverlogM \ll g$ it was conjectured in \cite{Alday:2007mf} that the generalized scaling function \eqref{vsf} can be computed using the  results of the ${O}(6)$ sigma model. Recently, it was confirmed \cite{Basso:2008tx} that the equation \eqref{endergebnis1} together with \eqref{allschleifennormierung} are equivalent to  the TBA equations of the non-linear ${O}(6)$ sigma model. Please refer to \cite{Gromov:2008en}-\cite{Bajnok:2008it} for the further development on this subject. 

\chapter{The Hubbard Model}
In this chapter we will explain an interesting relationship between the $\mathfrak{su}(2)$ asymptotic Bethe equations without the dressing factor and the spectral equations of a well-defined short-range integrable model, the Hubbard model. 
\section{The $\mathfrak{su}(2)$ Sector and the Hubbard Model}
The $\mathfrak{su}(2)$ sector is one of the simplest dynamical sectors in the $\N=4$ gauge theory. The operators of this sector are composed of two scalar fields only, which we will denote by $\fldZ$ and $\fldX$. Conventionally, we choose the reference vacuum to be
\beq
\vert \fldZ^L \rangle = \Tr \fldZ^L\,.
\eeq
Excited states are obtained by putting $\fldX$ fields into the trace
\beq
\Tr\lr{\fldX^M\,\fldZ^{L-M}}+\ldots \,,
\eeq
followed by the diagonalization of the $\mathfrak{su}(2)$ dilatation operator, which is a consistent truncation of the complete dilatation operator to this sector.
The Cartan weights of the primary state corresponding to the diagram \ref{fig:mulatadiagramm} are given by
\beq
\{\Delta_0, s_1, s_2, q_1, p, q_2, B, L\}=\{L,0,0,M,L-2M,M,0,L\}\,,
\eeq
from which, according to \eqref{allschleifenaz}, the following excitation pattern emerges
\beq
\{K_1, K_2, K_3, K_4, K_5, K_6, K_7\}=\{0,0,0,M,0,0,0\}\,.
\eeq
The corresponding all-loop asymptotic Bethe equations\footnote{In the literature these equations are known as the BDS equations \cite{Beisert:2004hm}.} take particularly simple form
\beq \label{allschleifensu2}
\left(\frac{x^+_k}{x^-_k}\right)^L=
\prod_{\substack{j=1,j \neq k}}^\M
\frac{u_k-u_j+i}{u_k-u_j-i}\,
\exp\left(2\,i\,\theta(u_k,u_j)\right)\,.
\eeq

At the one-loop order the above equations reduce to the spectral equations of the Heisenberg spin chain \eqref{XXX12bg} and consequently the dilatation operator of $\mathfrak{su}(2)$ may be identified with the Hamiltonian of the $XXX_{\frac{1}{2}}$ spin chain \eqref{HSKH}. The higher corrections to the dilatation operator were studied in \cite{Beisert:2003ys}, \cite{Beisert:2004hm} and \cite{Beisert:2007hz}, and under some assumptions were derived up to the five-loop order. It turns out that the $\ell$-loop dilatation operator acts simultaneously on the $(\ell+1)$ neighboring lattice sites. Moreover, these corrections become very complicated beyond the first few orders. Therefore it seems hopeless to guess its all-loop form solely from the perturbative expansion. In order to reveal the hidden connection to a different integrable model, we will investigate in detail the antiferromagnetic state of the $\mathfrak{su}(2)$ sector.\\

We start by noting that the S-matrix on the right hand side of \eqref{allschleifensu2}, upon neglecting the dressing factor, depends only on the difference of the rapidities. Since the dressing factor starts to contribute from the fourth-loop order, we will neglect it in what follows and discuss this approximation at the end of this section. The antiferromagnetic state is a maximally filled state. In the case of compact spin chains like $XXX_{\frac{1}{2}}$ this implies that $M_{\rm max}=\frac{L}{2}$ \footnote{Equations \eqref{allschleifensu2}, under the assumption that $g \ll 1$, may be considered as an all-loop deformation of the Heisenberg spin chain.}. The energy of this state in the case when the length of the chain becomes thermodynamically large was computed for the $XXX_{\frac{1}{2}}$ spin chain as early as in 1938 by Hulth\'en \cite{Hulthen}. He observed that the corresponding Bethe equations in the limit $L\to \infty$ can be written in the form of an integral equation, which in turn may be solved by Fourier transformation. Below we will apply this method to the deformed equations \eqref{allschleifensu2}. 

Taking the logarithm of the equations \eqref{allschleifensu2} and introducing the root denisty, one finds
\beq \label{rhosu2}
2\,\pi\,\rho(u)+ 2\,\int_{-\infty}^{\infty} du'\,
\frac{\rho(u')}{(u-u')^2+1}=i \frac{d}{du}\log\frac{x^+(u)}{x^-(u)}
\eeq
since in the $\rm Arctan$ branch of the logarithm the mode numbers are uniformly distributed. It follows directly from \eqref{uparametrisierung} that
\beq
i \frac{d}{du}\log\frac{x^+(u)}{x^-(u)}=\frac{i}{\sqrt{(u+i/2)^2-4g^2}}
-\frac{i}{\sqrt{(u-i/2)^2-4g^2}}\,.
\eeq
The integral kernel in \eqref{rhosu2} depends on the difference of the variables only, and the equation may thus be solved by Fourier transformation
\beq
\label{rhosu2explizit}
\rho(u)=
\int_0^{\infty}\,\frac{dt}{2 \pi}\,
\frac{\cos\left(t u\right)\,
J_0(2 g t)}{\cosh\left(\frac{t}{2}\right)}\, .
\eeq
The corresponding anomalous dimension may be found immediately from \eqref{allschleifenad}
\beqa \label{BDSad}
\frac{\gamma(g)}{2\,g^2\,L}&=&\int_{-\infty}^{\infty} du\, \rho(u)\,
\left(\frac{i}{x^+(u)}-\frac{i}{x^-(u)}\right)+\Op(L^{-1}) \nonumber \\
&=& 4
\int_0^{\infty}\,\frac{dt}{2 g t}\,
\frac{J_0(2 g t)\,J_1(2 g t)}{1+e^t}+\Op(L^{-1})\,.
\eeqa
It is now very interesting to note that equations \eqref{rhosu2explizit} and \eqref{BDSad} are well-known results of the solid state physics. They describe respectively the density and the energy of the antiferromagnetic state of the Hubbard model.  The integrability of this model was proven by Lieb and Wu in \cite{LW}, where also the corresponding Bethe equations were derived. 

The Hubbard model is a dynamical, short-range model of $N$ electrons on $L$ lattice sites. Due to the Pauli's exclusion principle, there are four possible states on each lattice site
\begin{enumerate}
\item no particles,
\item spin-up electron $\uparrow$,
\item spin down electron $\downarrow$ and
\item double occupied state with spin-up and spin-down electrons $\updownarrow:=\uparrow \downarrow$.
\end{enumerate}
In what follows, we will consider the half-filled case $N=L$. The Hamiltonian of the Hubbard model consists of the kinetic part that forces the electrons to jump between different sites and the potential part, which according to the value of $U$ corresponds to repulsive or attractive force
\beq
\label{H}
\hat{H}_{{\rm Hubbard}}=
-t\, \sum_{i=1}^L \sum_{\sigma=\uparrow,\downarrow}
\left(c^\dagger_{i,\sigma} c_{i+1,\sigma}+
c^\dagger_{i+1,\sigma} c_{i,\sigma}\right)+
t\,U\, \sum_{i=1}^L
c^\dagger_{i,\uparrow} c_{i,\uparrow}c^\dagger_{i,\downarrow} c_{i,\downarrow}\, .
\eeq
The operators $c^\dagger_{i,\sigma}$ and $c_{i,\sigma}$ are canonical Fermi operators obeying
\beqa \label{ccanti}
\{c_{i,\sigma},c_{j,\tau}\}&=&
\{c^\dagger_{i,\sigma},c^\dagger_{j,\tau}\}=0\, ,
\\ \label{ccdanti}
\{c_{i,\sigma},c^\dagger_{j,\tau}\}&=&
\delta_{i j}\,\delta_{\sigma \tau}\, .
\eeqa
We assume the system to be closed and thus we identify
\beq
c_{L+1,\sigma}=c_{1,\sigma}, \qquad \qquad c^\dagger _{L+1,\sigma}=c^\dagger _{1,\sigma}
\eeq
for $\sigma=\uparrow, \downarrow$. The Hamiltonian is invariant with respect to the $\mathfrak{su}(2)$ transformations
\beq
[\hat{H}_{\rm Hubbard},\hat{S}^{\,a}]=0 \qquad \qquad a=+,-,z \,,
\eeq
with $\hat{S}^{\,a}=\sum^{L}_{i=1} \hat{S}^{\,a}_i$. This allows to classify the spectrum according to the eigenvalues of the total spin an its $z$ component.\\

The precise correspondence between \eqref{rhosu2explizit}, \eqref{BDSad} and the corresponding results of the Hubbard model is established under following identification of the parameters
\beq \label{tUident}
t=-\frac{1}{2\,g}\,,
\qquad \qquad
U=\frac{1}{g}\, .
\eeq
This, however, suggests that perhaps all states of the BDS spin chain are up to the wrapping order also eigenstates of the Hubbard model Hamiltonian. It turns out that in the case of odd values of $L$ the BDS equations indeed diagonalize the Hamiltonian \eqref{H}, while for the even values the kinetic terms of the Hamiltonian need to be multiplied by additional phase factors. This is equivalent to coupling the Hubbard model to a homogeneous magnetic field and the aforementioned phase factors may be then considered as the Aharonov-Bohm phases
\beq \label{HABstrahl}
\hat{H}=
\frac{1}{2\,g}\, \sum_{j=1}^L \sum_{\sigma=\uparrow,\downarrow}
\left(e^{i \phi_\sigma}\,c^\dagger_{j,\sigma} c_{j+1,\sigma}+
e^{-i\phi_\sigma}\,c^\dagger_{j+1,\sigma} c_{j,\sigma}\right)-
\frac{1}{2g^2}\, \sum_{j=1}^L
c^\dagger_{j,\uparrow} c_{j,\uparrow}c^\dagger_{j,\downarrow} c_{j,\downarrow}\, ,
\eeq
with
\beqa \label{twist}
&\ & \phi_\uparrow=\phi_\downarrow=\phi\,, \qquad  \qquad \phi=\frac{\pi}{2L} \ \bigg\{\lr{L-1} \ \rm{mod} \ 2\bigg\}\,.
\eeqa
The perturbative expansion at small values of $g$ corresponds accordingly to the strongly coupled ($U \gg 1$) Hubbard model. It was shown in \cite{Rej:2005qt} that the perturbative expansion of the Hamiltonian \eqref{HABstrahl} at large values of $U=1/g$ coincides up to three orders in $g^2$ with the perturbative expansion of the dilatation operator found in \cite{Beisert:2004hm}. We will not repeat this calculation here, but instead show in the next section that the BDS equations \eqref{allschleifensu2} can be derived in the asymptotic region from the Lieb-Wu equations.

The vacuum state of the Hubbard model is annihilated by the $c_{i,\uparrow}, c_{i,\downarrow}$ operators and is a tensor product of $L$ empty lattice sites
\beq
\vert 0 \rangle_L = \underbrace{\vert 0 \rangle \otimes \vert 0 \rangle \otimes \ldots \vert 0 \rangle}_{L} \,.
\eeq
The elementary excitations on this vacuum are the spin-up and spin-down electrons. On the other hand, the double occupation must be considered as a composition of the elementary excitations. The interaction between the constituents of such a composed state is according to \eqref{HABstrahl} repulsive. Since we confine ourselves to the half-filled case, it is more convenient to choose the BPS vacuum as the reference vacuum
\beq
\label{BPS}
|Z^L\rangle = |\uparrow \uparrow \ldots \uparrow \uparrow \rangle=
c^\dagger_{1\uparrow} c^\dagger_{2\uparrow} \ldots
c^\dagger_{L-1 \uparrow} c^\dagger_{L \uparrow}\,|0\rangle_L \,.
\eeq
It is easy to show that this state is also a zero energy state of \eqref{HABstrahl}. The disadvantageous feature of the vacuum \eqref{BPS}, that is the fact that it is not annihilated by the annihilation operators, may be easily overcome by performing a particle-hole transformation
\beqa
\label{particlehole}
\circ &\Longleftrightarrow& \uparrow\,,
\\
\downarrow &\Longleftrightarrow& \updownarrow\,.
\eeqa
After this transformation the spin-up electrons are considered to be empty lattice sites, while the empty lattice sites become double occupied states. Explicitly, we identify
\beqa \label{Shiba}
c_{j,\circ}=c^\dagger_{j,\uparrow}\;, &\ &
\qquad c^\dagger_{j,\circ}=c_{j,\uparrow}\;,
\\
c_{j,\updownarrow}= c_{j,\downarrow}\;, &\ &
\qquad c^\dagger_{j,\updownarrow}= c^\dagger_{j,\downarrow}\;.
\eeqa
The algebraic relations between $c^\dagger _{j,\circ}, c_{j,\circ}, c^\dagger _{j,\updownarrow}$ and  $c _{j,\updownarrow}$ follow directly from the anitcommutation relations \eqref{ccanti} and \eqref{ccdanti}. Accordingly, the dual Hamiltonian can be written as 
\beq
\label{Hdual}
\hat{H}=
\frac{1}{2g}\, \sum_{j=1}^L \sum_{\sigma=\circ,\updownarrow}
\left(e^{i\phi_\sigma}\,c^\dagger_{j,\sigma} c_{j+1,\sigma}+
e^{-i\phi_\sigma}\,c^\dagger_{j+1,\sigma} c_{j,\sigma}\right)-
\frac{1}{2g^2}\, \sum_{j=1}^L
(1-c^\dagger_{j,\circ} c_{j,\circ})c^\dagger_{j,\updownarrow} c_{j,\updownarrow}\,,
\eeq
with the phases $\phi_\updownarrow=\phi_\downarrow$ and $\phi_\circ=\pi-\phi_\uparrow$ being different from the one for the original exciations. Comparing \eqref{HABstrahl} and \eqref{Hdual} one finds that performing the particle-hole transformation amounts to the substitution
\beq
\label{dualrelation}
 \hat{H}(g;\phi,\phi) \to -\hat{H}(-g;\pi-\phi,\phi)-\frac{M}{2g^2}\,,
\eeq
where $M$ denotes the number of the double occupancies. The sign in front of the potential part of the Hamiltonian \eqref{Hdual} is opposite to the one in \eqref{HABstrahl}, which means that the holes and the double occupancies attract each other and form bound states, the spin down electrons.

\section{The Lieb-Wu Equations}
The Hamiltonian \eqref{H}, as shown in \cite{LW}, is integrable and can be diagonalized with the help of the Bethe ansatz. The same holds for Hamiltonians with arbitrary Aharonov-Bohm phases, which were studied in \cite{YueDeguchi}. In the case of half-filling the Bethe equations corresponding to \eqref{HABstrahl} are given by (see \cite{YueDeguchi})
\beqa
\label{liebwu1}
&\ &e^{i\tilde q_nL}=\prod_{j=1}^M
\frac{u_j-2g\sin (\tilde q_n+\phi) -i/2}
{u_j-2g\sin  (\tilde q_n+\phi)+i/2}\, ,
\qquad n=1,\ldots, L
\\
\label{liebwu2}
&\ &\prod_{n=1}^{L}
\frac{u_k-2g\sin  (\tilde q_n+\phi) +i/2}
{u_k-2g\sin  (\tilde q_n+\phi)- i/2}=
\prod_{j=1,j\neq k}^M
\frac{u_k-u_j +i}{u_k-u_j-i}\, ,
\quad k=1,\ldots, M \,.
\eeqa
Here, the parameter $M$ denotes the number of the spin-down electrons. The energy of a state is a function of all $q_n$
\beq
\label{liebwueng}
E=\frac{1}{g}\;
\sum_{n=1}^{L}\cos  (\tilde q_n+\phi)\;.
\eeq
In the limit $g\to 0$ equations \eqref{liebwu1} and \eqref{liebwu2} become the momentum constraint (after setting $e^{i\tilde q_n L}=1$) and the Bethe equations of the $\mathfrak{su}(2)$ spin chain respectively. The $1/g$ correction to the energy \eqref{liebwueng} is zero due to
\beq
e^{i\tilde q_nL}=1
\qquad \Longrightarrow \qquad
\tilde q_n=\frac{2 \pi}{L} (n-1) + {\cal O}(g), \quad n=1,\ldots, L\, .
\eeq
To obtain the $\Op(g^0)$ term one must determine the $\Op(g)$ corrections to the roots $\tilde q_n$. Since the one-loop roots are known, this is merely a linear problem. Upon solving \eqref{liebwu1} to the order $\Op(g)$, one finds
\beq
\label{XXXeng}
E=\sum_{k=1}^M\,\frac{1}{u_k^2+\frac{1}{4}}\,+{\cal O}(g)\, ,
\eeq
which is the expression for the energy of the XXX spin chain, \textit{cf.} \eqref{Energie}. Knowing the one-loop solutions $\tilde q_n$ and $u_k$, one can solve equations \eqref{liebwu1} and \eqref{liebwu2} perturbatively. In general, for $1 \ll L < \infty$, it is very complicated to solve the Bethe equations exactly. On the other hand, the BDS equations are strictly speaking only valid for $L \to \infty$. Therefore it is convenient to dualize equation \eqref{liebwu1}. Introducing $x=e^{i \tilde q_n}$ one concludes that the equation \eqref{liebwu1} in its polynomial form has in total $L+2M$ solutions. Using the remaining $2M$ roots, one can rewrite \eqref{liebwu1} and \eqref{liebwu2} as
\beqa
\label{liebwudual1}
&\ &e^{i q_nL}=\prod_{j=1}^M
\frac{u_j-2g\sin (q_n-\phi) -i/2}
{u_j-2g\sin  (q_n-\phi)+i/2}\, ,
\qquad n=1,\ldots, 2 M
\\
\label{liebwudual2}
&\ &\prod_{n=1}^{2M}
\frac{u_k-2g\sin  ( q_n-\phi) +i/2}
{u_k-2g\sin  (q_n-\phi)- i/2}=
-\prod_{j=1,j\neq k}^M
\frac{u_k-u_j +i}{u_k-u_j-i}
\quad k=1,\ldots, M\,.
\eeqa
The energy in terms of the dual roots can be found to be
\beq \label{liebwudualeng}
E=-\frac{M}{2g^2}-\frac{1}{g}\;
\sum_{n=1}^{2M}\cos  (q_n-\phi)\;.
\eeq
Comparing \eqref{liebwudualeng} to \eqref{liebwueng} together with \eqref{dualrelation} one infers that the dual solutions $q_n, n=1,\ldots 2M$ diagonalize the dual Hamiltonian.

The advantage of the equations \eqref{liebwudual1} and \eqref{liebwudual2}, as opposed to \eqref{liebwu1} and \eqref{liebwu2}, is the independence of the former on the length $L$. Since the elementary dual excitations attract each other, it is natural to assume that bound states ($\downarrow=\circ\,+\updownarrow$) will be energetically favored. Usually, the bound states manifest theirselves in the exponential localization of the wave function. Therefore, for the lowest states in the energy band we make the following ansatz for the $2M$ dual roots\footnote{The separation of the $\frac{\pi}{2} {\rm sgn} (p_n)$ factor is made in order to simplify the calculations. The variable $p_n$ will correspond to the momentum of a magnon $p_n \in (-\pi, \pi)$ so that $(q_n-\phi) \in (\frac{\pi}{2}, \frac{3 \pi}{2})$. Interestingly, there exist no solutions of \eqref{liebwudual1} and \eqref{liebwudual2} with $(q_n- \phi)\in (-\pi/2, \pi/2)$. Such solutions would correspond to magnons with negative energies.}

\beqa \label{qn1}
 &\ & q_n- \phi =\frac{\pi}{2} {\rm sgn} (p_n)+\frac{p_n}{2}+i \beta_n\;, \\ \label{qn2}
&\ &  q_{n+M}- \phi =\frac{\pi}{2} {\rm sgn} (p_n)+\frac{p_n}{2}-i\beta_n\;,
\quad \beta_n>0\;, \quad n=1,\ldots ,M\,, \nonumber
\eeqa
with $p_n \in \mathbb{R}$ denoting the bound state momentum. Substituting \eqref{qn1} and \eqref{qn2} into \eqref{liebwudual1} one finds that when $L \gg 1$ the left hand side of this equation vanishes like $e^{-\beta_n L}$ for $n=1,\ldots, M$ and diverges like $e^{\beta_n L}$ for $n=M+1,\ldots,2M$.  This is true independently of the value of $M$, and therefore the following relations must hold up to exponential corrections 
\beq \label{unun}
 u_n-i/2 =2g \sin ( q_n -\phi) +\Op(e^{-\beta_n L})\;, \qquad  u_n+i/2 =2g\,
\sin (  q_{n+M} -\phi)+\Op(e^{-\beta_n L})\,.
\eeq
Putting \eqref{unun} into the equation \eqref{liebwudual2} one finds that the latter is trivially satisfied. The both conditions \eqref{qn1} and \eqref{qn2} may be compactly written as
\beq \label{unzupnbn}
u_n\pm i/2=2 g\; {\rm sgn} (p_n) \cos ( \frac{p_n}{2} \mp i\beta_n)+\Op(e^{-\beta_n L})\;.
\eeq
Solving \eqref{unzupnbn} for $p_n$ and $\beta_n$ results in
\beq
 \label{betapn}
  \sinh \beta_n=\frac{1}{4 g \; {\rm sgn} (p_n) \sin \frac{p_n}{2}}=
\frac{1}{4 g\; | \sin \frac{p_n}{2}|}\;,
\eeq
\beq
  u_n=2 g\; {\rm sgn} (p_n)\, \cos \frac{p_n}{2}\,\cosh \beta_n=
\frac{1}{2}\cot \frac{p_n}{2}\; \sqrt{1+16g^2\sin^2 \frac{p_n}{2}}\;.
\eeq
The second relation is equivalent to the all-loop dispersion relation \eqref{allschleifenbg}
\beq \label{dispersionsrelation}
e^{i p}= \frac{x^{+}(u)}{x^{-}(u)}
\eeq
In order to eliminate $q_n$ from \eqref{liebwudual1} one substitutes the relation \eqref{unun} into \eqref{liebwudual1} and subsequently multiplies the equation for $u_n$ with the equation for $u_{n+M}$. The divergencies appearing for $L \to \infty$ cancel each other and the finite part can be written as
\beq \label{bdsbis}
 e^{i(p_n+2 \phi+s_n\pi)L}=
-\prod_{j=1,j\neq n}^M\frac{u_n-u_j+i}{u_n-u_j-i}\;.
\eeq
Using \eqref{twist} and \eqref{dispersionsrelation} one confirms that \eqref{bdsbis} is equivalent to the BDS equations \eqref{allschleifensu2}. The energy \eqref{liebwudualeng} reads
\beqa \label{BDSeng}
E&=&-\frac{1}{g} \sum_{n=1}^M\left(\cos(q_n- \phi)+
\cos(q_{n+M}- \phi)\right)-\frac{M}{2g^2}\\ \nonumber
&=&\frac{2}{g}\sum_{n=1}^M\left |
\sin\frac{p_n}{2} \right |
\cosh \beta_n-\frac{M}{2g^2}
=\sum_{n=1}^M \lr{\frac{1}{x^{+}(u_n)}-\frac{1}{x^{-}(u_n)}}=\frac{\gamma(g)}{2g^2}\,,
\eeqa
where we have made use of \eqref{betapn} and \eqref{dispersionsrelation}. The above construction can be generalized to the case of complex momenta $p_n$. The argument of the sigma function in \eqref{qn1} and \eqref{unzupnbn} must then be replaced by the real part of $p_n$.\\

For the purpose of derivation of the BDS equation we have assumed that the holes $\circ$ and the double occupancies $\updownarrow$ form bound states. Though this is true for the low-energy states, there exist states for which one or more bound states are split up and the corresponding momenta are real. Their presence may be also noted by a simple counting. In the Hubbard model there are roughly $4^L/L$ cyclic states, while in the Heisenberg or BDS spin chain the number of cyclic states amounts to $2^L/L$. This discrepancy is explained by the existence of the excited states. One should note, however, that the energies of these states are according to  \eqref{liebwudualeng} non-perturbative
\beq
E \sim \frac{1}{g^2}\,,
\eeq
and therefore the corresponding operator cannot be defined in the perturbative gauge theory.  The accuracy of the equation \eqref{bdsbis} is related to the accuracy of the solution \eqref{unun}. It follows immediately from \eqref{betapn} that for perturbative values of $g$
\beq \label{betazug}
\beta_n \simeq -\log g + \Op(g^0)\,,
\eeq
and thus the solutions \eqref{unun} receive corrections starting from the order $e^{-\beta_n L}\simeq g^{L}$. The energy \eqref{BDSeng}, on the other hand, due to the expansion
\beq
\cos (x+\epsilon)=\cos (x)-\sin (x) \epsilon-\frac{1}{2} \cos (x) \epsilon ^2+\Op(\epsilon^3)\,,
\eeq
and the fact that $q_n$ and $q_{n+M}$ are complex conjugated to each other, leeds to a correct result up to the order $\Op(g^{2L-2})$. According to the relation $\gamma(g)=2\,g^2\,E(g)$ this corresponds precisely to the wrapping order. This is a strong indication that there exist an integrable short range model, which correctly captures the wrapping interactions and in the asymptotic region leeds to the asymptotic Bethe equations \eqref{allschleifensu2}. It should be noted, however, that the usual Hubbard model may not be a candidate for such model since it leads to the trivial dressing factor $\sigma(u,v)=1$, \textit{cf.} \eqref{bdsbis}. This contradicts both the proposed crossing symmetry \cite{Janik:2006dc} ($\sigma(u,v)=1$ does not satisfy the crossing equation) and the explicit perturbative calculation \cite{Beisert:2007hz}. It is, however, likely that a suitable deformation of the Hubbard model would lead to the ``dressed'' BDS equations.

The hypothesis that the dilatation operator of the $\N=4$ SYM theory is equivalent to a Hamiltonian of a short-range integrable model and that the observed long-rangeness is only an artifact of the perturbative expansion is very appealing, although several issues need still to be understood. Most importantly, it is not known whether such a short-range system with both bosonic and fermionic elementary excitations can be found since, in general, unlimited number of bosons can occupy each lattice site preventing the factorization of the scattering into a sequence of two-body processes. Another complication is the presence of the dressing phase, which has a very complicated transcendental structure, see \cite{Beisert:2006ez}. 

It is also interesting to note, as it was found in \cite{Beisert:2006qh}, that the $S_{\suzz}$ S-matrix may be identified with the R-matrix of the Hubbard model. However, this observation does not seem to be related with the context in which the Hubbard model was introduced in this section. 

\chapter*{Acknowledgements}
First of all, I would like to express my gratitude to my PhD advisor Matthias Staudacher for his constant interest and many interesting discussions. Without his encouragement and support this work would not have been possible. I would also like to thank to Lisa Freyhult, Anatoly Kotikov, Lev Lipatov, Didina Serban, Fabian Spill, Arkady Tseytlin,  Vitaly Velizhanin  and Stefan Zieme for the fruitful collaboration on the projects on which this review is based in part.
\newpage

\end{document}